\documentclass[pra,twocolumn,superscriptaddress,nofootinbib,noshowpacs,preprintnumbers,longbibliography,floatfix]{revtex4-2}
\usepackage[utf8]{inputenc}
\usepackage{graphicx}
\usepackage{float}
\usepackage{amssymb}
\usepackage{amsmath}  
\usepackage{adjustbox}
\usepackage{mathtools}
\usepackage{dsfont}
\usepackage{array}
\usepackage{bm,fixmath}
\usepackage{mathrsfs}
\usepackage{pifont}
\usepackage{enumitem}
\usepackage{multirow}
\usepackage{upgreek}
\usepackage{subfigure}
\usepackage{xcolor}
\usepackage{bm}
\usepackage{bbm}
\usepackage{physics}
\usepackage{comment}
\usepackage{tikz}
\usetikzlibrary{quantikz}
\usepackage[colorlinks = true,
            linkcolor = black,
            urlcolor  = blue,
            citecolor = violet,
            anchorcolor = blue]{hyperref}

\usepackage{orcidlink} 
\begin{document}

\title{Analysis of the confinement string in \texorpdfstring{$(2+1)$}{(2+1)}-dimensional Quantum Electrodynamics with a trapped-ion quantum computer
}

\author{Arianna~Crippa~\orcidlink{0000-0003-2376-5682}}
\email{arianna.crippa@desy.de}
\affiliation{Deutsches Elektronen-Synchrotron DESY, Platanenallee 6, 15738 Zeuthen, Germany
}
\affiliation{Institut für Physik, Humboldt-Universität zu Berlin, Newtonstr. 15, 12489 Berlin, Germany}
\author{Karl~Jansen~\orcidlink{0000-0002-1574-7591}}
\email{karl.jansen@desy.de}
\affiliation{
 Computation-Based Science and Technology Research Center, The Cyprus Institute, 20 Kavafi Street,
2121 Nicosia, Cyprus
}
\affiliation{Deutsches Elektronen-Synchrotron DESY, Platanenallee 6, 15738 Zeuthen, Germany
}

\author{Enrico~Rinaldi~\orcidlink{0000-0003-4134-809X}}
\email{enrico.rinaldi@quantinuum.com}
\affiliation{Quantinuum K.K.,\\
Financial City Grand Cube 3F,
1-9-2 Otemachi,Chiyoda-ku, Tokyo, Japan}
\affiliation{Interdisciplinary Theoretical and Mathematical Sciences (iTHEMS) Program, RIKEN,\\
Wako, Saitama 351-0198, Japan}
\affiliation{Center for Quantum Computing (RQC), RIKEN,\\ Wako, Saitama 351-0198, Japan}
\affiliation{Theoretical Quantum Physics Laboratory, Cluster of Pioneering Research, RIKEN\\
Wako, Saitama 351-0198, Japan}

\date{\today}

\begin{abstract}
Compact lattice Quantum Electrodynamics is a complex quantum field theory with dynamical gauge and matter fields and it has similarities with Quantum Chromodynamics, in particular asymptotic freedom and confinement.
We consider a (2+1)-dimensional lattice discretization of Quantum Electrodynamics with the inclusion of dynamical fermionic matter.
We define a suitable quantum algorithm to measure the static potential as a function of the distance between two charges on the lattice and we use a variational quantum calculation to explore the Coulomb, confinement and string breaking regimes.
A symmetry-preserving and resource-efficient variational quantum circuit is employed to prepare the ground state of the theory at various values of the coupling constant, corresponding to different physical distances, allowing the accurate extraction of the static potential from a quantum computer.
We demonstrate that results from quantum experiments on the Quantinuum H1-1 trapped-ion device and emulator, with full connectivity between qubits, agree with classical noiseless simulations using circuits with 10 and 24 qubits.
Moreover, we visualize the electric field flux configurations that mostly contribute in the wave-function of the quantum ground state in the different regimes of the potential, thus giving insights into the mechanisms of confinement and string breaking.
These results are a promising step forward in the grand challenge of solving higher dimensional lattice gauge theory problems with quantum computing algorithms.
\end{abstract}

\maketitle

\section{Introduction\label{sec:intro}}

One of the most prominent examples of non-perturbative physics is the confinement of constituent particles in gauge theories. 
In fact, this has been one of the main motivations for Wilson to introduce lattice gauge theories (LGT)s~\cite{Wilson:1974sk}, see e.g. Refs.~\cite{rothe2012lattice,gattringer2009quantum} for introductions to LGT.
In Quantum Chromodynamics (QCD)~\cite{Gross:2022hyw} the confinement phenomenon is responsible for the binding of quarks and gluons into hadrons at low energies (large distances). 
Confinement between static charges also plays a vital role in Quantum Electrodynamics (QED) in $(2+1)$ dimensions, where it is related to the physics of instantons, as first discussed in Ref.~\cite{Polyakov:1976fu}.
For an overview we refer to Ref.~\cite{Wen:2004book}. 

A very interesting situation arises when a confining gauge theory is coupled to matter fields.
When the energy of the confining string becomes too large, it is energetically more favourable to form heavy-light meson states between a static charge and a particle excitation from the dynamical matter field.
This is the celebrated phenomenon of {\em string breaking} and it has been studied in LGT with Monte Carlo (MC) from the pioneering work of Refs.~\cite{Knechtli_2000,Philipsen:1998de}.
Ref.~\cite{Bali:2000gf} is a review of LGT studies of string breaking in the action formulation.

\begin{figure}[htp!]
    \centering
    \includegraphics[width=1\columnwidth]{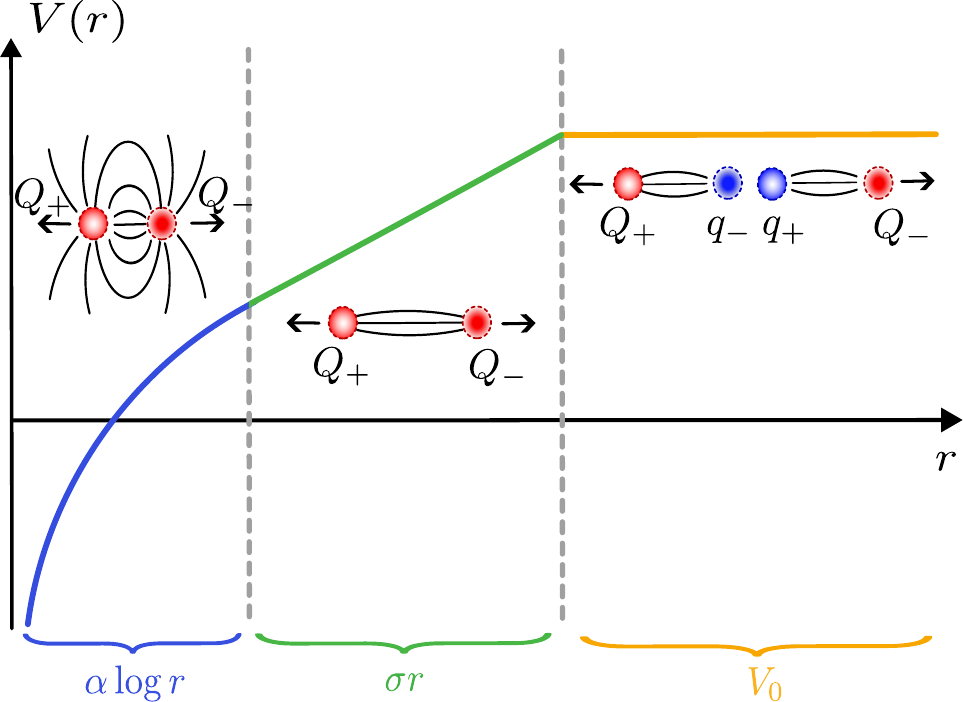}
    \caption{\textbf{Static potential of two charges:} The expected behavior of the static potential $V(r)$ for this model as a function of the distance of two static charges, $r$. For small $r$ (\textit{blue section}) there is a Coulomb potential. Then, an electric flux tube forms between the charges when the distance increases (\textit{green section}) dominating the potential in this regime. At a certain $r$ the flux tube breaks and a new pair of charge/anticharge forms (\textit{orange section}), and hence the linear part of the potential is not continuing. This is the qualitative non-perturbative picture of the transition between confinement and charge screening in QED with light matter fields, and it is similar to the one in QCD in $(3+1)$ dimensions.}
    \label{fig:vrplot}
\end{figure}

To be more concrete, in $(2+1)$ dimensions the QED static potential between two static charges at distance $r$ has a Coulomb logarithmic term, a confining linear part and a string breaking regime~\cite{loan2003path,PhysRevResearch.2.043145}:
\begin{equation}
    V(r)=V_0 +\alpha \log r + \sigma r ,
\end{equation}
where $\alpha$ is the coupling, $\sigma$ the \textit{string tension} and $V_0$ refers to a constant term.
The form of the static potential is illustrated in Fig.~\ref{fig:vrplot}.
At small $r$, $V(r)$ is a logarithm function representing the Coulomb potential in two space dimensions.
The coupling which determines the strength of the Coulomb potential becomes perturbative with decreasing distance, due to asymptotic freedom.
At intermediate distances, the electric field between a pair of static charges forms a flux tube (or string) between them, leading to a linear behaviour of the potential as a function of the distance and hence to confinement of the static charges.
However, when dynamical matter fields are included, the linear potential does not extend to indefinitely large distances.  
For sufficiently large separations it is energetically favourable to pair-produce a particle and antiparticle with opposite charges, thereby \emph{breaking} the string.
The static charges are screened by the dynamical matter fields and now bound in heavy-light mesons.

With the advent of quantum technologies, and quantum computing in particular, the study of confinement, and in general lattice gauge theories, has become one of the exciting areas of discovery and development.
In recent years, lower dimensional LGTs are helping to explore the potential of applying quantum computing to high energy physics, to develop quantum algorithms and are opening new ways of computations to tackle physical problems, see the reviews in Refs.~\cite{banuls2020simulating,Bauer:2022hpo,PRXQuantum.5.037001,Funcke:2023jbq}.
For example, the phenomenon of string breaking has been considered in the context of quantum simulators in Refs.~\cite{PhysRevLett.109.175302,wiese2013ultracold,zohar2015quantum}, using tensor networks in Refs.~\cite{magnifico2021lattice,zohar2011confinement,cochran2024visualizingdynamicschargesstrings} and, more recently using quantum hardware in Refs.~\cite{De:2024smi,cochran2024visualizingdynamicschargesstrings,gonzalezcuadra2024observationstringbreaking2}.

While the focus in these latter papers is more on the dynamics of the string in some specific LGTs, in our work we will study static flux configurations in the different regimes of the static potential as well as 
the probability of the states contributing at different bare gauge couplings.
Moreover, our work is the first one addressing the static potential of QED in $(2+1)$ dimensions by preparing the ground state of the QED Hamiltonian at multiple couplings across a variety of distances.
In this work, we perform a qualitative analysis of the static potential, in the regimes described above, by considering QED in $(2+1)$ dimensions and a quantum computing approach with ion-trap devices.
We consider two fixed static charges and we increase the electric energy between them by varying the bare coupling $g$.
This is applicable since the lattice spacing $a$, defining the physical distance $r_{\rm ph}=a r_{\rm latt}$, is implicitly and non-perturbatively determined by the coupling, $g=g(a)$.
Thus, by changing $g$, we change the physical distance $r_{\rm ph}(g)$, scanning the static potential $V(r_{\rm ph})=V(r)$.

One advantage we want to point out is our ability to visualize the electric fluxes appearing in the ground state and thus obtain direct information about the flux configurations in the different regimes of the static potential. 
We observe this phenomenon experimentally on the quantum computer H-series System Model H1-1 at Quantinuum~\cite{quantinuum_H1_1}.
This not only allows to achieve new insights in the physics of the considered model but also sets the basis for future quantum analyses on this interesting topic and also showing the precision of the results with ion-trapped devices.

\vspace{0.2cm}

The paper is structured as follows: in Sec.~\ref{sec:hamilt} we introduce the QED Hamiltonian for a $(2+1)$-dimensional ($(2+1)$-d) lattice and give a description of the truncation technique for the gauge fields. 
Sec~\ref{sec:encoding} defines the encoding of gauge fields and fermionic degrees of freedom for numerical calculations and describes the parameterized circuits developed for the variational quantum computation.
We also give a concise description of the Quantinuum ion-trap hardware and the noise mitigation techniques applied in the analysis. 
In Sec.~\ref{sec:res3x2}, we discuss the results of the static potential at different couplings for a $3 \times 2$ lattice (10 qubits), and in Sec.~\ref{sec:4x3system} we consider a larger system, with $4 \times 3$ fermionic sites (24 qubits). 
In Sec.~\ref{sec:conclusion}, we report our summary and conclusions, and we give an outlook on future possible extensions of this work.
Appendix~\ref{app:quantumcirc} shows the quantum circuit developed for the $3 \times 2$ system with a brief description of the mutual information and how it was considered to build the circuit entanglement.
In Appendix~\ref{app:noisemodel} we consider an analysis of the computation of the static potential with a selected range of shots (measurements of the quantum circuit) on the emulator H1-1E.
We study the dependence on the truncation applied to the gauge fields and the effects when applying Gauss's law in Appendix~\ref{app:truncation}.
Lastly, in Appendix~\ref{app:4x3circuit} we give the explicit form of the parameterized quantum circuit for the $4\times 3$ lattice.

\begin{figure}[htp!]
    \centering
\includegraphics[width=0.8\columnwidth]{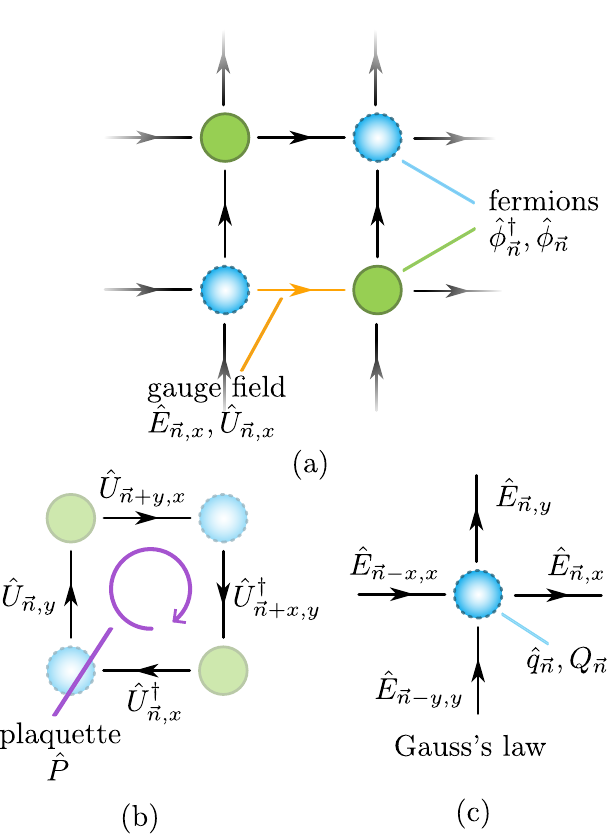}
    \caption{\textbf{Lattice structure for $(2+1)$-dimensional QED:} (\textit{panel (a)}) The gauge fields live on the links which connect the fermionic sites on the lattice. (\textit{panel (b)}) The plaquette operator is the product of four link operators $\hat{P}_{\vec{n}} =\hat{U}_{\vec{n},y} \hat{U}_{\vec{n}+y,x} \hat{U}^{\dag}_{\vec{n}+x,y} \hat{U}_{\vec{n},x}^{\dag}$. (\textit{panel (c)}) Gauss's law for the fermionic site $\vec{n}$ control the balance between ingoing/outgoing electric field and dynamical charge $\hat{q}_{\vec{n}}$ (and eventual static charge $Q_{\vec{n}}$), Eq.~\eqref{gausslaw}. }
    \label{fig:lattices_p_gauss}
\end{figure}

\section{QED Hamiltonian}
\label{sec:hamilt}

In this work, we consider a lattice discretization of $(2+1)$-d QED using Kogut-Susskind staggered fermions~\cite{PhysRevD.11.395,robson1980gauge,ligterink2000many}. 
This formulation has been introduced in order to deal with the so-called \textit{doubling problem}~\cite{rothe2012lattice,Nielsen:1981hk,PhysRevD.16.3031},
i.e. an incorrect continuum limit of the theory, that arises with a naive lattice discretization of the fermionic degrees of freedom.

The spinor components are distributed on different lattice sites, thus excluding the additional (unphysical) degrees of freedom. 
In Fig.~\ref{fig:lattices_p_gauss}a we depict the basic components that build the lattice structure for $(2+1)$-dimensional QED. In particular, we describe how the gauge and fermionic degrees of freedom are represented on the lattice. The fermions are onto the sites (\textit{dashed circles} describe matter fields, \textit{solid circles} antimatter fields), while gauge fields are the links connecting the sites (\textit{arrows}).
The Hamiltonian can be written as,
\begin{subequations}\label{eq:fullH}
\begin{align}
\hat{H}_\text{QED}&= \frac{g^{2}}{2} \sum_{\vec{n}}\left(\hat{E}^{2}_{\vec{n}, x} 
+ \hat{E}^{2}_{\vec{n}, y}\right)\label{eq:hel}\\
&-\frac{1}{2a^2g^{2}} \sum_{\vec{n}} \left(\hat{P}_{\vec{n}} + \hat{P}_{\vec{n}}^{\dag}
    \right)\label{eq:hb}\\
   &+m \sum_{\vec{n}} (-1)^{n_x+n_y} \hat{\phi}^\dag_{\vec{n}} \hat{\phi}_{\vec{n}}\label{eq:hmass}\\
    &+\frac{i}{2a}\sum_{\vec{n}}(\hat{\phi}_{\vec{n}}^{\dagger}\hat{U}^{\dagger}_{\vec{n},x}\hat{\phi}_{\vec{n}+x} - \text{h.c.}) \nonumber\\ 
        &-\frac{(-1)^{n_x+n_y}}{2a}\sum_{\vec{n}}(\hat{\phi}_{\vec{n}}^{\dagger}\hat{U}^{\dagger}_{\vec{n},y}\hat{\phi}_{\vec{n}+y} + \text{h.c.}).\label{eq:hkin}
\end{align}
\end{subequations}
The electric energy in Eq.~\eqref{eq:hel}, is built with $\hat{E}_{\vec{n}, \mu}$, the dimensionless electric field operator that acts on the link with initial coordinates $\vec{n}=(n_x,n_y)$ and direction $\mu\in\{x,y\}$. 
The bare coupling $g$, is also present in the magnetic term, Eq.~\eqref{eq:hb}. Here, the \textit{plaquette operator} $\hat{P}_{\vec{n}} =\hat{U}_{\vec{n},y} \hat{U}_{\vec{n}+y,x} \hat{U}^{\dag}_{\vec{n}+x,y} \hat{U}_{\vec{n},x}^{\dag}$, Fig.~\ref{fig:lattices_p_gauss}b, defines the strength of the interaction (with the notation $\vec{n}+x\equiv (n_x+1,n_y)$ or $\vec{n}+y\equiv (n_x,n_y+1)$). 
The unitary link operators $\hat{U}_{\vec{n},\mu}$ represent the gauge connection between the fermionic fields and are related to the discretized vector field $\hat{A}_{\vec{n},\mu}$ as
\begin{equation}
    \hat{U}_{\vec{n},\mu}=e^{iag\hat{A}_{\vec{n},\mu}},
\end{equation}
where $ag\hat{A}_{\vec{n},\mu}$ is restricted to $[0,2\pi)$, thus the group of gauge transformations is the compact $U(1)$ group.
In the following, the coupling $g$ will be defined as a function of the \textit{lattice spacing} $a$, $g\mapsto g(a)$. 
We then set $a=1$ without loss of generality.

The electric field $\hat{E}_{\vec{n},\nu}$ and the link operator $\hat{U}_{\vec{n}',\mu}$ are connected through the commutation relations,
\begin{align}
    [\hat{E}_{\vec{n},\nu},\hat{U}_{\vec{n}',\mu}]&=\delta_{\vec{n},\vec{n}'}\delta_{\nu,\mu} \hat{U}_{\vec{n},\nu},\label{eq:U_E_commutation_relation1}\\
    [\hat{E}_{\vec{n},\nu},\hat{U}^{\dag}_{\vec{n}',\mu}]&=-\delta_{\vec{n},\vec{n}'}\delta_{\nu,\mu}\hat{U}^{\dag}_{\vec{n}',\nu}.
    \label{eq:U_E_commutation_relation2}
\end{align}

The last two terms in the Hamiltonian describe the fermionic degrees of freedom. 
Starting from a continuum formulation with two-component Dirac spinors, we discretize the Hamiltonian with the staggered formulation.
The fermionic mass term, involving the bare lattice fermion mass $m$, Eq.~\eqref{eq:hmass}, has a single-component fermionic field ($\hat{\phi}_{\vec{n}}$) residing on the site $\vec{n}$.  
The kinetic term, in Eq.~\eqref{eq:hkin}, describes a process in which a fermion moves between two neighbouring lattice sites, causing an associated alteration of the electric field along the link connecting these sites.

The states that fulfill Gauss's law at each site $\vec{n}$, Fig.~\ref{fig:lattices_p_gauss}c,
\begin{align}\label{gausslaw}
    \begin{aligned}
    \Bigg[\sum\limits_{\mu=x,y}
\left(\hat{E}_{\vec{n}- \mu, \mu}-\hat{E}_{\vec{n}, \mu}  \right) - \hat{q}_{\vec{n}} &- Q_{\vec{n}}\Bigg] \ket{\Phi} = 0 \\
&\iff \ket{\Phi} \in \mathcal{H}_{\text{ph}},
    \end{aligned}
\end{align}
belong to a gauge invariant subspace $\mathcal{H}_{\text{ph}}$. 
In this equation,
\begin{equation}\label{dyncharge}
    \hat{q}_{\vec{n}}=\hat{\phi}^\dag_{\vec{n}} \hat{\phi}_{\vec{n}}-\frac{1}{2}\left[1+(-1)^{n_x+n_y+1}\right]
\end{equation}
are \textit{dynamical charges}, and $Q_{\vec{n}}$ represent \textit{static charges}. 

In this work, we impose Gauss's law and study only the physically relevant subspace. By applying this method, we reduce the number of links to a subset of dynamical ones, i.e. we can rewrite some of them in terms of a reduced set of independent variables, by solving Eq.~\eqref{gausslaw}. 
In general, the number of dynamical links, before Gauss's law has been applied follows the rule $l=n$ ($l=nd$) total number of links in Open Boundary Conditions (OBC) (Periodic Boundary Conditions, (PBC)) system, with $n=$ number of sites and $d=$ spatial dimensions. With $n-1$ constraint from Gauss's law, we have a total of $\tilde{l}=l-(n-1)$ dynamical links, e.g. a $3\times 2$ OBC system has a subset of $\tilde{l}=7-(6-1)=2$ dynamical links.
Note that the choice of this subset may affect the configurations on the lattice, especially for small couplings. 
See Appendix~\ref{app:truncation} for an extensive analysis.

\subsection{Numerical implementation of gauge fields}\label{subs:gaugefields}
 
We have seen that the compact $U(1)$ group describes QED. 
However, for a numerical implementation of the Hamiltonian on finite computational resources, we need to consider a correspondingly finite set of possible solutions: this is achieved with a truncation of the infinite-dimensional gauge Hilbert space. 
Here, we follow the truncation of $U(1)$, in the electric basis, to $\mathbb{Z}_{2l+1}$, where $l$ defines the truncation and sets the Hilbert space dimension~\cite{Haase2021resourceefficient}.
With this method, the unbounded gauge degrees of freedom are truncated to a finite dimension within the range $[-l,l]$, resulting in a total Hilbert space dimension of $(2l+1)^N$, where $N$ denotes the number of gauge fields in the system.
After the truncation, the eigenstates of the electric field operator, $\hat{E}_{\vec{n}, \mu}$, form a basis for the link degrees of freedom. 
From Eqs.~\eqref{eq:U_E_commutation_relation1},~\eqref{eq:U_E_commutation_relation2}, the link operators $\hat{U}_{\vec{n},\mu}$ ($\hat{U}^{\dag}_{\vec{n},\mu}$) act as a raising (lowering) operator on the electric field eigenstates,
\begin{align}
    &\hat{E}_{\vec{n}, \mu} \ket{e_{\vec{n}, \mu}}=e_{\vec{n}, \mu}\ket{e_{\vec{n}, \mu}} , \ \ \ \text{with} \ \ e_{\vec{n}, \mu} \in [-l,l]
    \, ,\\
    &\hat{U}_{\vec{n},\mu} \ket{e_{\vec{n}, \mu}}=\ket{e_{\vec{n}, \mu}+1},\ \ \  \hat{U}^{\dag}_{\vec{n},\mu}\ket{e_{\vec{n}, \mu}}=\ket{e_{\vec{n}, \mu}-1}.
\end{align}

Alternative ways to provide a suitable formulation for a numerical analysis can be with quantum link models, with a cyclic group, or also encoding the gauge fields with qudits, see e.g. Refs.~\cite{PhysRevD.102.094501,chandrasekharan1997quantum,wiese2013ultracold,Hashizume_2022,notarnicola2015discrete,meth2023simulating}.

\section{Quantum computation setup}
\label{sec:encoding}

This work will consider a variational quantum approach to study the static potential.
A parameterized quantum circuit is used to efficiently prepare the ground state of the Hamiltonian at various values of the coupling $g$ and the expectation value of the Hamiltonian itself will allow us to map out the static potential function $V(r)$.
To employ this method we prepare a quantum circuit with parameterized gates as our ground state Ansatz.
The main property we endow on our Ansatz is the ability to explore only the physical Hilbert space, thus being more efficient in representing the allowed quantum states of the theory.
In particular, we restrict the space of states to the truncated Hilbert space for the gauge fields, introduced in Sec.~\ref{subs:gaugefields}, and to the fermionic sector with zero total charge. 
This section describes how we encode the gauge and fermionic degrees of freedom on the lattice into qubits and the set of quantum gates utilized in the variational algorithm. 
The \texttt{python} code used in this paper to build the $(2+1)$-d QED lattice Hamiltonian and the parameterized quantum circuit is available at Ref.~\cite{QEDHamiltrepo}.

\subsection{Encoding of gauge fields}
\label{sec:encoding-gauge-fields}
\begin{figure}[htp!]
    \centering
    \includegraphics[width=0.5\columnwidth]{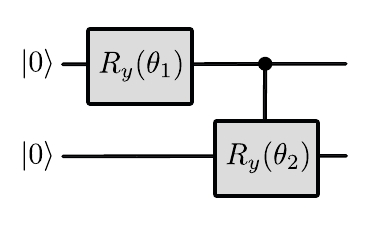}
    \caption{\textbf{Variational circuit for Gray encoding with $l=1$:} \textit{Vacuum state} is $\ket{01}$, and state $\ket{10}$ is excluded.}
    \label{fig:graycircuit}
\end{figure}

For the implementation on a quantum circuit it is advantageous to employ a suitable encoding that accurately represents the physical values of gauge fields.
Examples can be the \textit{linear encoding}~\cite{PRXQuantum.2.030334}, where gauge physical states are mapped onto $2l+1$ qubits, or the \textit{logarithm encoding}~\cite{PhysRevD.102.094501}. 
With the latter formulation, the minimum number of qubits required for each gauge variable is $q_\text{min} = \lceil \log_2(2l+1) \rceil$.
In this work we consider the \textit{Gray encoding} (see e.g. Ref.~\cite{PhysRevA.103.042405}) to represent the physical values of the gauge fields in a quantum simulation.
With this approach, the encoded gauge fields are chosen in such a way that the difference in the bit string representation of the states, when applying lowering and raising operators, is just a single bit.
In addition, since our objective is to do a qualitative analysis, in this project we will mainly consider the truncation $l=1$, with additional details about other truncation values in Appendix~\ref{app:truncation}.
With $l=1$ we have the following states\footnote{Here, and in the rest of the paper, we follow the \textit{right-left} ($\ket{..q_2q_1q_0}$), \textit{top-bottom} ordering of the qubits.},
\begin{subequations}
\begin{align}
    &\ket{-1}_{\text{ph}} \mapsto \ket{00},\label{eq:ket00} \\
    &\ket{0}_{\text{ph}} \mapsto \ket{01}, \label{eq:ket01}\\
    &\ket{1}_{\text{ph}} \mapsto \ket{11}.\label{eq:ket11}
\end{align}
\end{subequations}
The circuit, depicted in Fig.~\ref{fig:graycircuit}, can be understood as follows:
\begin{itemize}[label={\ding{228}}]
\item Beginning with the state $ \ket{00} $, setting both parameters $ \theta_1 $ and $ \theta_2 $ to zero enables the representation of the physical state $ \ket{-1}_{\text{ph}} $, Eq.~\eqref{eq:ket00}.
\item When a non-zero value is assigned to $ \theta_1 $, the state transitions to $ \ket{01}$, representing the vacuum state (Eq.~\eqref{eq:ket01}), with a certain probability. 
\item A full rotation occurs when $ \theta_1 = \pi $, ensuring that the second state is achieved with a probability of 1.0. 
\item After this, the second controlled gate is activated only if the first qubit is $ \ket{1} $, allowing the exploration of $ \ket{11} $ (Eq.~\eqref{eq:ket11}) and excluding $\ket{10}$.
\end{itemize}

\subsection{Encoding of fermionic fields}
\label{sec:encoding-fermionic-fields}
\begin{figure}[htp!]
    \centering
    \includegraphics[width=0.9\columnwidth]{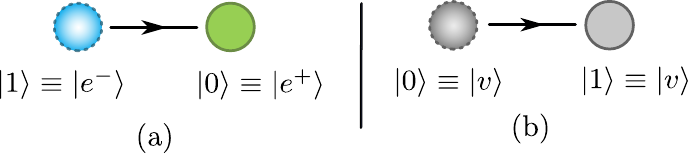}
    \caption{\textbf{Fermionic sites configurations:} If the state on even sites (\textit{dotted circles}) is $\ket{1}$ (\textit{panel (a)}) ($\ket{0}$, \textit{panel (b)}), there is a particle, $e^-$, (vacuum, \textit{v}) on that site. In the case of odd sites (\textit{solid circles}) if $\ket{0}$ (\textit{panel (a)}) ($\ket{1}$, \textit{panel (b)}), we have an antiparticle, $e^+$, (vacuum, \textit{v}).}
    \label{fig:charges_dyn}
\end{figure}

The fermionic degrees of freedom at site $\vec{n}$ can be mapped to spins using a Jordan-Wigner transformation~\cite{jordan1993paulische},
\begin{subequations}
\begin{align}
    \hat{\phi}_{\vec{n}}          &= \Big[\prod\limits_{{\vec{k}}<\vec{n}}(-i\sigma^z_{\vec{k}})\Big] \sigma^{+}_{\vec{n}},\label{eq:jwr_f1}\\
    \hat{\phi}_{\vec{n}}^\dagger  &= \Big[\prod\limits_{{\vec{k}}<\vec{n}}(i\sigma^z_{\vec{k}})\Big] \sigma^{-}_{\vec{n}}\label{eq:jwr_f2},
\end{align}
\end{subequations}
where $\sigma_{\vec{n}}^\pm = \frac{\sigma_{\vec{n}}^x\pm i \sigma_{\vec{n}}^y}{2}$ ($\sigma_{\vec{n}}^x,\sigma_{\vec{n}}^y,\sigma_{\vec{n}}^z$ are Pauli matrices, and $I_{\vec{n}}$ is the identity matrix, acting on the spin at site $\vec{n}$). 
The relation between site coordinates ${\vec{k}}<\vec{n}$ is defined to
satisfy the fermionic anticommutation relations.
The dynamical charges, Eq.~\eqref{dyncharge}, can be written as
\begin{equation}
    \hat{q}_{\vec{n}}\ \mapsto \begin{cases}
			\frac{I_{\vec{n}}-\sigma^z_{\vec{n}}}{2} & \text{if $\vec{n}$ even},\\
            -\frac{I_{\vec{n}}+\sigma^z_{\vec{n}}}{2} & \text{if $\vec{n}$ odd}.
		 \end{cases}
\end{equation}
The mass term Eq.~\eqref{eq:hmass} in the Hamiltonian identifies the Dirac vacuum with the state where the odd fermionic sites are occupied, and creating a particle at an even site is equivalent to creating a charged $q=1$ \textit{fermion} in the Dirac vacuum. 
Destroying a particle at odd sites is thus equivalent to creating an \textit{antifermion} with charge $q=-1$. 
With the vacuum state on the fermionic sites represented as $\ket{..1010}$, we can summarize the configurations in Table~\ref{tab:particles_config} (the first site is even $\vec{n}=(n_x,n_y)=(0,0)$).
\begin{table}[htp!]
    \centering
    \begin{tabular}{|c||c|c|}
    \hline
        & $\ket{0}$ & $\ket{1}$ \\
        \hline
        $n_x+n_y=$even & $q=0 $ \textit{vacuum} & $q=1$ \textit{fermion}\\
        $n_x+n_y=$odd & $q=-1$ \textit{antifermion} & $q=0 $ \textit{vacuum}\\
        \hline
    \end{tabular}
    \caption{\textbf{Particle configurations on a $2D$ lattice:} The vacuum state is $\ket{..1010}$ and the configuration with particles/antiparticles at every site is $\ket{..0101}$. (Note that these configurations depend on the choice of the Jordan-Wigner transformation). }
    \label{tab:particles_config}
\end{table}
These configurations are visualized also in Fig.~\ref{fig:charges_dyn}, where we have in \textit{panel (a)} the case with a particle, $e^-$, on the even site and an antiparticle, $e^+$, on the odd site, corresponding to charges $q=1,-1$ respectively. \textit{Panel (b)} shows the situation with vacuum states, \textit{v}, both on even/odd sites.

We now define a quantum circuit that excludes the states with non-zero total charge. 
This can be achieved with a set of parameterized $i\text{SWAP}_{j,k}(\theta)=e^{-i\frac{\theta}{4}(\sigma^x_j\sigma^x_k+\sigma^y_j\sigma^y_k)}$ gates, where $j$ and $k$ are the qubits on which the gate acts and $\theta$ an angle parameter~\cite{PRXQuantum.2.030334}. 
They can be realized with a combination of parameterized rotational gates $R_{xx}(\theta)=e^{-i\frac{\theta}{2}\sigma_x\sigma_x}$ and $R_{yy}(\theta)=e^{-i\frac{\theta}{2}\sigma_y\sigma_y}$ on two qubits.
The action of the $i$SWAP gate is swapping the values of two qubits, i.e. if we start from a state $\ket{10}$,  $R_{xx}(\theta/2)R_{yy}(\theta/2)\ket{10}$ with $\theta=\frac{\pi}{2}$ will give us $\ket{01}$. 
With these gates, we can explore the fermionic states in the Hilbert space with zero total charge\footnote{If we choose the NFT optimizer~\cite{PhysRevResearch.2.043158}, we need to satisfy a set of requirements, one being that the gates in the variational circuits must be of the form $R(\theta)=e^{-i\frac{\theta}{2}A}$ with $A^2=I$. However, $[\frac{1}{2}(\sigma^x_j\sigma^x_k+\sigma^y_j\sigma^y_k)]^2\neq I$. To solve this issue, we can extend the gate to $\frac{1}{2}(\sigma^x_j\sigma^x_k + \sigma^y_j\sigma^y_k+ \sigma^z_j\sigma^z_k +I_jI_k)$, which satisfies the condition. Note that we can discard the identity and we only need to implement the $R_{zz}(\theta/2)=e^{-i\frac{\theta}{4}\sigma_z\sigma_z}$.}. 

The states of a generic system will be written as the tensor product of gauge fields states and fermionic states: $\ket{\Psi}=\ket{\psi_f}\otimes \ket{\psi_g}$. 
With this ordering we can read the quantum variational solutions and identify the corresponding configuration of gauge and fermionic degrees of freedom. 
For example, the vacuum state for a $3\times 2$ system, will correspond to $\ket{01}$ states for each gauge field (truncation $l=1$) and $\ket{101010}$ for the six fermionic sites. 
Combining them, we get that the vacuum state is $\ket{v}=\ket{101010}\otimes \ket{0101}$.

\subsection{Quantinuum Hardware}
\label{sec:quantinuum}
The optimal quantum circuits, resulting from the variational quantum parameters, that prepare the ground state at various couplings are run on Quantinuum H-series System Model H1-1, both in emulation and in real hardware~\cite{quantinuum_H1_1}.
The quantum job submission workflow is supported by the Quantinuum Nexus cloud platform~\cite{quantinuum_nexus}.

The quantum device we utilize is based on the QCCD architecture~\cite{Pino:2020mku} and it shuttles \textit{Ytterbium-171} ions\footnote{Each Ytterbium ion is paired in a crystal with a Barium ion used for sympathetic laser cooling.} along a linear trap, with the qubit information stored in the ion's atomic hyperfine states.
A total of up to 20 qubits can be manipulated across 5 parallel gate zones, realizing an effective all-to-all connectivity between qubits that is advantageous for the circuits representing our Ansatz.
The H1-1 device can be emulated with an accurate physical noise model using the H1-1E emulator~\cite{pecos,RyanAnderson2021}.
We use the emulator in its statevector configuration.

\subsection{Noise mitigation}
\label{subsec:noise_mitig}
In the present paper we employ two types of noise mitigation techniques to post-process the resulting shot counts and expectation values. 
The first one is the Partition Measurement Symmetry Verification (PMSV)~\cite{Yamamoto_2022} method, which uses global symmetries of the Hamiltonian to validate measurements, before combining shots to compute expectation values across multiple circuits.
Another approach involves mitigating state preparation and measurement (SPAM) noise~\cite{PhysRevA.92.042312}. 
With SPAM, the noise-induced errors are considered only to occur during the state preparation and measurement steps. 
This method uses the density matrix to first get the noise profile of the device when it comes to readout operations: this is achieved with the submission of a calibration circuit.
Then, it computes the inverse of this matrix, suppressing the errors caused by the noise channels in the readout.
Both PMSV and SPAM are implemented in the quantum computational software \texttt{InQuanto}~\cite{inquanto}.

Moreover, since we are only interested in the zero-charge sector for the fermionic sites and the truncated Hilbert space for the gauge fields, we apply a simple symmetry-based error detection post-processing step during the sampling in the computational basis.
We exclude the shots whose bitstrings do not satisfy the fermionic symmetry constraints and that fall outside of the physical Hilbert space of the gauge links.
After the selection of the physical bitstrings, the probability distribution is computed on the  renormalized counts.
We will consider this post-selection for the sampling analysis, while we will use PMSV and SPAM to compute the Hamiltonian expectation values.

\section{Results: $3\times2$ lattice}
\label{sec:res3x2}

\begin{figure}[htp!]
    \centering
    \includegraphics[width=0.35\textwidth]{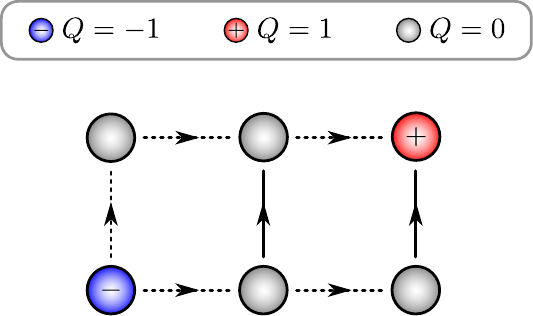}
    \caption{\textbf{Lattice system $3\times2$:} Two static charges with values $Q\pm1$ are placed onto two sites $(n_x,n_y)$: $Q=-1\mapsto (0,0),\ Q=1\mapsto (2,1)$. The solid arrows represent the link operator that remains dynamical after Gauss's law is applied.}
    \label{fig:lattice3x2}
\end{figure}

In this work, we fix the position of two static charges on the lattice, at a fixed distance of $r=\sqrt{5}$ lattice sites, and vary the bare coupling $g$.
The first lattice system studied is depicted in Fig.~\ref{fig:lattice3x2}. 
The total number of qubits for the quantum circuit is 10 (4 for the gauge fields with truncation $l=1$ and 6 for the fermions), for further details see Appendix~\ref{app:quantumcirc}.
Table~\ref{tab:res3x2obc} shows the resource estimation for this system size and three values of the truncation, $l = 1, 3, 7$. 
In particular, we show the total number of qubits used, the number of variational parameters, CNOT gates and their depth, which refers to the amount of CNOT layers\footnote{Every layer is represented by parallel CNOTs in the quantum circuit, e.g. if a first CNOT acts on qubit $q_1$ and $q_2$ and a second CNOT gate on $q_3$ and $q_4$, they belong to the same layer. 
Two different layers are counted if the second gate acts also on $q_1$ or $q_2$.}.

\begin{table}[htp!]
\centering
\begin{tabular}{ |p{0.4cm}||p{1.5cm}|p{1.6cm}|p{2.2cm}|p{2.0cm}|  }
 \hline
 \multicolumn{5}{|c|}{Resource Estimation $3\times 2$ OBC system} \\
 \hline
 $l$ & $\#$ Qubits & $\#$ CNOTs & CNOT Depth & $\#$ Parameters\\
 \hline
 1 & 10  &  152   & 60 & 30  \\
 3 & 12  &  200   & 88 & 41  \\
 7 & 14  &  252   & 122 & 54  \\
 \hline
\end{tabular}
\caption{\textbf{Resources required for the variational circuit for Gray encoding:} In a $3\times 2$ OBC system with fermions, the two dynamical gauge fields and fermionic sites can be simulated with the specified total number of qubits. In particular, the number of qubits for the fermions is fixed to 6. Additionally, we quantify the total count of CNOT gates and the CNOT depth, representing the layers of CNOT gates in the circuit. The rightmost column displays the total number of parameters in the variational Ansatz.}
\label{tab:res3x2obc}
\end{table}

\begin{figure}[htp!]
    \centering
    \includegraphics[width=1\columnwidth]{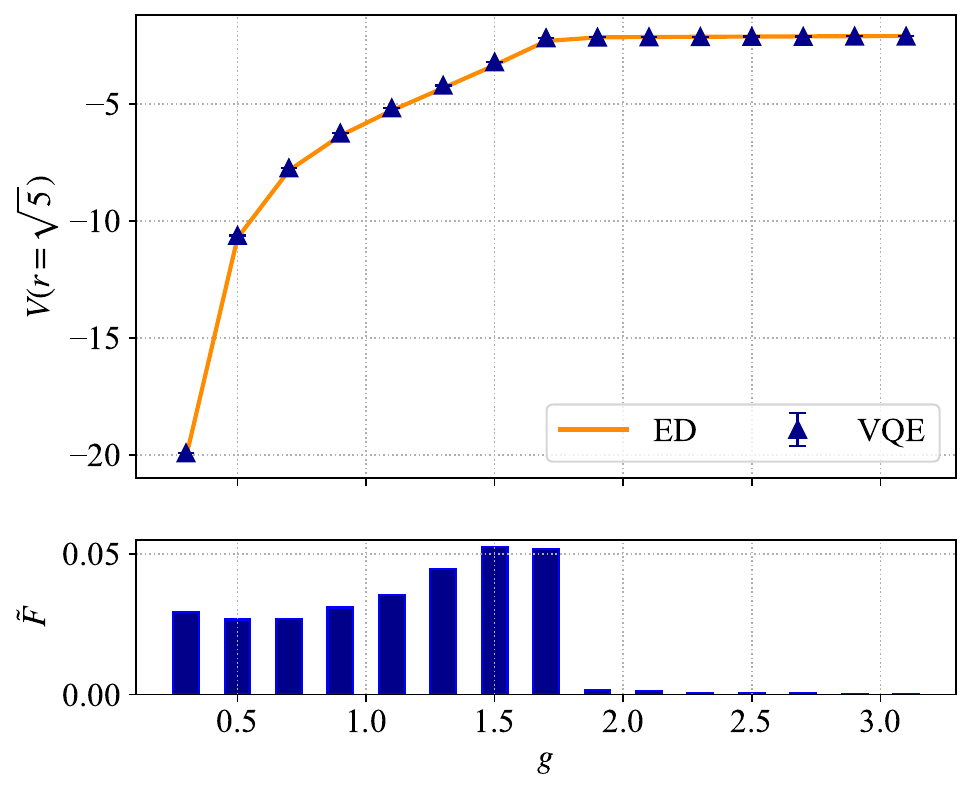}
    \caption{\textbf{Variational quantum results $3 \times 2$ system:} (\textit{top panel}) Static potential at different coupling $g$ with ED (\textit{solid line}) and quantum variational results (\textit{triangles}), performed with NFT optimizer and $10^4$ shots. (\textit{bottom panel}) Infidelity ($1-$fidelity) between variational quantum data and ED. The error bars are smaller than the markers.}
    \label{fig:vqe_res3x2}
\end{figure}

We first consider a noiseless analysis with the Variational Quantum Eigensolver (VQE)~\cite{peruzzo2014variational}. 
The top panel of Fig.~\ref{fig:vqe_res3x2} shows the comparison between the static potential $V(r)$ with exact diagonalization (ED) (\textit{solid line}) and the quantum variational results (\textit{triangles}), performed with NFT optimizer~\cite{PhysRevResearch.2.043158,Ostaszewski2021structure} and $10^4$ shots. 
In the bottom panel, we show the infidelity of the results,
\begin{equation}\label{eq:infidelity}
    \tilde{F}\equiv1-F=1-|\bra{\psi_{\text{VQE}}}\ket{\psi_{\text{ED}}}|^2 ,
\end{equation}
where $F$ is the fidelity.
We see that the infidelity $\tilde{F}$ is $<5\%$ at almost every coupling $g$ considered, and we are able to reproduce the expected behavior of the static potential with our variational ground state Ansatz. 
For the analysis, we consider optimization results coming from two different and independent initial points in parameter space:
\begin{enumerate}
    \item We consider a set of initial parameters that correspond to the preparation of the vacuum state and the electric strings. 
    Then we perturb them with additive Gaussian noise before starting the optimization. 
    This ensures that we have a large probability of reaching configurations corresponding to the vacuum and an electric flux tube.
    \item Alternatively, we consider parameters corresponding to the preparation of the string breaking configuration.
    We add an additive Gaussian noise perturbation and then start the optimization procedure.
\end{enumerate}
It is possible to define both the initial states above by directly inspecting the lattice structure and the encoding utilized in this project.
The results from both initial points are compared and we select the set of optimized parameters that gives the state with the highest fidelity, which usually depends on the value of the coupling constant $g$.
If the fidelity cannot be computed, there are protocols to test for convergence of the variational optimization and to decide which variational parameters to use, see e.g. Ref.~\cite{Haase2021resourceefficient}.

The uncertainties (standard deviation) are computed with the combination of the variances of the Pauli terms $P_i$ in the Hamiltonian, which is a sum of Pauli strings, $\hat{H}=\sum_i c_i P_i $, (with $c_i$ coefficients)~\cite{PhysRevResearch.4.033173}. 
The expectation value of $\hat{H}$ can be written as
\begin{equation}
  \bra{\psi}\hat{H}\ket{\psi}=  \sum_i c_i \bra{\psi}P_i \ket{\psi}.
\end{equation}
If we perform $n$ times of measurements (shots) for each Pauli string $P_i$, the variance of the estimated $\bra{\psi}P_i \ket{\psi}$ due to a finite $n$ is
\begin{equation}
    \sigma_{P_i}^2= \bra{\psi}P_i^2 \ket{\psi} - \bra{\psi}P_i \ket{\psi}^2 =1- \bra{\psi}P_i \ket{\psi} ^2.
\end{equation}
Then, the final standard deviation error is
\begin{equation}\label{eq:std_dev_paulis}
    \sigma=\sqrt{\sum_i |c_i|^2 \frac{1 - \bra{\psi}P_i \ket{\psi} ^2 }{n}}.
\end{equation}

\subsection{Sampling in the computational basis}
\label{sec:sampling-computational-states}
We select three values of the coupling representing the main regimes in the static potential: Coulomb, linear electric strings and string breaking with $g=0.3,1.1,1.9$ respectively.
With the optimal parameters obtained from the solutions of the quantum variational approach, we build the quantum circuit to prepare the ground state and we run it on the emulator H1-1E and on the real quantum hardware H1-1.
Note that before running on these devices, the circuits are rebased to the native operations of the H-series machines and are optimized using the \texttt{pytket} default optimization.
This results in a two-qubit gate count of approximately 80 $R_{zz}$ arbitrary angle operations, a significant reduction compared to the resources of Table~\ref{tab:res3x2obc}.

\begin{figure}[htp!]
    \centering
    \includegraphics[width=1\columnwidth]{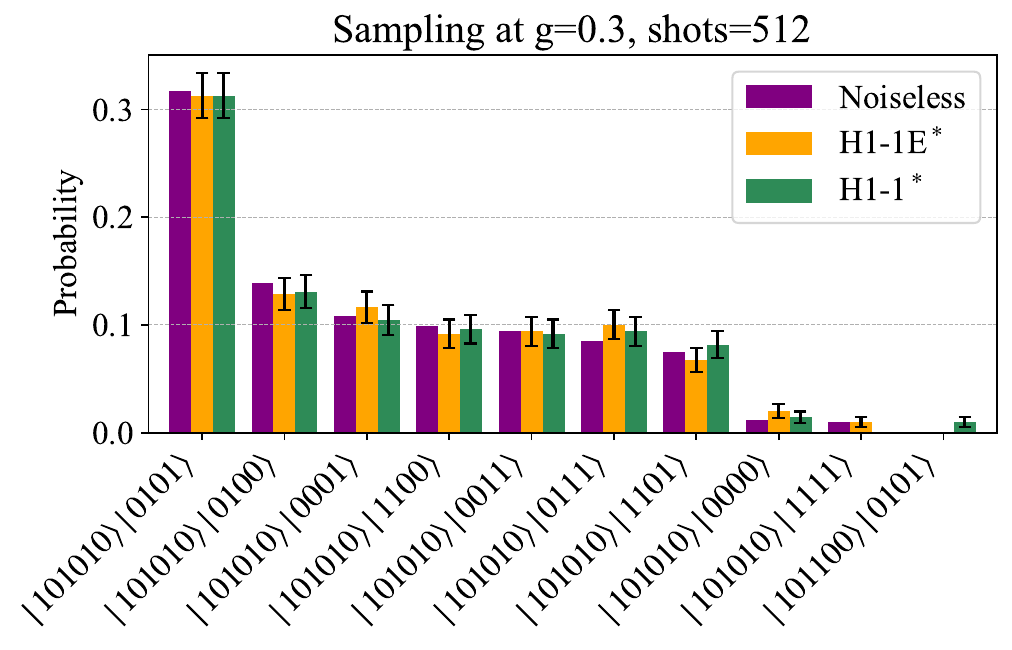}
    \caption{\textbf{Ground state probabilities at $\mathbf{g=0.3}$:} (\textit{bars from left to right}) Noiseless results (state vector calculation with the optimal parameter from VQE), emulator H1-1E and real quantum hardware H1-1. The emulator and hardware results were performed in a single run with 512 shots. The data mitigated by excluding the unphysical bitstrings are indicated by ($^*$). }
    \label{fig:probabg0_3}
\end{figure}

\begin{figure*}[htp!]
    \centering
    \includegraphics[width=0.76\textwidth]{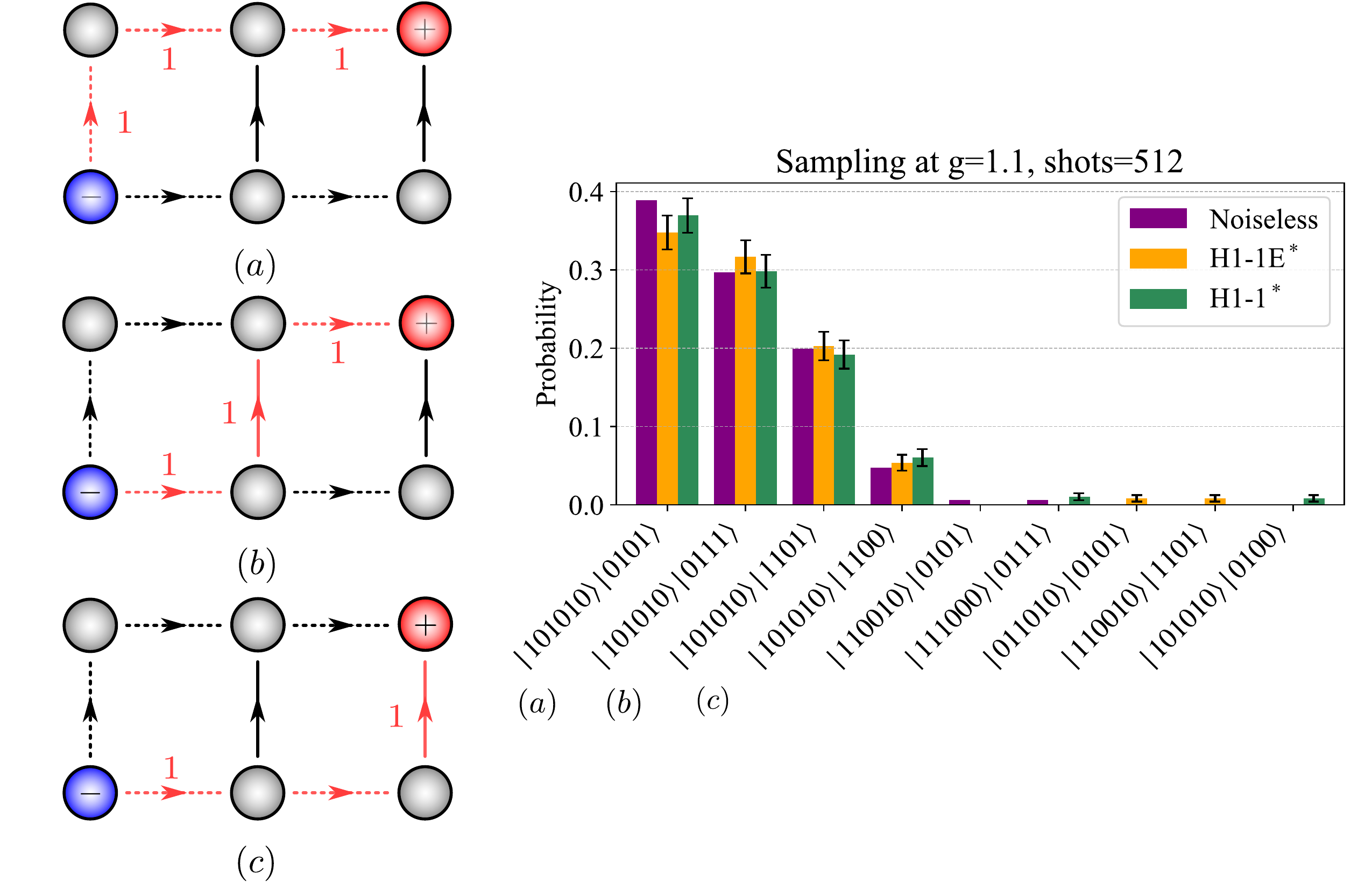}
    \caption{\textbf{Ground state probabilities and lattice configurations at $\mathbf{g=1.1}$:} (\textit{bars from left to right}) Noiseless results (state vector calculation with the optimal parameter from VQE), emulator H1-1E and real quantum hardware H1-1. The emulator and hardware results were performed in a single run with 512 shots. The data mitigated by excluding the unphysical bitstrings are indicated by ($^*$). (\textit{panel (a), (b)} and \textit{(c)}) For this intermediate value of $g$ we observe that the states with the highest probability form three configurations with electric strings between the static charges. The links and sites (values of the dynamical charges) without a number are equal to zero.}
    \label{fig:probabg1_1}
\end{figure*}

\begin{figure}[htp!]
    \centering
    \includegraphics[width=1\columnwidth]{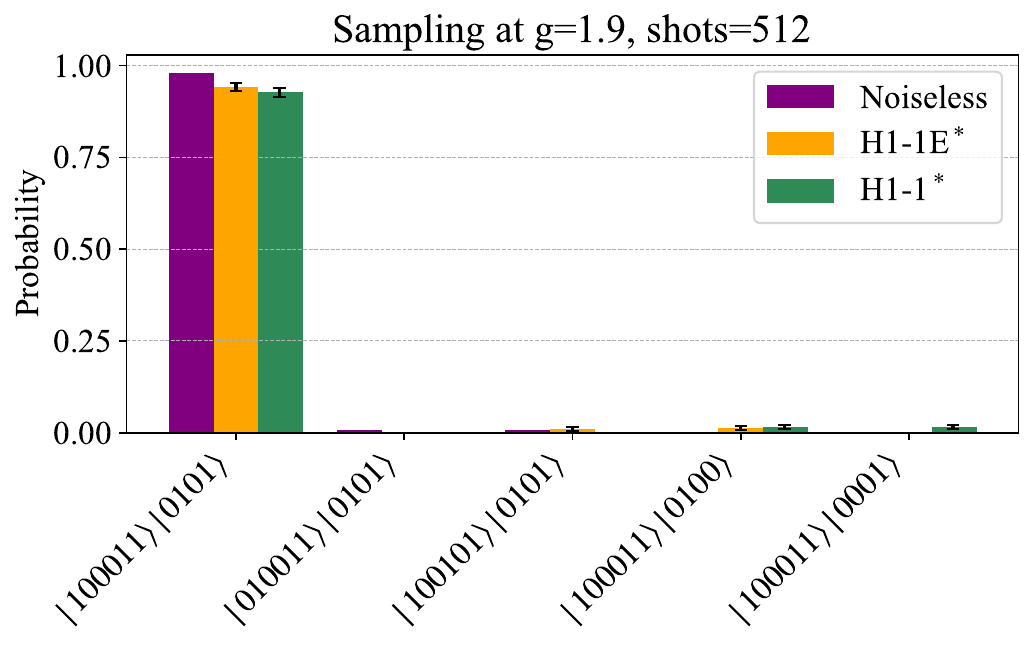}
    \includegraphics[width=0.5\columnwidth]{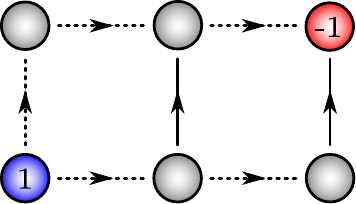}
    \caption{\textbf{Ground state probabilities and lattice configuration at $\mathbf{g=1.9}$:} (\textit{bars from left to right}) Noiseless results (state vector calculation with the optimal parameter from VQE), emulator H1-1E and real quantum hardware H1-1. The emulator and hardware results were performed in a single run with 512 shots. The data mitigated by excluding the unphysical bitstrings are indicated by ($^*$). The most favorable configuration corresponds to the electric strings breaking with the formation of two particle/antiparticle pairs. This state refers to the lattice where links and sites without a number are equal to zero, while the values $\pm1$ that appear on the sites correspond to two non-zero dynamical charges $\hat{q}$. }
    \label{fig:probabg1_9}
\end{figure}

We sample the final state of the circuit and show the configurations with the highest probability in Figs.~\ref{fig:probabg0_3},~\ref{fig:probabg1_1},~\ref{fig:probabg1_9}.
On the $x$-axis, the states $\ket{\psi_f}\otimes \ket{\psi_g}$ assume numerical values corresponding to the sampled bitstrings from measuring the quantum state in the computational basis.
In all the figures, the error bars on the probability are obtained by considering the shots to be drawn from a Bernoulli distribution.

\begin{figure}[htp!]
    \centering
    \includegraphics[width=0.9\columnwidth]{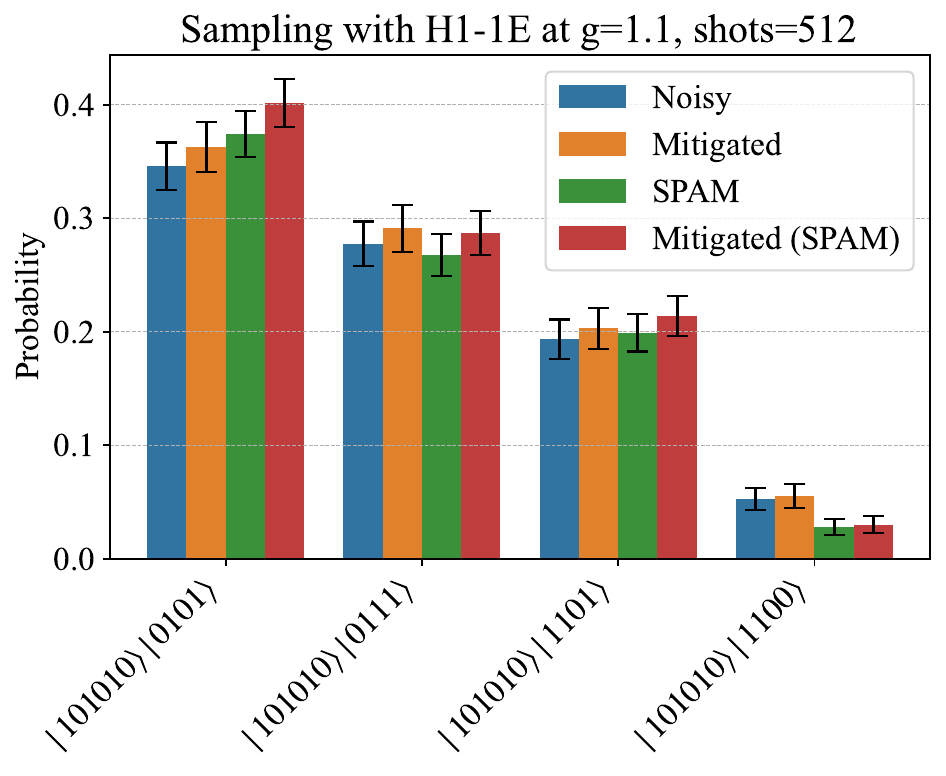}
    \caption{\textbf{Ground state probabilities at $\mathbf{g=1.1}$ with emulator H1-1E:} (\textit{bars from left to right}) Comparison between noisy results and mitigated via the exclusion of unphysical bitstrings. SPAM error mitigation and subsequent bistrings exclusion are also considered.}
    \label{fig:sampling_g11_spam}
\end{figure}

\begin{figure}[htp!]
    \centering
    \includegraphics[width=1\columnwidth]{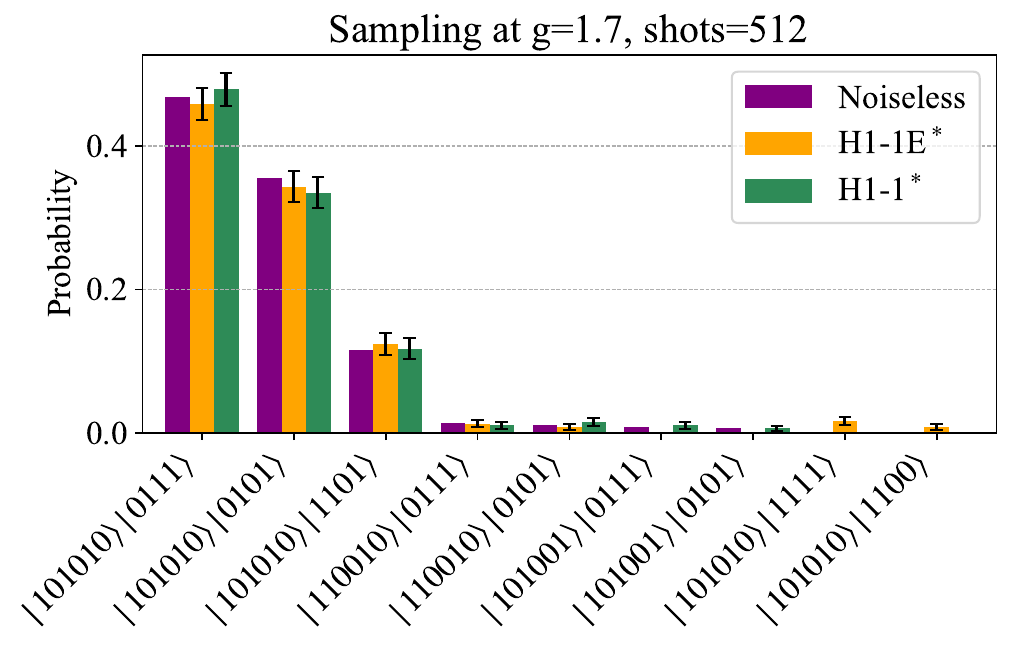}
    \caption{\textbf{Ground state probabilities at $\mathbf{g=1.7}$:} (\textit{bars from left to right}) Noiseless results (state vector calculation with the optimal parameter from VQE), emulator H1-1E and real quantum hardware H1-1. The emulator and hardware results were performed in a single run with 512 shots. The data mitigated by excluding the unphysical bitstrings are indicated by ($^*$).}
    \label{fig:probabg1_7}
\end{figure}
In the weak coupling regime, there are many basis states with non-negligible amplitudes, as depicted in Fig.~\ref{fig:probabg0_3}, thus the ground state is represented by a superposition of a lot of different possible configurations.
The basis state with the highest probability corresponds to a vacuum configuration for both gauge fields and fermionic sites, $\ket{v}=\ket{101010}\otimes \ket{0101}$\footnote{This result corresponds to Fig.~\ref{fig:probabg1_1}a and depends on the choice of dynamical links during the application of Gauss's law. 
The interpretation is discussed extensively in Appendix~\ref{app:truncation}.}.
From the left, the bars define the noiseless results, computed with a state vector calculation with the optimal parameters from the variational quantum analysis. 
The bars in the center and right are the probabilities of the states obtained with the H1-1E emulator and on the real quantum hardware H1-1, respectively. 
The runs on Quantinuum devices were performed with a fixed number of shots of $2^9$. 

When $g$ increases, we have the probabilities and the corresponding configurations depicted in Fig.~\ref{fig:probabg1_1}.
We also test the coupling $g=1.1$ on H1-1E with the application of the SPAM mitigation technique. 
In Fig.~\ref{fig:sampling_g11_spam}, we plot the noisy results and their corresponding mitigated values from left to right by excluding the unphysical bitstrings. 
Then we consider a run with the SPAM method and apply the same mitigation.
We can see that the results are not highly affected by the SPAM mitigation and for every case they could reach the desired noiseless configurations of Fig.~\ref{fig:probabg1_1}.

\begin{figure}[htp!]
    \centering
    \includegraphics[width=1\columnwidth]{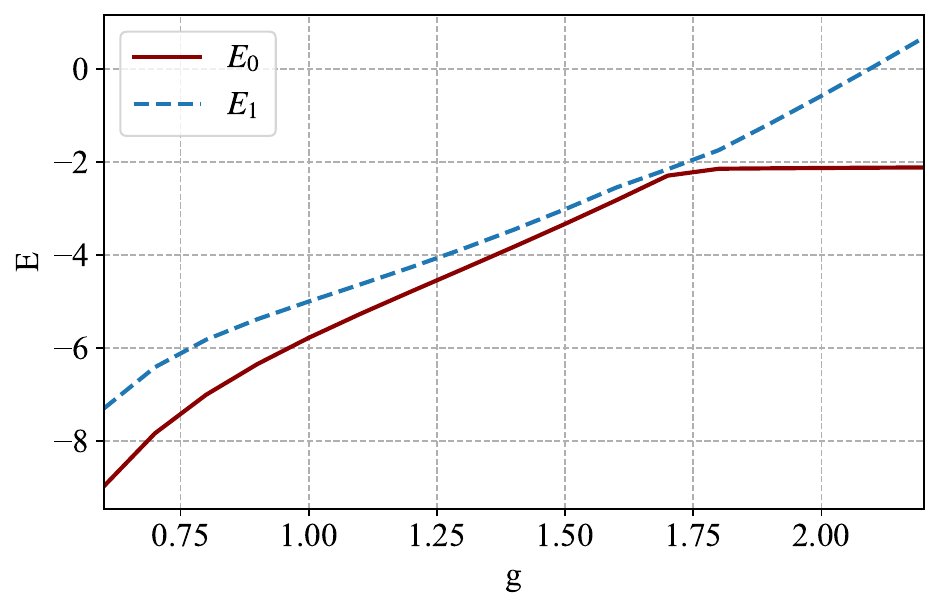}
    \caption{\textbf{Results of energy gap:} Data with exact diagonalization for the ground state ($E_0$ \textit{solid line}) and first excited state ($E_1$ \textit{dashed line}) at different couplings $g$. The gap $E_1-E_0$ closes when approaching $g\sim 1.7$.}
    \label{fig:ED_g_E0_E1}
\end{figure}

At strong $g$ we have a shift to a different regime with basically a single configuration, Fig.~\ref{fig:probabg1_9}. 
In this case, we do not have electric strings between the static charges, but we see the formation of two dynamical charges, representing string breaking and the creation of a particle/antiparticle pair.
We also compute the coupling where we see the transition between linear and string breaking regime, $g\sim 1.7$. 
Fig.~\ref{fig:probabg1_7} illustrates again accurate results both with H1-1E and H1-1, with the most probable states as in Fig.~\ref{fig:probabg1_1}a,~\ref{fig:probabg1_1}b,~\ref{fig:probabg1_1}c.
This data point corresponds to a region where the energy gap, i.e. between the ground state and the first excited state, becomes small, as depicted in Fig.~\ref{fig:ED_g_E0_E1}.

\subsection{Static potential results}
\label{sec:static-potential-results}
\begin{figure}[htp!]
    \centering
    \includegraphics[width=1\columnwidth]{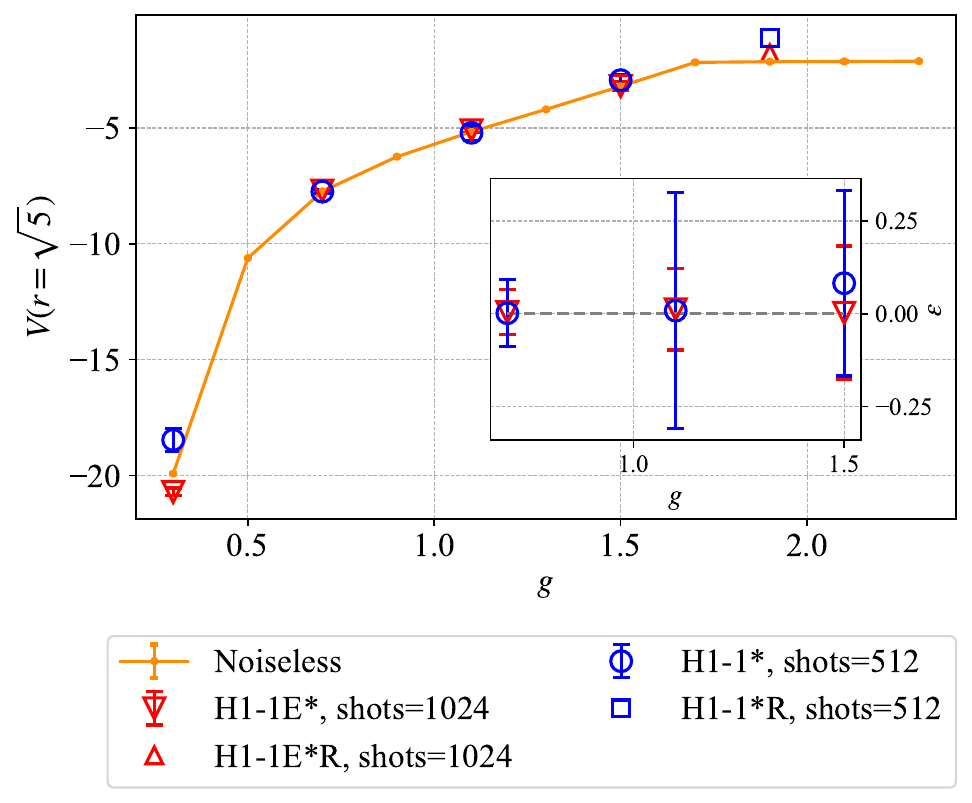}
    \caption{\textbf{Static potential results with H1-1E and H1-1:} The \textit{orange curve} represents the noiseless results (error bars smaller than the markers). For the results of the emulator (\textit{triangles}) we used 1024 shots for each coupling, while for runs on H1-1 (\textit{circles} and \textit{square}) we used 512 shots.  PMSV and SPAM mitigations are considered for the first four couplings $g=0.3,0.7,1.1,1.5$, (data indicated by $(^*)$). The last data point, at $g=1.9$ has been found via the basis sampling approach, selecting $R=4$ dominant states, (data indicated by $(^*R)$) and it is a variational bound on the energy. The \textit{inserted plot} highlights the relative error $\varepsilon$ between the data points computed with H1-1E or H1-1 and the noiseless results at $g=0.7,1.1,1.5$.}
    \label{fig:vr_emu_real}
\end{figure}

We now consider the calculation of the static potential for five values of the bare coupling. 
Fig.~\ref{fig:vr_emu_real} shows a comparison between the data with H1-1E (emulator) and H1-1 (quantum hardware).
For the couplings $g=0.3,0.7,1.1,1.5$ we used a combination of PMSV and SPAM error mitigation methods, included in the software \texttt{InQuanto}~\cite{inquanto}.
For the last coupling $g=1.9$ we considered a different algorithm based on sampling in the computational basis~\cite{PhysRevResearch.4.033173}, where only the $R=4$ most probable states are used, and we applied an error mitigation technique consisting in the post-processing of the sampled bitstrings to remove the ones with unphysical constraints.
The emulator results (\textit{triangles}) have been computed with $2^{10}$ shots, while for the hardware runs (\textit{circles}), we used $2^{9}$ shots for each $g$ using only a single run. 
The \textit{inserted plot} highlights the relative error $\varepsilon$ between the data points computed with H1-1E or H1-1 and the noiseless results at $g=0.7,1.1,1.5$.
The uncertainties are computed with Eq.~\eqref{eq:std_dev_paulis}.
From Fig.~\ref{fig:vr_emu_real}, we can see that, generally, both the emulator and hardware results can reproduce the expected behavior.
In the case of the smallest coupling, $g=0.3$, we have a good agreement between the noiseless result and the result from H1-1E, and expect to reach a better understanding of the systematic errors if we consider multiple runs of the hardware experiment.
For other couplings, we have good agreement on both the emulator and the hardware (\textit{blue circles}). 
Lastly, we note that the last point $g=1.9$ is only a variational bound on the expectation value, since it is obtained by sampling in the computational basis and considering only a subset of states for the calculation of the energy~\cite{PhysRevResearch.4.033173}.
For this reason we do not report the statistical error due to the shots.

\section{Results: $4\times3$ lattice}
\label{sec:4x3system}

\begin{figure}[htp!]
    \centering
    \includegraphics[width=0.7\columnwidth]{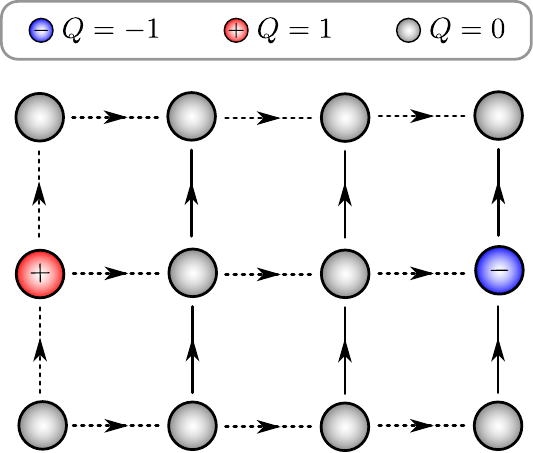}
    \caption{\textbf{Lattice system $4\times 3$:} Two static charges with values $Q\pm1$ are placed onto two sites $(n_x,n_y)$: $Q=1\mapsto (0,1),\ Q=-1\mapsto (3,1)$. The solid arrows represent the link operator that remains dynamical after Gauss's law is applied.}
    \label{fig:chargeconfig4x3}
\end{figure}

\begin{figure}[htp!]
    \centering
    \includegraphics[width=1\columnwidth]{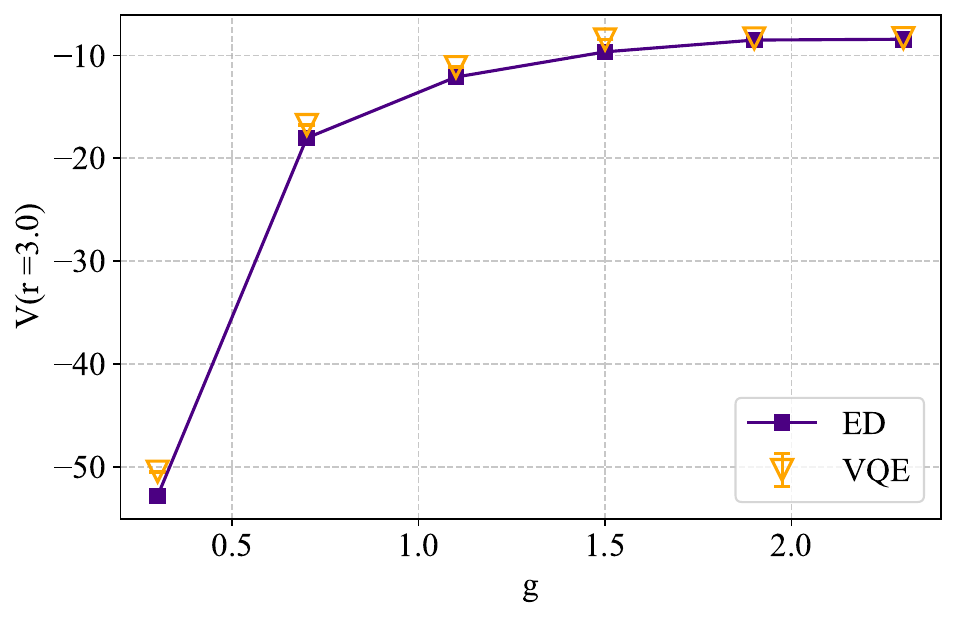}
    \caption{\textbf{Variational quantum results $4\times 3$ system:} Static potential at different coupling $g$ at truncation $l=1$ with ED (\textit{solid line}) and quantum variational results (\textit{triangles}), performed with NFT optimizer and $10^4$ shots. The error bars are smaller than the markers.}
    \label{fig:4x3vr_results}
\end{figure}
This section studies the static potential for a larger $4\times3$ lattice, as depicted in Fig.~\ref{fig:chargeconfig4x3}, where the static charges are placed at a distance $r=3$ onto the two fermionic sites $(n_x,n_y)$: $Q=1\mapsto (0,1),\ Q=-1\mapsto (3,1)$.

\begin{table}[htp!]
\centering
\begin{tabular}{ |p{0.4cm}||p{1.5cm}|p{1.6cm}|p{2.2cm}|p{2.0cm}|   }
 \hline
 \multicolumn{5}{|c|}{Resource Estimation $4\times 3$ OBC system} \\
 \hline
 $l$ & $\#$ Qubits & $\#$ CNOTs & CNOT Depth & $\#$ Parameters\\
 \hline
 1 & 24 &  450   & 136 & 81  \\
 3 & 30  &  582  & 186 & 123  \\
 7 &  36 &  738   & 238 & 177  \\
 \hline
\end{tabular}
\caption{\textbf{Resources required for the variational circuit for Gray encoding:} In a $4\times 3$ OBC system with fermions, the six dynamical gauge fields and fermionic sites can be simulated with the specified total number of qubits. In particular, the number of qubits for the fermions is fixed to 12. Additionally, we quantify the total count of CNOT gates and the CNOT depth, representing the layers of CNOT gates in the circuit. The rightmost column displays the total number of parameters in the variational Ansatz.}
\label{tab:res4x3obc}
\end{table}

\begin{figure*}[htp!]
    \centering
    \includegraphics[width=0.8 \textwidth]{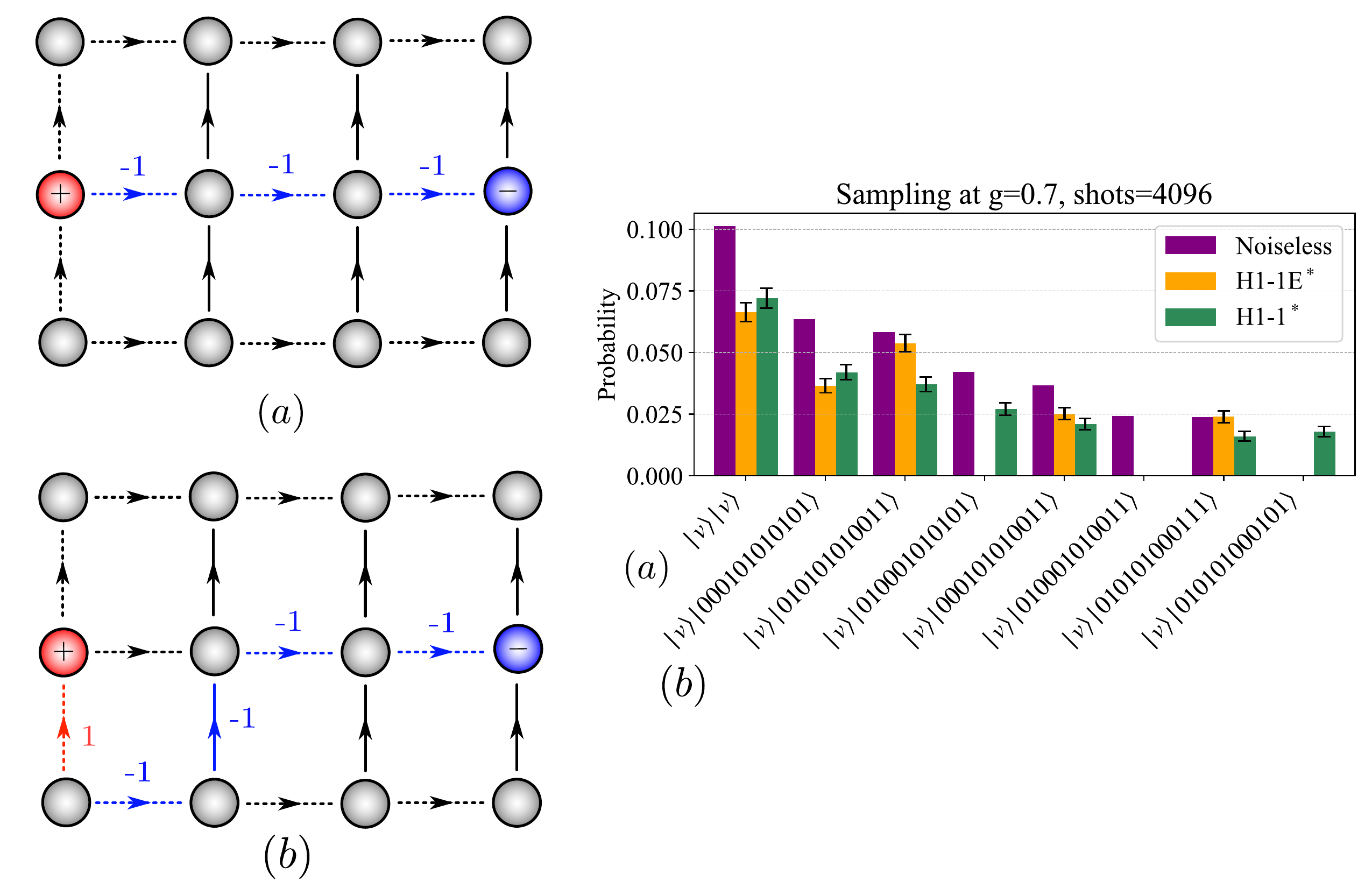}
    \caption{\textbf{Ground state probabilities with reduced circuit at $\mathbf{g=0.7}$:}  (\textit{bars from left to right}) Noiseless results (state vector calculation with the optimal parameter from VQE), emulator H1-1E and real quantum hardware H1-1. The emulator and hardware results were performed in a single run with 4096 shots. The data mitigated by excluding the unphysical bitstrings are indicated by ($^*$). On the $x$-axis, the bit strings are written as $\ket{\Psi}=\ket{\psi_f}\otimes \ket{\psi_g}$, where $\ket{v}$ is the vacuum, i.e. no dynamical charges on the sites or zero values for the dynamical links. At small couplings, the electric string can propagate through the lattice, as in this case at $g=0.7$. At stronger $g$, the dominant configuration becomes the straight string between the static charges, until it breaks and two dynamical charges form.}
    \label{fig:4x3_m2_omega1_l1_g0.7}
\end{figure*}

\begin{figure}[htp!]
    \centering
    \includegraphics[width=1\columnwidth]{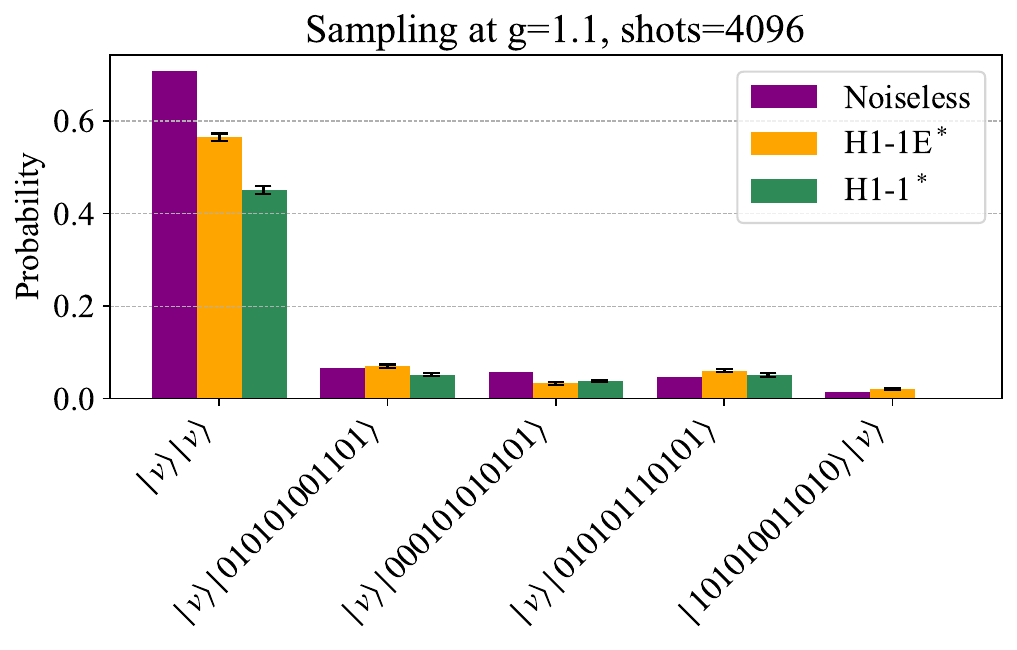}
    \caption{\textbf{Ground state probabilities with reduced circuit at $\mathbf{g=1.1}$:}  (\textit{bars from left to right}) Noiseless results (state vector calculation with the optimal parameter from VQE), emulator H1-1E and real quantum hardware H1-1. The emulator and hardware results were performed in a single run with 4096 shots. The data mitigated by excluding the unphysical bitstrings are indicated by ($^*$). On the $x$-axis, the bit strings are written as $\ket{\Psi}=\ket{\psi_f}\otimes \ket{\psi_g}$, where $\ket{v}$ is the vacuum, i.e. no dynamical charges on the sites or zero values for the dynamical links. The main configuration corresponds to the lattice in Fig.~\ref{fig:4x3_m2_omega1_l1_g0.7}a.}
    \label{fig:4x3_m2_omega1_l1_g1.1}
\end{figure}

For this system we used a total of 24 qubits: 12 for the fermionic sites and $2\cdot 6$ for the six dynamical gauge links (with truncation $l=1$). 
We built the quantum circuit, with a similar structure of the smaller lattice $3\times 2$ (see Fig.~\ref{fig:qcircuit4x3} in Appendix~\ref{app:4x3circuit}). 
Table~\ref{tab:res4x3obc} shows the resources needed for this quantum circuit. Note that the CNOT depth for this system is more than doubled compared to the $3\times2$ lattice, and the raw number of CNOT operations is three times the one on the small system.

In Fig.~\ref{fig:4x3vr_results} we illustrate the first attempt to compute the static potential with this larger system and a quantum variational approach. The uncertainties are computed with Eq.~\eqref{eq:std_dev_paulis}.
The quantum variational results (\textit{triangles}), performed with $10^4$ shots and the NFT optimizer, are able to qualitatively reproduce the static potential curve (\textit{solid line and squares}), when simulated without the presence of noise.
We also measure the fidelity to be $65-95\%$ for the couplings $g=0.7,1.1,1.9$, suggesting that we have not reached convergence for some of them.
In order to run a circuit with 24 qubits on the H1-1E emulator and the H1-1 quantum computer (with 20 qubits), we employ the automatic qubit reuse compilation~\cite{PhysRevX.13.041057} made possible by the mid-circuit measurement and reset capabilities of Quantinuum H-series devices and implemented in the \texttt{TKET} quantum compiler~\cite{Sivarajah:2020lfo}.
By measuring 9 qubits in the middle of the circuit executions and resetting them to be reused in the same circuit, we obtain an equivalent 15-qubit circuit that we use in our experiments.
The circuit with 15 qubits is then rebased on the native gates of H-series and optimized: the total number of two-qubit $R_{zz}$ gate operations is approximately 270 at all couplings, a major reduction compared to the resource in Table~\ref{tab:res4x3obc}.

\begin{figure}[htp!]
    \centering
    \includegraphics[width=1\columnwidth]{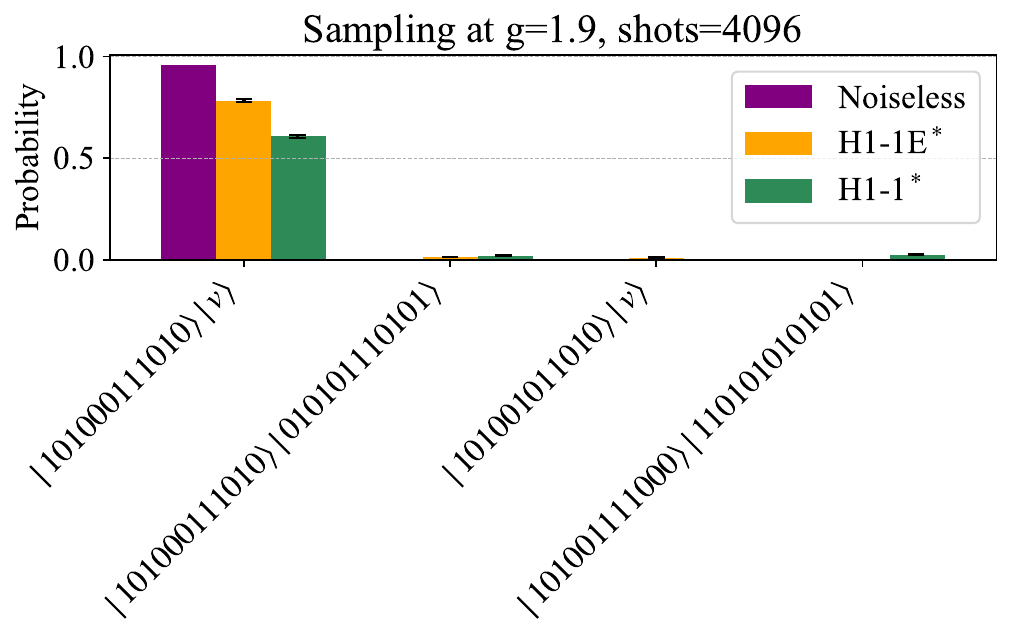}
    \includegraphics[width=0.75\columnwidth]{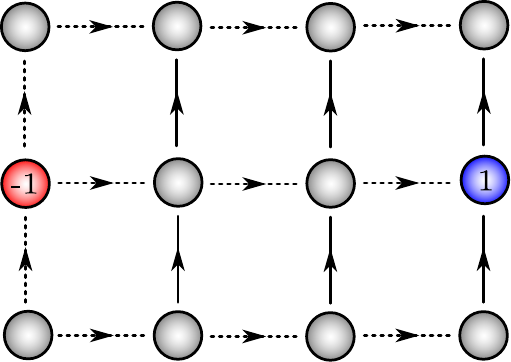}
    \caption{\textbf{Ground state probabilities with reduced circuit at $\mathbf{g=1.9}$:}  (\textit{bars from left to right}) Noiseless results (state vector calculation with the optimal parameter from VQE), emulator H1-1E and real quantum hardware H1-1. The emulator and hardware results were performed in a single run with 4096 shots. The data mitigated by excluding the unphysical bitstrings are indicated by ($^*$). On the $x$-axis, the bit strings are written as $\ket{\Psi}=\ket{\psi_f}\otimes \ket{\psi_g}$, where $\ket{v}$ is the vacuum, i.e. no dynamical charges on the sites or zero values for the dynamical links. The main configuration corresponds to the lattice, where the electric string is broken, and two dynamical charges are formed.}
    \label{fig:4x3_m2_omega1_l1_g1.9}
\end{figure}

We study some of the interesting lattice configurations that arise in this larger $4\times 3$ system. For example, at coupling $g=0.7$, the most probable computational basis state is $\ket{\psi_f}\otimes \ket{\psi_g}=\ket{v} \otimes\ket{v}$, shown in panel $(a)$ of Fig.~\ref{fig:4x3_m2_omega1_l1_g0.7}.
Here $\ket{v}$ is the vacuum, i.e. no dynamical charges on the sites and zero values for dynamical links.
The second most probable state is $\ket{\psi_f}\otimes \ket{\psi_g}=\ket{v}\otimes \ket{000101010101}$ and it corresponds to the configuration illustrated in panel $(b)$.
The probability associated with these additional states having a snake-like pattern of the flux tube vanishes when going to stronger couplings and the straight flux tube (Fig.~\ref{fig:4x3_m2_omega1_l1_g0.7}$(a)$) dominates.
We see this at the stronger coupling $g=1.1$ in Fig.~\ref{fig:4x3_m2_omega1_l1_g1.1}. 
At $g=1.9$, on the other hand, we have the breaking of the electric string and the formation of mesons pairs, shown in Fig~\ref{fig:4x3_m2_omega1_l1_g1.9}.

\section{Conclusion and outlook}
\label{sec:conclusion}
In this paper we have performed a qualitative analysis of the static potential between two static charges, exploring the Coulomb, confinement and string breaking regimes where we determined the electric flux configurations and the probabilities of the contributing states. 
To this end, we developed a symmetry-preserving variational quantum circuit and employed a variational quantum algorithm to create the ground state of the Hamiltonian, corresponding to the static potential. 
Additionally, in the design of the Ansatz, we employed the mutual information between qubits, which led to a reduction of the depth of the quantum circuit. 
In order to explore the different regimes of the potential we used selected values of the coupling constant, corresponding to different physical distances.

We have focused our studies on a $3\times 2$ lattice with open boundary conditions and demonstrated that results from quantum experiments on a trapped-ion emulator, H1-1E, and a real quantum device, H1-1, agreed with classical noiseless simulations for the static potential, obtained with the application of the mentioned quantum variational approach.
The relevant electric flux configurations, which contribute to the quantum ground state in the different distance regimes of the static potential, were visualized.
We could clearly identify flux configurations which correspond to the Coulomb, confinement and string breaking regimes, gaining insights on the flux tube structure of the ground state.

We also considered experiments on a larger system, of $4\times 3$ fermionic sites, with a 24 qubit variational circuit.
An implementation on the 20 qubits of the H1-1 quantum device becomes possible with the reduction of the number of qubits from 24 to 15.
In the current mutual-information adapted Ansatz, this was achieved by using mid-circuit measurements, resetting selected qubits and reusing them in the quantum computation.

Considering further hardware results with the largest Quantinuum ion-trap devices, a possibility is to study a $6\times 4$ lattice, which requires up to a total of 54 qubits for the quantum computation, thus suitable for the H2 device~\cite{quantinuum_H2_1}.
This exciting outlook to go to larger system sizes in the future offers new possibilities.
First, it will allow to study the static potential as a function of the distance in lattice units, which provides the opportunity to fit the anticipated analytical form of the potential and extract the values of the coupling, the string tension and the distance, where string breaking occurs, on a quantitative level. 
Second, it will become possible to determine the properties of the confining string, such as its width and the fluctuations, quantitatively.
By combining these Hamiltonian calculations with Monte Carlo simulations, which will provide a physical value of the lattice spacing, see Ref.~\cite{Crippa:2024cqr}, we can eventually give results in physical units, which could be relevant for experiments described by $(2+1)$ dimensional QED.   
However, we believe that to achieve this goal, further improvements of quantum circuit design as well as advances in quantum hardware are needed.
For example, it would become infeasible to train the variational parameters of the Ansatz, whose number will also scale with the size of the system: this can be circumvented by scalable variational approaches such as SC-ADAPT-VQE~\cite{Farrell:2023fgd} or by adiabatic evolution based on a Trotterized Hamiltonian with reduced Trotter errors~\cite{Granet:2024kuu}.
We also mention that there are corrections to the linear potential, originating from fluctuations of the electric flux string, see Refs.~\cite{luescher1981symmetry,luscher2002quark,nogueira2005quantum,chagdaa2017width} and a recent work in Ref.~\cite{caselle2024numerical}. It would be very interesting to determine this correction within our setup by considering different lattices geometries allowing larger separations of the static charges.
We remark that recently also a Hamiltonian formulation of Maxwell-Chern-Simons theory has been developed for compact $U(1)$ gauge theory on the lattice~\cite{Peng:2024xbl}, which combines confinement and topology and opens new avenues to look at confinement properties in a non-perturbative fashion.

\begin{acknowledgments}
We acknowledge Henrik Dreyer and David Zsolt Manrique for a careful review of the manuscript, and Irfan Khan for discussions on the qubit reuse automatic compilation.
We thank the entire Quantinuum NEXUS team.
We thank Davide Materia for sharing his work on the employment of the mutual information in quantum chemistry.
We are grateful to Enrique Rico Ortega, Francesco Di Marcantonio and Maria Cristina Diamantini for fruitful discussions on the topic.

This work is supported with funds from the Ministry of Science, Research and Culture of the State of Brandenburg within the Centre for Quantum Technologies and Applications (CQTA). 

\begin{figure}[htp!]
\centering
    \includegraphics[width = 0.08\textwidth]{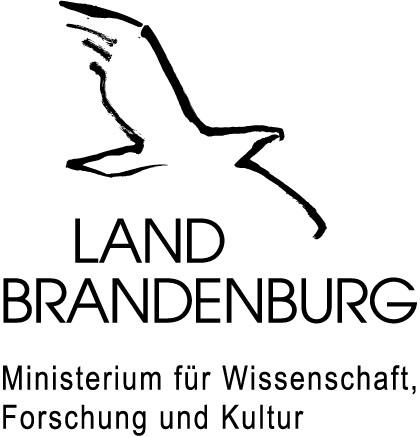}
\end{figure}

A. C. is supported in part by the Helmholtz Association —“Innopool Project Variational Quantum Computer Simulations (VQCS).”
This work is funded by the European Union’s Horizon Europe Frame-work Programme (HORIZON) under the ERA Chair scheme with grant agreement no.~{101087126}.

\end{acknowledgments}

\bibliography{references}

\begin{thebibliography}{69}%
\makeatletter
\providecommand \@ifxundefined [1]{%
 \@ifx{#1\undefined}
}%
\providecommand \@ifnum [1]{%
 \ifnum #1\expandafter \@firstoftwo
 \else \expandafter \@secondoftwo
 \fi
}%
\providecommand \@ifx [1]{%
 \ifx #1\expandafter \@firstoftwo
 \else \expandafter \@secondoftwo
 \fi
}%
\providecommand \natexlab [1]{#1}%
\providecommand \enquote  [1]{``#1''}%
\providecommand \bibnamefont  [1]{#1}%
\providecommand \bibfnamefont [1]{#1}%
\providecommand \citenamefont [1]{#1}%
\providecommand \href@noop [0]{\@secondoftwo}%
\providecommand \href [0]{\begingroup \@sanitize@url \@href}%
\providecommand \@href[1]{\@@startlink{#1}\@@href}%
\providecommand \@@href[1]{\endgroup#1\@@endlink}%
\providecommand \@sanitize@url [0]{\catcode `\\12\catcode `\$12\catcode `\&12\catcode `\#12\catcode `\^12\catcode `\_12\catcode `\%12\relax}%
\providecommand \@@startlink[1]{}%
\providecommand \@@endlink[0]{}%
\providecommand \url  [0]{\begingroup\@sanitize@url \@url }%
\providecommand \@url [1]{\endgroup\@href {#1}{\urlprefix }}%
\providecommand \urlprefix  [0]{URL }%
\providecommand \Eprint [0]{\href }%
\providecommand \doibase [0]{https://doi.org/}%
\providecommand \selectlanguage [0]{\@gobble}%
\providecommand \bibinfo  [0]{\@secondoftwo}%
\providecommand \bibfield  [0]{\@secondoftwo}%
\providecommand \translation [1]{[#1]}%
\providecommand \BibitemOpen [0]{}%
\providecommand \bibitemStop [0]{}%
\providecommand \bibitemNoStop [0]{.\EOS\space}%
\providecommand \EOS [0]{\spacefactor3000\relax}%
\providecommand \BibitemShut  [1]{\csname bibitem#1\endcsname}%
\let\auto@bib@innerbib\@empty
%</preamble>
\bibitem [{\citenamefont {Wilson}(1974)}]{Wilson:1974sk}%
  \BibitemOpen
  \bibfield  {author} {\bibinfo {author} {\bibfnamefont {K.~G.}\ \bibnamefont {Wilson}},\ }\bibfield  {title} {\bibinfo {title} {{Confinement of Quarks}},\ }\href {https://doi.org/10.1103/PhysRevD.10.2445} {\bibfield  {journal} {\bibinfo  {journal} {Phys. Rev. D}\ }\textbf {\bibinfo {volume} {10}},\ \bibinfo {pages} {2445} (\bibinfo {year} {1974})}\BibitemShut {NoStop}%
\bibitem [{\citenamefont {Rothe}(2012)}]{rothe2012lattice}%
  \BibitemOpen
  \bibfield  {author} {\bibinfo {author} {\bibfnamefont {H.~J.}\ \bibnamefont {Rothe}},\ }\href@noop {} {\emph {\bibinfo {title} {Lattice gauge theories: an introduction}}}\ (\bibinfo  {publisher} {World Scientific Publishing Company},\ \bibinfo {year} {2012})\BibitemShut {NoStop}%
\bibitem [{\citenamefont {Gattringer}\ and\ \citenamefont {Lang}(2009)}]{gattringer2009quantum}%
  \BibitemOpen
  \bibfield  {author} {\bibinfo {author} {\bibfnamefont {C.}~\bibnamefont {Gattringer}}\ and\ \bibinfo {author} {\bibfnamefont {C.}~\bibnamefont {Lang}},\ }\href {https://doi.org/10.1007/978-3-642-01850-3} {\emph {\bibinfo {title} {Quantum chromodynamics on the lattice: an introductory presentation}}},\ Vol.\ \bibinfo {volume} {788}\ (\bibinfo  {publisher} {Springer Science \& Business Media},\ \bibinfo {year} {2009})\BibitemShut {NoStop}%
\bibitem [{\citenamefont {Gross}\ \emph {et~al.}(2023)\citenamefont {Gross} \emph {et~al.}}]{Gross:2022hyw}%
  \BibitemOpen
  \bibfield  {author} {\bibinfo {author} {\bibfnamefont {F.}~\bibnamefont {Gross}} \emph {et~al.},\ }\bibfield  {title} {\bibinfo {title} {{50 Years of Quantum Chromodynamics}},\ }\href {https://doi.org/10.1140/epjc/s10052-023-11949-2} {\bibfield  {journal} {\bibinfo  {journal} {Eur. Phys. J. C}\ }\textbf {\bibinfo {volume} {83}},\ \bibinfo {pages} {1125} (\bibinfo {year} {2023})},\ \Eprint {https://arxiv.org/abs/2212.11107} {arXiv:2212.11107 [hep-ph]} \BibitemShut {NoStop}%
\bibitem [{\citenamefont {Polyakov}(1977)}]{Polyakov:1976fu}%
  \BibitemOpen
  \bibfield  {author} {\bibinfo {author} {\bibfnamefont {A.~M.}\ \bibnamefont {Polyakov}},\ }\bibfield  {title} {\bibinfo {title} {{Quark Confinement and Topology of Gauge Groups}},\ }\href {https://doi.org/10.1016/0550-3213(77)90086-4} {\bibfield  {journal} {\bibinfo  {journal} {Nucl. Phys. B}\ }\textbf {\bibinfo {volume} {120}},\ \bibinfo {pages} {429} (\bibinfo {year} {1977})}\BibitemShut {NoStop}%
\bibitem [{\citenamefont {Wen}(2004)}]{Wen:2004book}%
  \BibitemOpen
  \bibfield  {author} {\bibinfo {author} {\bibfnamefont {X.-G.}\ \bibnamefont {Wen}},\ }\href@noop {} {\emph {\bibinfo {title} {Quantum Field Theory of Many-body Systems}}}\ (\bibinfo  {publisher} {Oxford University Press},\ \bibinfo {address} {New York, NY},\ \bibinfo {year} {2004})\BibitemShut {NoStop}%
\bibitem [{\citenamefont {Knechtli}\ and\ \citenamefont {Sommer}(2000)}]{Knechtli_2000}%
  \BibitemOpen
  \bibfield  {author} {\bibinfo {author} {\bibfnamefont {F.}~\bibnamefont {Knechtli}}\ and\ \bibinfo {author} {\bibfnamefont {R.}~\bibnamefont {Sommer}},\ }\bibfield  {title} {\bibinfo {title} {String breaking as a mixing phenomenon in the su(2) higgs model},\ }\href {https://doi.org/10.1016/s0550-3213(00)00470-3} {\bibfield  {journal} {\bibinfo  {journal} {Nuclear Physics B}\ }\textbf {\bibinfo {volume} {590}},\ \bibinfo {pages} {309–328} (\bibinfo {year} {2000})}\BibitemShut {NoStop}%
\bibitem [{\citenamefont {Philipsen}\ and\ \citenamefont {Wittig}(1998)}]{Philipsen:1998de}%
  \BibitemOpen
  \bibfield  {author} {\bibinfo {author} {\bibfnamefont {O.}~\bibnamefont {Philipsen}}\ and\ \bibinfo {author} {\bibfnamefont {H.}~\bibnamefont {Wittig}},\ }\bibfield  {title} {\bibinfo {title} {{String breaking in nonAbelian gauge theories with fundamental matter fields}},\ }\href {https://doi.org/10.1103/PhysRevLett.81.4056} {\bibfield  {journal} {\bibinfo  {journal} {Phys. Rev. Lett.}\ }\textbf {\bibinfo {volume} {81}},\ \bibinfo {pages} {4056} (\bibinfo {year} {1998})},\ \bibinfo {note} {[Erratum: Phys.Rev.Lett. 83, 2684 (1999)]},\ \Eprint {https://arxiv.org/abs/hep-lat/9807020} {arXiv:hep-lat/9807020} \BibitemShut {NoStop}%
\bibitem [{\citenamefont {Bali}(2001)}]{Bali:2000gf}%
  \BibitemOpen
  \bibfield  {author} {\bibinfo {author} {\bibfnamefont {G.~S.}\ \bibnamefont {Bali}},\ }\bibfield  {title} {\bibinfo {title} {{QCD forces and heavy quark bound states}},\ }\href {https://doi.org/10.1016/S0370-1573(00)00079-X} {\bibfield  {journal} {\bibinfo  {journal} {Phys. Rept.}\ }\textbf {\bibinfo {volume} {343}},\ \bibinfo {pages} {1} (\bibinfo {year} {2001})},\ \Eprint {https://arxiv.org/abs/hep-ph/0001312} {arXiv:hep-ph/0001312} \BibitemShut {NoStop}%
\bibitem [{\citenamefont {Loan}\ \emph {et~al.}(2003)\citenamefont {Loan}, \citenamefont {Brunner}, \citenamefont {Sloggett},\ and\ \citenamefont {Hamer}}]{loan2003path}%
  \BibitemOpen
  \bibfield  {author} {\bibinfo {author} {\bibfnamefont {M.}~\bibnamefont {Loan}}, \bibinfo {author} {\bibfnamefont {M.}~\bibnamefont {Brunner}}, \bibinfo {author} {\bibfnamefont {C.}~\bibnamefont {Sloggett}},\ and\ \bibinfo {author} {\bibfnamefont {C.}~\bibnamefont {Hamer}},\ }\bibfield  {title} {\bibinfo {title} {Path integral monte carlo approach to the u (1) lattice gauge theory in 2+ 1 dimensions},\ }\href@noop {} {\bibfield  {journal} {\bibinfo  {journal} {Physical Review D}\ }\textbf {\bibinfo {volume} {68}},\ \bibinfo {pages} {034504} (\bibinfo {year} {2003})}\BibitemShut {NoStop}%
\bibitem [{\citenamefont {Bender}\ \emph {et~al.}(2020)\citenamefont {Bender}, \citenamefont {Emonts}, \citenamefont {Zohar},\ and\ \citenamefont {Cirac}}]{PhysRevResearch.2.043145}%
  \BibitemOpen
  \bibfield  {author} {\bibinfo {author} {\bibfnamefont {J.}~\bibnamefont {Bender}}, \bibinfo {author} {\bibfnamefont {P.}~\bibnamefont {Emonts}}, \bibinfo {author} {\bibfnamefont {E.}~\bibnamefont {Zohar}},\ and\ \bibinfo {author} {\bibfnamefont {J.~I.}\ \bibnamefont {Cirac}},\ }\bibfield  {title} {\bibinfo {title} {Real-time dynamics in $2+1d$ compact qed using complex periodic gaussian states},\ }\href {https://doi.org/10.1103/PhysRevResearch.2.043145} {\bibfield  {journal} {\bibinfo  {journal} {Phys. Rev. Res.}\ }\textbf {\bibinfo {volume} {2}},\ \bibinfo {pages} {043145} (\bibinfo {year} {2020})}\BibitemShut {NoStop}%
\bibitem [{\citenamefont {Banuls}\ \emph {et~al.}(2020)\citenamefont {Banuls}, \citenamefont {Blatt}, \citenamefont {Catani}, \citenamefont {Celi}, \citenamefont {Cirac}, \citenamefont {Dalmonte}, \citenamefont {Fallani}, \citenamefont {Jansen}, \citenamefont {Lewenstein}, \citenamefont {Montangero} \emph {et~al.}}]{banuls2020simulating}%
  \BibitemOpen
  \bibfield  {author} {\bibinfo {author} {\bibfnamefont {M.~C.}\ \bibnamefont {Banuls}}, \bibinfo {author} {\bibfnamefont {R.}~\bibnamefont {Blatt}}, \bibinfo {author} {\bibfnamefont {J.}~\bibnamefont {Catani}}, \bibinfo {author} {\bibfnamefont {A.}~\bibnamefont {Celi}}, \bibinfo {author} {\bibfnamefont {J.~I.}\ \bibnamefont {Cirac}}, \bibinfo {author} {\bibfnamefont {M.}~\bibnamefont {Dalmonte}}, \bibinfo {author} {\bibfnamefont {L.}~\bibnamefont {Fallani}}, \bibinfo {author} {\bibfnamefont {K.}~\bibnamefont {Jansen}}, \bibinfo {author} {\bibfnamefont {M.}~\bibnamefont {Lewenstein}}, \bibinfo {author} {\bibfnamefont {S.}~\bibnamefont {Montangero}}, \emph {et~al.},\ }\bibfield  {title} {\bibinfo {title} {Simulating lattice gauge theories within quantum technologies},\ }\href@noop {} {\bibfield  {journal} {\bibinfo  {journal} {The European physical journal D}\ }\textbf {\bibinfo {volume} {74}},\ \bibinfo {pages} {1} (\bibinfo {year} {2020})}\BibitemShut {NoStop}%
\bibitem [{\citenamefont {Bauer}\ \emph {et~al.}(2023)\citenamefont {Bauer} \emph {et~al.}}]{Bauer:2022hpo}%
  \BibitemOpen
  \bibfield  {author} {\bibinfo {author} {\bibfnamefont {C.~W.}\ \bibnamefont {Bauer}} \emph {et~al.},\ }\bibfield  {title} {\bibinfo {title} {{Quantum Simulation for High-Energy Physics}},\ }\href {https://doi.org/10.1103/PRXQuantum.4.027001} {\bibfield  {journal} {\bibinfo  {journal} {PRX Quantum}\ }\textbf {\bibinfo {volume} {4}},\ \bibinfo {pages} {027001} (\bibinfo {year} {2023})},\ \Eprint {https://arxiv.org/abs/2204.03381} {arXiv:2204.03381 [quant-ph]} \BibitemShut {NoStop}%
\bibitem [{\citenamefont {Di~Meglio}\ \emph {et~al.}(2024)\citenamefont {Di~Meglio}, \citenamefont {Jansen}, \citenamefont {Tavernelli}, \citenamefont {Alexandrou}, \citenamefont {Arunachalam}, \citenamefont {Bauer}, \citenamefont {Borras}, \citenamefont {Carrazza}, \citenamefont {Crippa}, \citenamefont {Croft}, \citenamefont {de~Putter}, \citenamefont {Delgado}, \citenamefont {Dunjko}, \citenamefont {Egger}, \citenamefont {Fern\'andez-Combarro}, \citenamefont {Fuchs}, \citenamefont {Funcke}, \citenamefont {Gonz\'alez-Cuadra}, \citenamefont {Grossi}, \citenamefont {Halimeh}, \citenamefont {Holmes}, \citenamefont {K\"uhn}, \citenamefont {Lacroix}, \citenamefont {Lewis}, \citenamefont {Lucchesi}, \citenamefont {Martinez}, \citenamefont {Meloni}, \citenamefont {Mezzacapo}, \citenamefont {Montangero}, \citenamefont {Nagano}, \citenamefont {Pascuzzi}, \citenamefont {Radescu}, \citenamefont {Ortega}, \citenamefont {Roggero}, \citenamefont {Schuhmacher}, \citenamefont {Seixas}, \citenamefont {Silvi}, \citenamefont
  {Spentzouris}, \citenamefont {Tacchino}, \citenamefont {Temme}, \citenamefont {Terashi}, \citenamefont {Tura}, \citenamefont {T\"uys\"uz}, \citenamefont {Vallecorsa}, \citenamefont {Wiese}, \citenamefont {Yoo},\ and\ \citenamefont {Zhang}}]{PRXQuantum.5.037001}%
  \BibitemOpen
  \bibfield  {author} {\bibinfo {author} {\bibfnamefont {A.}~\bibnamefont {Di~Meglio}}, \bibinfo {author} {\bibfnamefont {K.}~\bibnamefont {Jansen}}, \bibinfo {author} {\bibfnamefont {I.}~\bibnamefont {Tavernelli}}, \bibinfo {author} {\bibfnamefont {C.}~\bibnamefont {Alexandrou}}, \bibinfo {author} {\bibfnamefont {S.}~\bibnamefont {Arunachalam}}, \bibinfo {author} {\bibfnamefont {C.~W.}\ \bibnamefont {Bauer}}, \bibinfo {author} {\bibfnamefont {K.}~\bibnamefont {Borras}}, \bibinfo {author} {\bibfnamefont {S.}~\bibnamefont {Carrazza}}, \bibinfo {author} {\bibfnamefont {A.}~\bibnamefont {Crippa}}, \bibinfo {author} {\bibfnamefont {V.}~\bibnamefont {Croft}}, \bibinfo {author} {\bibfnamefont {R.}~\bibnamefont {de~Putter}}, \bibinfo {author} {\bibfnamefont {A.}~\bibnamefont {Delgado}}, \bibinfo {author} {\bibfnamefont {V.}~\bibnamefont {Dunjko}}, \bibinfo {author} {\bibfnamefont {D.~J.}\ \bibnamefont {Egger}}, \bibinfo {author} {\bibfnamefont {E.}~\bibnamefont {Fern\'andez-Combarro}}, \bibinfo {author}
  {\bibfnamefont {E.}~\bibnamefont {Fuchs}}, \bibinfo {author} {\bibfnamefont {L.}~\bibnamefont {Funcke}}, \bibinfo {author} {\bibfnamefont {D.}~\bibnamefont {Gonz\'alez-Cuadra}}, \bibinfo {author} {\bibfnamefont {M.}~\bibnamefont {Grossi}}, \bibinfo {author} {\bibfnamefont {J.~C.}\ \bibnamefont {Halimeh}}, \bibinfo {author} {\bibfnamefont {Z.}~\bibnamefont {Holmes}}, \bibinfo {author} {\bibfnamefont {S.}~\bibnamefont {K\"uhn}}, \bibinfo {author} {\bibfnamefont {D.}~\bibnamefont {Lacroix}}, \bibinfo {author} {\bibfnamefont {R.}~\bibnamefont {Lewis}}, \bibinfo {author} {\bibfnamefont {D.}~\bibnamefont {Lucchesi}}, \bibinfo {author} {\bibfnamefont {M.~L.}\ \bibnamefont {Martinez}}, \bibinfo {author} {\bibfnamefont {F.}~\bibnamefont {Meloni}}, \bibinfo {author} {\bibfnamefont {A.}~\bibnamefont {Mezzacapo}}, \bibinfo {author} {\bibfnamefont {S.}~\bibnamefont {Montangero}}, \bibinfo {author} {\bibfnamefont {L.}~\bibnamefont {Nagano}}, \bibinfo {author} {\bibfnamefont {V.~R.}\ \bibnamefont {Pascuzzi}}, \bibinfo
  {author} {\bibfnamefont {V.}~\bibnamefont {Radescu}}, \bibinfo {author} {\bibfnamefont {E.~R.}\ \bibnamefont {Ortega}}, \bibinfo {author} {\bibfnamefont {A.}~\bibnamefont {Roggero}}, \bibinfo {author} {\bibfnamefont {J.}~\bibnamefont {Schuhmacher}}, \bibinfo {author} {\bibfnamefont {J.}~\bibnamefont {Seixas}}, \bibinfo {author} {\bibfnamefont {P.}~\bibnamefont {Silvi}}, \bibinfo {author} {\bibfnamefont {P.}~\bibnamefont {Spentzouris}}, \bibinfo {author} {\bibfnamefont {F.}~\bibnamefont {Tacchino}}, \bibinfo {author} {\bibfnamefont {K.}~\bibnamefont {Temme}}, \bibinfo {author} {\bibfnamefont {K.}~\bibnamefont {Terashi}}, \bibinfo {author} {\bibfnamefont {J.}~\bibnamefont {Tura}}, \bibinfo {author} {\bibfnamefont {C.}~\bibnamefont {T\"uys\"uz}}, \bibinfo {author} {\bibfnamefont {S.}~\bibnamefont {Vallecorsa}}, \bibinfo {author} {\bibfnamefont {U.-J.}\ \bibnamefont {Wiese}}, \bibinfo {author} {\bibfnamefont {S.}~\bibnamefont {Yoo}},\ and\ \bibinfo {author} {\bibfnamefont {J.}~\bibnamefont {Zhang}},\ }\bibfield
   {title} {\bibinfo {title} {Quantum computing for high-energy physics: State of the art and challenges},\ }\href {https://doi.org/10.1103/PRXQuantum.5.037001} {\bibfield  {journal} {\bibinfo  {journal} {PRX Quantum}\ }\textbf {\bibinfo {volume} {5}},\ \bibinfo {pages} {037001} (\bibinfo {year} {2024})}\BibitemShut {NoStop}%
\bibitem [{\citenamefont {Funcke}\ \emph {et~al.}(2023)\citenamefont {Funcke}, \citenamefont {Hartung}, \citenamefont {Jansen},\ and\ \citenamefont {K\"uhn}}]{Funcke:2023jbq}%
  \BibitemOpen
  \bibfield  {author} {\bibinfo {author} {\bibfnamefont {L.}~\bibnamefont {Funcke}}, \bibinfo {author} {\bibfnamefont {T.}~\bibnamefont {Hartung}}, \bibinfo {author} {\bibfnamefont {K.}~\bibnamefont {Jansen}},\ and\ \bibinfo {author} {\bibfnamefont {S.}~\bibnamefont {K\"uhn}},\ }\bibfield  {title} {\bibinfo {title} {{Review on Quantum Computing for Lattice Field Theory}},\ }\href {https://doi.org/10.22323/1.430.0228} {\bibfield  {journal} {\bibinfo  {journal} {PoS}\ }\textbf {\bibinfo {volume} {LATTICE2022}},\ \bibinfo {pages} {228} (\bibinfo {year} {2023})},\ \Eprint {https://arxiv.org/abs/2302.00467} {arXiv:2302.00467 [hep-lat]} \BibitemShut {NoStop}%
\bibitem [{\citenamefont {Banerjee}\ \emph {et~al.}(2012)\citenamefont {Banerjee}, \citenamefont {Dalmonte}, \citenamefont {M\"uller}, \citenamefont {Rico}, \citenamefont {Stebler}, \citenamefont {Wiese},\ and\ \citenamefont {Zoller}}]{PhysRevLett.109.175302}%
  \BibitemOpen
  \bibfield  {author} {\bibinfo {author} {\bibfnamefont {D.}~\bibnamefont {Banerjee}}, \bibinfo {author} {\bibfnamefont {M.}~\bibnamefont {Dalmonte}}, \bibinfo {author} {\bibfnamefont {M.}~\bibnamefont {M\"uller}}, \bibinfo {author} {\bibfnamefont {E.}~\bibnamefont {Rico}}, \bibinfo {author} {\bibfnamefont {P.}~\bibnamefont {Stebler}}, \bibinfo {author} {\bibfnamefont {U.-J.}\ \bibnamefont {Wiese}},\ and\ \bibinfo {author} {\bibfnamefont {P.}~\bibnamefont {Zoller}},\ }\bibfield  {title} {\bibinfo {title} {Atomic quantum simulation of dynamical gauge fields coupled to fermionic matter: From string breaking to evolution after a quench},\ }\href {https://doi.org/10.1103/PhysRevLett.109.175302} {\bibfield  {journal} {\bibinfo  {journal} {Phys. Rev. Lett.}\ }\textbf {\bibinfo {volume} {109}},\ \bibinfo {pages} {175302} (\bibinfo {year} {2012})}\BibitemShut {NoStop}%
\bibitem [{\citenamefont {Wiese}(2013)}]{wiese2013ultracold}%
  \BibitemOpen
  \bibfield  {author} {\bibinfo {author} {\bibfnamefont {U.-J.}\ \bibnamefont {Wiese}},\ }\bibfield  {title} {\bibinfo {title} {Ultracold quantum gases and lattice systems: quantum simulation of lattice gauge theories},\ }\href@noop {} {\bibfield  {journal} {\bibinfo  {journal} {Annalen der Physik}\ }\textbf {\bibinfo {volume} {525}},\ \bibinfo {pages} {777} (\bibinfo {year} {2013})}\BibitemShut {NoStop}%
\bibitem [{\citenamefont {Zohar}\ \emph {et~al.}(2015)\citenamefont {Zohar}, \citenamefont {Cirac},\ and\ \citenamefont {Reznik}}]{zohar2015quantum}%
  \BibitemOpen
  \bibfield  {author} {\bibinfo {author} {\bibfnamefont {E.}~\bibnamefont {Zohar}}, \bibinfo {author} {\bibfnamefont {J.~I.}\ \bibnamefont {Cirac}},\ and\ \bibinfo {author} {\bibfnamefont {B.}~\bibnamefont {Reznik}},\ }\bibfield  {title} {\bibinfo {title} {Quantum simulations of lattice gauge theories using ultracold atoms in optical lattices},\ }\href@noop {} {\bibfield  {journal} {\bibinfo  {journal} {Reports on Progress in Physics}\ }\textbf {\bibinfo {volume} {79}},\ \bibinfo {pages} {014401} (\bibinfo {year} {2015})}\BibitemShut {NoStop}%
\bibitem [{\citenamefont {Magnifico}\ \emph {et~al.}(2021)\citenamefont {Magnifico}, \citenamefont {Felser}, \citenamefont {Silvi},\ and\ \citenamefont {Montangero}}]{magnifico2021lattice}%
  \BibitemOpen
  \bibfield  {author} {\bibinfo {author} {\bibfnamefont {G.}~\bibnamefont {Magnifico}}, \bibinfo {author} {\bibfnamefont {T.}~\bibnamefont {Felser}}, \bibinfo {author} {\bibfnamefont {P.}~\bibnamefont {Silvi}},\ and\ \bibinfo {author} {\bibfnamefont {S.}~\bibnamefont {Montangero}},\ }\bibfield  {title} {\bibinfo {title} {Lattice quantum electrodynamics in (3+ 1)-dimensions at finite density with tensor networks},\ }\href@noop {} {\bibfield  {journal} {\bibinfo  {journal} {Nature communications}\ }\textbf {\bibinfo {volume} {12}},\ \bibinfo {pages} {3600} (\bibinfo {year} {2021})}\BibitemShut {NoStop}%
\bibitem [{\citenamefont {Zohar}\ and\ \citenamefont {Reznik}(2011)}]{zohar2011confinement}%
  \BibitemOpen
  \bibfield  {author} {\bibinfo {author} {\bibfnamefont {E.}~\bibnamefont {Zohar}}\ and\ \bibinfo {author} {\bibfnamefont {B.}~\bibnamefont {Reznik}},\ }\bibfield  {title} {\bibinfo {title} {Confinement and lattice quantum-electrodynamic electric flux tubes simulated with ultracold atoms},\ }\href@noop {} {\bibfield  {journal} {\bibinfo  {journal} {Physical review letters}\ }\textbf {\bibinfo {volume} {107}},\ \bibinfo {pages} {275301} (\bibinfo {year} {2011})}\BibitemShut {NoStop}%
\bibitem [{\citenamefont {Cochran}\ \emph {et~al.}(2024)\citenamefont {Cochran} \emph {et~al.}}]{cochran2024visualizingdynamicschargesstrings}%
  \BibitemOpen
  \bibfield  {author} {\bibinfo {author} {\bibfnamefont {T.~A.}\ \bibnamefont {Cochran}} \emph {et~al.},\ }\href {https://arxiv.org/abs/2409.17142} {\bibinfo {title} {Visualizing dynamics of charges and strings in (2+1)d lattice gauge theories}} (\bibinfo {year} {2024}),\ \Eprint {https://arxiv.org/abs/2409.17142} {arXiv:2409.17142 [quant-ph]} \BibitemShut {NoStop}%
\bibitem [{\citenamefont {De}\ \emph {et~al.}(2024)\citenamefont {De} \emph {et~al.}}]{De:2024smi}%
  \BibitemOpen
  \bibfield  {author} {\bibinfo {author} {\bibfnamefont {A.}~\bibnamefont {De}} \emph {et~al.},\ }\bibfield  {title} {\bibinfo {title} {{Observation of string-breaking dynamics in a quantum simulator}},\ }\href@noop {} {\  (\bibinfo {year} {2024})},\ \Eprint {https://arxiv.org/abs/2410.13815} {arXiv:2410.13815 [quant-ph]} \BibitemShut {NoStop}%
\bibitem [{\citenamefont {Gonzalez-Cuadra}\ \emph {et~al.}(2024)\citenamefont {Gonzalez-Cuadra}, \citenamefont {Hamdan}, \citenamefont {Zache}, \citenamefont {Braverman}, \citenamefont {Kornjaca}, \citenamefont {Lukin}, \citenamefont {Cantu}, \citenamefont {Liu}, \citenamefont {Wang}, \citenamefont {Keesling}, \citenamefont {Lukin}, \citenamefont {Zoller},\ and\ \citenamefont {Bylinskii}}]{gonzalezcuadra2024observationstringbreaking2}%
  \BibitemOpen
  \bibfield  {author} {\bibinfo {author} {\bibfnamefont {D.}~\bibnamefont {Gonzalez-Cuadra}}, \bibinfo {author} {\bibfnamefont {M.}~\bibnamefont {Hamdan}}, \bibinfo {author} {\bibfnamefont {T.~V.}\ \bibnamefont {Zache}}, \bibinfo {author} {\bibfnamefont {B.}~\bibnamefont {Braverman}}, \bibinfo {author} {\bibfnamefont {M.}~\bibnamefont {Kornjaca}}, \bibinfo {author} {\bibfnamefont {A.}~\bibnamefont {Lukin}}, \bibinfo {author} {\bibfnamefont {S.~H.}\ \bibnamefont {Cantu}}, \bibinfo {author} {\bibfnamefont {F.}~\bibnamefont {Liu}}, \bibinfo {author} {\bibfnamefont {S.-T.}\ \bibnamefont {Wang}}, \bibinfo {author} {\bibfnamefont {A.}~\bibnamefont {Keesling}}, \bibinfo {author} {\bibfnamefont {M.~D.}\ \bibnamefont {Lukin}}, \bibinfo {author} {\bibfnamefont {P.}~\bibnamefont {Zoller}},\ and\ \bibinfo {author} {\bibfnamefont {A.}~\bibnamefont {Bylinskii}},\ }\href {https://arxiv.org/abs/2410.16558} {\bibinfo {title} {Observation of string breaking on a (2 + 1)d rydberg quantum simulator}} (\bibinfo {year} {2024}),\
  \Eprint {https://arxiv.org/abs/2410.16558} {arXiv:2410.16558 [quant-ph]} \BibitemShut {NoStop}%
\bibitem [{qua(2024{\natexlab{a}})}]{quantinuum_H1_1}%
  \BibitemOpen
  \href {https://www.quantinuum.com/products-solutions/system-model-h1-series} {\bibinfo {title} {Quantinuum {H1-1}}} (\bibinfo {year} {2024}{\natexlab{a}})\BibitemShut {NoStop}%
\bibitem [{\citenamefont {Kogut}\ and\ \citenamefont {Susskind}(1975)}]{PhysRevD.11.395}%
  \BibitemOpen
  \bibfield  {author} {\bibinfo {author} {\bibfnamefont {J.}~\bibnamefont {Kogut}}\ and\ \bibinfo {author} {\bibfnamefont {L.}~\bibnamefont {Susskind}},\ }\bibfield  {title} {\bibinfo {title} {Hamiltonian formulation of wilson's lattice gauge theories},\ }\href {https://doi.org/10.1103/PhysRevD.11.395} {\bibfield  {journal} {\bibinfo  {journal} {Phys. Rev. D}\ }\textbf {\bibinfo {volume} {11}},\ \bibinfo {pages} {395} (\bibinfo {year} {1975})}\BibitemShut {NoStop}%
\bibitem [{\citenamefont {Robson}\ and\ \citenamefont {Webber}(1980)}]{robson1980gauge}%
  \BibitemOpen
  \bibfield  {author} {\bibinfo {author} {\bibfnamefont {D.}~\bibnamefont {Robson}}\ and\ \bibinfo {author} {\bibfnamefont {D.}~\bibnamefont {Webber}},\ }\bibfield  {title} {\bibinfo {title} {Gauge theories on a small lattice},\ }\href {https://doi.org/doi.org/10.1007/BF01577321} {\bibfield  {journal} {\bibinfo  {journal} {Zeitschrift f{\"u}r Physik C Particles and Fields}\ }\textbf {\bibinfo {volume} {7}},\ \bibinfo {pages} {53} (\bibinfo {year} {1980})}\BibitemShut {NoStop}%
\bibitem [{\citenamefont {Ligterink}\ \emph {et~al.}(2000)\citenamefont {Ligterink}, \citenamefont {Walet},\ and\ \citenamefont {Bishop}}]{ligterink2000many}%
  \BibitemOpen
  \bibfield  {author} {\bibinfo {author} {\bibfnamefont {N.}~\bibnamefont {Ligterink}}, \bibinfo {author} {\bibfnamefont {N.}~\bibnamefont {Walet}},\ and\ \bibinfo {author} {\bibfnamefont {R.}~\bibnamefont {Bishop}},\ }\bibfield  {title} {\bibinfo {title} {A many-body treatment of hamiltonian lattice gauge theory},\ }\href {https://doi.org/10.1006/aphy.2000.6070} {\bibfield  {journal} {\bibinfo  {journal} {Nuclear physics. A, Nuclear and hadronic physics}\ }\textbf {\bibinfo {volume} {663}},\ \bibinfo {pages} {983c} (\bibinfo {year} {2000})}\BibitemShut {NoStop}%
\bibitem [{\citenamefont {Nielsen}\ and\ \citenamefont {Ninomiya}(1981)}]{Nielsen:1981hk}%
  \BibitemOpen
  \bibfield  {author} {\bibinfo {author} {\bibfnamefont {H.~B.}\ \bibnamefont {Nielsen}}\ and\ \bibinfo {author} {\bibfnamefont {M.}~\bibnamefont {Ninomiya}},\ }\bibfield  {title} {\bibinfo {title} {{No Go Theorem for Regularizing Chiral Fermions}},\ }\href {https://doi.org/10.1016/0370-2693(81)91026-1} {\bibfield  {journal} {\bibinfo  {journal} {Phys. Lett. B}\ }\textbf {\bibinfo {volume} {105}},\ \bibinfo {pages} {219} (\bibinfo {year} {1981})}\BibitemShut {NoStop}%
\bibitem [{\citenamefont {Susskind}(1977)}]{PhysRevD.16.3031}%
  \BibitemOpen
  \bibfield  {author} {\bibinfo {author} {\bibfnamefont {L.}~\bibnamefont {Susskind}},\ }\bibfield  {title} {\bibinfo {title} {Lattice fermions},\ }\href {https://doi.org/10.1103/PhysRevD.16.3031} {\bibfield  {journal} {\bibinfo  {journal} {Phys. Rev. D}\ }\textbf {\bibinfo {volume} {16}},\ \bibinfo {pages} {3031} (\bibinfo {year} {1977})}\BibitemShut {NoStop}%
\bibitem [{\citenamefont {Haase}\ \emph {et~al.}(2021)\citenamefont {Haase}, \citenamefont {Dellantonio}, \citenamefont {Celi}, \citenamefont {Paulson}, \citenamefont {Kan}, \citenamefont {Jansen},\ and\ \citenamefont {Muschik}}]{Haase2021resourceefficient}%
  \BibitemOpen
  \bibfield  {author} {\bibinfo {author} {\bibfnamefont {J.~F.}\ \bibnamefont {Haase}}, \bibinfo {author} {\bibfnamefont {L.}~\bibnamefont {Dellantonio}}, \bibinfo {author} {\bibfnamefont {A.}~\bibnamefont {Celi}}, \bibinfo {author} {\bibfnamefont {D.}~\bibnamefont {Paulson}}, \bibinfo {author} {\bibfnamefont {A.}~\bibnamefont {Kan}}, \bibinfo {author} {\bibfnamefont {K.}~\bibnamefont {Jansen}},\ and\ \bibinfo {author} {\bibfnamefont {C.~A.}\ \bibnamefont {Muschik}},\ }\bibfield  {title} {\bibinfo {title} {A resource efficient approach for quantum and classical simulations of gauge theories in particle physics},\ }\href {https://doi.org/10.22331/q-2021-02-04-393} {\bibfield  {journal} {\bibinfo  {journal} {{Quantum}}\ }\textbf {\bibinfo {volume} {5}},\ \bibinfo {pages} {393} (\bibinfo {year} {2021})}\BibitemShut {NoStop}%
\bibitem [{\citenamefont {Mathis}\ \emph {et~al.}(2020)\citenamefont {Mathis}, \citenamefont {Mazzola},\ and\ \citenamefont {Tavernelli}}]{PhysRevD.102.094501}%
  \BibitemOpen
  \bibfield  {author} {\bibinfo {author} {\bibfnamefont {S.~V.}\ \bibnamefont {Mathis}}, \bibinfo {author} {\bibfnamefont {G.}~\bibnamefont {Mazzola}},\ and\ \bibinfo {author} {\bibfnamefont {I.}~\bibnamefont {Tavernelli}},\ }\bibfield  {title} {\bibinfo {title} {Toward scalable simulations of lattice gauge theories on quantum computers},\ }\href {https://doi.org/10.1103/PhysRevD.102.094501} {\bibfield  {journal} {\bibinfo  {journal} {Phys. Rev. D}\ }\textbf {\bibinfo {volume} {102}},\ \bibinfo {pages} {094501} (\bibinfo {year} {2020})}\BibitemShut {NoStop}%
\bibitem [{\citenamefont {Chandrasekharan}\ and\ \citenamefont {Wiese}(1997)}]{chandrasekharan1997quantum}%
  \BibitemOpen
  \bibfield  {author} {\bibinfo {author} {\bibfnamefont {S.}~\bibnamefont {Chandrasekharan}}\ and\ \bibinfo {author} {\bibfnamefont {U.-J.}\ \bibnamefont {Wiese}},\ }\bibfield  {title} {\bibinfo {title} {Quantum link models: A discrete approach to gauge theories},\ }\href@noop {} {\bibfield  {journal} {\bibinfo  {journal} {Nuclear Physics B}\ }\textbf {\bibinfo {volume} {492}},\ \bibinfo {pages} {455} (\bibinfo {year} {1997})}\BibitemShut {NoStop}%
\bibitem [{\citenamefont {Hashizume}\ \emph {et~al.}(2022)\citenamefont {Hashizume}, \citenamefont {Halimeh}, \citenamefont {Hauke},\ and\ \citenamefont {Banerjee}}]{Hashizume_2022}%
  \BibitemOpen
  \bibfield  {author} {\bibinfo {author} {\bibfnamefont {T.}~\bibnamefont {Hashizume}}, \bibinfo {author} {\bibfnamefont {J.}~\bibnamefont {Halimeh}}, \bibinfo {author} {\bibfnamefont {P.}~\bibnamefont {Hauke}},\ and\ \bibinfo {author} {\bibfnamefont {D.}~\bibnamefont {Banerjee}},\ }\bibfield  {title} {\bibinfo {title} {Ground-state phase diagram of quantum link electrodynamics in $(2+1)$-d},\ }\bibfield  {journal} {\bibinfo  {journal} {SciPost Physics}\ }\textbf {\bibinfo {volume} {13}},\ \href {https://doi.org/10.21468/scipostphys.13.2.017} {10.21468/scipostphys.13.2.017} (\bibinfo {year} {2022})\BibitemShut {NoStop}%
\bibitem [{\citenamefont {Notarnicola}\ \emph {et~al.}(2015)\citenamefont {Notarnicola}, \citenamefont {Ercolessi}, \citenamefont {Facchi}, \citenamefont {Marmo}, \citenamefont {Pascazio},\ and\ \citenamefont {Pepe}}]{notarnicola2015discrete}%
  \BibitemOpen
  \bibfield  {author} {\bibinfo {author} {\bibfnamefont {S.}~\bibnamefont {Notarnicola}}, \bibinfo {author} {\bibfnamefont {E.}~\bibnamefont {Ercolessi}}, \bibinfo {author} {\bibfnamefont {P.}~\bibnamefont {Facchi}}, \bibinfo {author} {\bibfnamefont {G.}~\bibnamefont {Marmo}}, \bibinfo {author} {\bibfnamefont {S.}~\bibnamefont {Pascazio}},\ and\ \bibinfo {author} {\bibfnamefont {F.~V.}\ \bibnamefont {Pepe}},\ }\bibfield  {title} {\bibinfo {title} {Discrete abelian gauge theories for quantum simulations of qed},\ }\href@noop {} {\bibfield  {journal} {\bibinfo  {journal} {Journal of Physics A: Mathematical and Theoretical}\ }\textbf {\bibinfo {volume} {48}},\ \bibinfo {pages} {30FT01} (\bibinfo {year} {2015})}\BibitemShut {NoStop}%
\bibitem [{\citenamefont {Meth}\ \emph {et~al.}(2023)\citenamefont {Meth}, \citenamefont {Haase}, \citenamefont {Zhang}, \citenamefont {Edmunds}, \citenamefont {Postler}, \citenamefont {Steiner}, \citenamefont {Jena}, \citenamefont {Dellantonio}, \citenamefont {Blatt}, \citenamefont {Zoller} \emph {et~al.}}]{meth2023simulating}%
  \BibitemOpen
  \bibfield  {author} {\bibinfo {author} {\bibfnamefont {M.}~\bibnamefont {Meth}}, \bibinfo {author} {\bibfnamefont {J.~F.}\ \bibnamefont {Haase}}, \bibinfo {author} {\bibfnamefont {J.}~\bibnamefont {Zhang}}, \bibinfo {author} {\bibfnamefont {C.}~\bibnamefont {Edmunds}}, \bibinfo {author} {\bibfnamefont {L.}~\bibnamefont {Postler}}, \bibinfo {author} {\bibfnamefont {A.}~\bibnamefont {Steiner}}, \bibinfo {author} {\bibfnamefont {A.~J.}\ \bibnamefont {Jena}}, \bibinfo {author} {\bibfnamefont {L.}~\bibnamefont {Dellantonio}}, \bibinfo {author} {\bibfnamefont {R.}~\bibnamefont {Blatt}}, \bibinfo {author} {\bibfnamefont {P.}~\bibnamefont {Zoller}}, \emph {et~al.},\ }\bibfield  {title} {\bibinfo {title} {Simulating 2d lattice gauge theories on a qudit quantum computer},\ }\href@noop {} {\bibfield  {journal} {\bibinfo  {journal} {arXiv preprint arXiv:2310.12110}\ } (\bibinfo {year} {2023})}\BibitemShut {NoStop}%
\bibitem [{\citenamefont {Crippa}(2024)}]{QEDHamiltrepo}%
  \BibitemOpen
  \bibfield  {author} {\bibinfo {author} {\bibfnamefont {A.}~\bibnamefont {Crippa}},\ }\href {https://doi.org/10.5281/zenodo.11046038} {\bibinfo {title} {\text{QED\_Hamiltonian/v0.0.2}}} (\bibinfo {year} {2024})\BibitemShut {NoStop}%
\bibitem [{\citenamefont {Paulson}\ \emph {et~al.}(2021)\citenamefont {Paulson}, \citenamefont {Dellantonio}, \citenamefont {Haase}, \citenamefont {Celi}, \citenamefont {Kan}, \citenamefont {Jena}, \citenamefont {Kokail}, \citenamefont {van Bijnen}, \citenamefont {Jansen}, \citenamefont {Zoller},\ and\ \citenamefont {Muschik}}]{PRXQuantum.2.030334}%
  \BibitemOpen
  \bibfield  {author} {\bibinfo {author} {\bibfnamefont {D.}~\bibnamefont {Paulson}}, \bibinfo {author} {\bibfnamefont {L.}~\bibnamefont {Dellantonio}}, \bibinfo {author} {\bibfnamefont {J.~F.}\ \bibnamefont {Haase}}, \bibinfo {author} {\bibfnamefont {A.}~\bibnamefont {Celi}}, \bibinfo {author} {\bibfnamefont {A.}~\bibnamefont {Kan}}, \bibinfo {author} {\bibfnamefont {A.}~\bibnamefont {Jena}}, \bibinfo {author} {\bibfnamefont {C.}~\bibnamefont {Kokail}}, \bibinfo {author} {\bibfnamefont {R.}~\bibnamefont {van Bijnen}}, \bibinfo {author} {\bibfnamefont {K.}~\bibnamefont {Jansen}}, \bibinfo {author} {\bibfnamefont {P.}~\bibnamefont {Zoller}},\ and\ \bibinfo {author} {\bibfnamefont {C.~A.}\ \bibnamefont {Muschik}},\ }\bibfield  {title} {\bibinfo {title} {Simulating 2d effects in lattice gauge theories on a quantum computer},\ }\href {https://doi.org/10.1103/PRXQuantum.2.030334} {\bibfield  {journal} {\bibinfo  {journal} {PRX Quantum}\ }\textbf {\bibinfo {volume} {2}},\ \bibinfo {pages} {030334} (\bibinfo {year}
  {2021})}\BibitemShut {NoStop}%
\bibitem [{\citenamefont {Di~Matteo}\ \emph {et~al.}(2021)\citenamefont {Di~Matteo}, \citenamefont {McCoy}, \citenamefont {Gysbers}, \citenamefont {Miyagi}, \citenamefont {Woloshyn},\ and\ \citenamefont {Navr\'atil}}]{PhysRevA.103.042405}%
  \BibitemOpen
  \bibfield  {author} {\bibinfo {author} {\bibfnamefont {O.}~\bibnamefont {Di~Matteo}}, \bibinfo {author} {\bibfnamefont {A.}~\bibnamefont {McCoy}}, \bibinfo {author} {\bibfnamefont {P.}~\bibnamefont {Gysbers}}, \bibinfo {author} {\bibfnamefont {T.}~\bibnamefont {Miyagi}}, \bibinfo {author} {\bibfnamefont {R.~M.}\ \bibnamefont {Woloshyn}},\ and\ \bibinfo {author} {\bibfnamefont {P.}~\bibnamefont {Navr\'atil}},\ }\bibfield  {title} {\bibinfo {title} {Improving hamiltonian encodings with the gray code},\ }\href {https://doi.org/10.1103/PhysRevA.103.042405} {\bibfield  {journal} {\bibinfo  {journal} {Phys. Rev. A}\ }\textbf {\bibinfo {volume} {103}},\ \bibinfo {pages} {042405} (\bibinfo {year} {2021})}\BibitemShut {NoStop}%
\bibitem [{\citenamefont {Jordan}\ and\ \citenamefont {Wigner}(1993)}]{jordan1993paulische}%
  \BibitemOpen
  \bibfield  {author} {\bibinfo {author} {\bibfnamefont {P.}~\bibnamefont {Jordan}}\ and\ \bibinfo {author} {\bibfnamefont {E.~P.}\ \bibnamefont {Wigner}},\ }\href@noop {} {\emph {\bibinfo {title} {{\"U}ber das paulische {\"a}quivalenzverbot}}}\ (\bibinfo  {publisher} {Springer},\ \bibinfo {year} {1993})\BibitemShut {NoStop}%
\bibitem [{\citenamefont {Nakanishi}\ \emph {et~al.}(2020)\citenamefont {Nakanishi}, \citenamefont {Fujii},\ and\ \citenamefont {Todo}}]{PhysRevResearch.2.043158}%
  \BibitemOpen
  \bibfield  {author} {\bibinfo {author} {\bibfnamefont {K.~M.}\ \bibnamefont {Nakanishi}}, \bibinfo {author} {\bibfnamefont {K.}~\bibnamefont {Fujii}},\ and\ \bibinfo {author} {\bibfnamefont {S.}~\bibnamefont {Todo}},\ }\bibfield  {title} {\bibinfo {title} {Sequential minimal optimization for quantum-classical hybrid algorithms},\ }\href {https://doi.org/10.1103/PhysRevResearch.2.043158} {\bibfield  {journal} {\bibinfo  {journal} {Phys. Rev. Res.}\ }\textbf {\bibinfo {volume} {2}},\ \bibinfo {pages} {043158} (\bibinfo {year} {2020})}\BibitemShut {NoStop}%
\bibitem [{qua(2024{\natexlab{b}})}]{quantinuum_nexus}%
  \BibitemOpen
  \href {https://nexus.quantinuum.com/} {\bibinfo {title} {Quantinuum nexus}} (\bibinfo {year} {2024}{\natexlab{b}})\BibitemShut {NoStop}%
\bibitem [{\citenamefont {Pino}\ \emph {et~al.}(2021)\citenamefont {Pino} \emph {et~al.}}]{Pino:2020mku}%
  \BibitemOpen
  \bibfield  {author} {\bibinfo {author} {\bibfnamefont {J.~M.}\ \bibnamefont {Pino}} \emph {et~al.},\ }\bibfield  {title} {\bibinfo {title} {{Demonstration of the trapped-ion quantum CCD computer architecture}},\ }\href {https://doi.org/10.1038/s41586-021-03318-4} {\bibfield  {journal} {\bibinfo  {journal} {Nature}\ }\textbf {\bibinfo {volume} {592}},\ \bibinfo {pages} {209} (\bibinfo {year} {2021})},\ \Eprint {https://arxiv.org/abs/2003.01293} {arXiv:2003.01293 [quant-ph]} \BibitemShut {NoStop}%
\bibitem [{\citenamefont {Ryan-Anderson}(2018)}]{pecos}%
  \BibitemOpen
  \bibfield  {author} {\bibinfo {author} {\bibfnamefont {C.}~\bibnamefont {Ryan-Anderson}},\ }\href {https://github.com/PECOS-packages/PECOS} {\bibinfo {title} {Pecos: Performance estimator of codes on surfaces}},\ \bibinfo {howpublished} {\url{https://github.com/PECOS-packages/PECOS}} (\bibinfo {year} {2018})\BibitemShut {NoStop}%
\bibitem [{\citenamefont {Ryan-Anderson}\ \emph {et~al.}(2021)\citenamefont {Ryan-Anderson}, \citenamefont {Bohnet}, \citenamefont {Lee}, \citenamefont {Gresh}, \citenamefont {Hankin}, \citenamefont {Gaebler}, \citenamefont {Francois}, \citenamefont {Chernoguzov}, \citenamefont {Lucchetti}, \citenamefont {Brown}, \citenamefont {Gatterman}, \citenamefont {Halit}, \citenamefont {Gilmore}, \citenamefont {Gerber}, \citenamefont {Neyenhuis}, \citenamefont {Hayes},\ and\ \citenamefont {Stutz}}]{RyanAnderson2021}%
  \BibitemOpen
  \bibfield  {author} {\bibinfo {author} {\bibfnamefont {C.}~\bibnamefont {Ryan-Anderson}}, \bibinfo {author} {\bibfnamefont {J.~G.}\ \bibnamefont {Bohnet}}, \bibinfo {author} {\bibfnamefont {K.}~\bibnamefont {Lee}}, \bibinfo {author} {\bibfnamefont {D.}~\bibnamefont {Gresh}}, \bibinfo {author} {\bibfnamefont {A.}~\bibnamefont {Hankin}}, \bibinfo {author} {\bibfnamefont {J.~P.}\ \bibnamefont {Gaebler}}, \bibinfo {author} {\bibfnamefont {D.}~\bibnamefont {Francois}}, \bibinfo {author} {\bibfnamefont {A.}~\bibnamefont {Chernoguzov}}, \bibinfo {author} {\bibfnamefont {D.}~\bibnamefont {Lucchetti}}, \bibinfo {author} {\bibfnamefont {N.~C.}\ \bibnamefont {Brown}}, \bibinfo {author} {\bibfnamefont {T.~M.}\ \bibnamefont {Gatterman}}, \bibinfo {author} {\bibfnamefont {S.~K.}\ \bibnamefont {Halit}}, \bibinfo {author} {\bibfnamefont {K.}~\bibnamefont {Gilmore}}, \bibinfo {author} {\bibfnamefont {J.~A.}\ \bibnamefont {Gerber}}, \bibinfo {author} {\bibfnamefont {B.}~\bibnamefont {Neyenhuis}}, \bibinfo {author}
  {\bibfnamefont {D.}~\bibnamefont {Hayes}},\ and\ \bibinfo {author} {\bibfnamefont {R.~P.}\ \bibnamefont {Stutz}},\ }\bibfield  {title} {\bibinfo {title} {Realization of real-time fault-tolerant quantum error correction},\ }\href {https://doi.org/10.1103/PhysRevX.11.041058} {\bibfield  {journal} {\bibinfo  {journal} {Phys. Rev. X}\ }\textbf {\bibinfo {volume} {11}},\ \bibinfo {pages} {041058} (\bibinfo {year} {2021})}\BibitemShut {NoStop}%
\bibitem [{\citenamefont {Yamamoto}\ \emph {et~al.}(2022)\citenamefont {Yamamoto}, \citenamefont {Manrique}, \citenamefont {Khan}, \citenamefont {Sawada},\ and\ \citenamefont {Ramo}}]{Yamamoto_2022}%
  \BibitemOpen
  \bibfield  {author} {\bibinfo {author} {\bibfnamefont {K.}~\bibnamefont {Yamamoto}}, \bibinfo {author} {\bibfnamefont {D.~Z.}\ \bibnamefont {Manrique}}, \bibinfo {author} {\bibfnamefont {I.~T.}\ \bibnamefont {Khan}}, \bibinfo {author} {\bibfnamefont {H.}~\bibnamefont {Sawada}},\ and\ \bibinfo {author} {\bibfnamefont {D.~M.}\ \bibnamefont {Ramo}},\ }\bibfield  {title} {\bibinfo {title} {Quantum hardware calculations of periodic systems with partition-measurement symmetry verification: Simplified models of hydrogen chain and iron crystals},\ }\bibfield  {journal} {\bibinfo  {journal} {Physical Review Research}\ }\textbf {\bibinfo {volume} {4}},\ \href {https://doi.org/10.1103/physrevresearch.4.033110} {10.1103/physrevresearch.4.033110} (\bibinfo {year} {2022})\BibitemShut {NoStop}%
\bibitem [{\citenamefont {Jackson}\ and\ \citenamefont {van Enk}(2015)}]{PhysRevA.92.042312}%
  \BibitemOpen
  \bibfield  {author} {\bibinfo {author} {\bibfnamefont {C.}~\bibnamefont {Jackson}}\ and\ \bibinfo {author} {\bibfnamefont {S.~J.}\ \bibnamefont {van Enk}},\ }\bibfield  {title} {\bibinfo {title} {Detecting correlated errors in state-preparation-and-measurement tomography},\ }\href {https://doi.org/10.1103/PhysRevA.92.042312} {\bibfield  {journal} {\bibinfo  {journal} {Phys. Rev. A}\ }\textbf {\bibinfo {volume} {92}},\ \bibinfo {pages} {042312} (\bibinfo {year} {2015})}\BibitemShut {NoStop}%
\bibitem [{\citenamefont {Tranter}\ \emph {et~al.}(2022)\citenamefont {Tranter}, \citenamefont {Di~Paola}, \citenamefont {Mu\~{n}oz Ramo}, \citenamefont {Manrique}, \citenamefont {Gowland}, \citenamefont {Plekhanov}, \citenamefont {Greene-Diniz}, \citenamefont {Christopoulou}, \citenamefont {Prokopiou}, \citenamefont {Keen}, \citenamefont {Polyak}, \citenamefont {Khan}, \citenamefont {Pilipczuk}, \citenamefont {Kirsopp}, \citenamefont {Yamamoto}, \citenamefont {Tudorovskaya}, \citenamefont {Krompiec}, \citenamefont {Sze}, \citenamefont {Fitzpatrick}, \citenamefont {Anderson},\ and\ \citenamefont {Bhasker}}]{inquanto}%
  \BibitemOpen
  \bibfield  {author} {\bibinfo {author} {\bibfnamefont {A.}~\bibnamefont {Tranter}}, \bibinfo {author} {\bibfnamefont {C.}~\bibnamefont {Di~Paola}}, \bibinfo {author} {\bibfnamefont {D.}~\bibnamefont {Mu\~{n}oz Ramo}}, \bibinfo {author} {\bibfnamefont {D.~Z.}\ \bibnamefont {Manrique}}, \bibinfo {author} {\bibfnamefont {D.}~\bibnamefont {Gowland}}, \bibinfo {author} {\bibfnamefont {E.}~\bibnamefont {Plekhanov}}, \bibinfo {author} {\bibfnamefont {G.}~\bibnamefont {Greene-Diniz}}, \bibinfo {author} {\bibfnamefont {G.}~\bibnamefont {Christopoulou}}, \bibinfo {author} {\bibfnamefont {G.}~\bibnamefont {Prokopiou}}, \bibinfo {author} {\bibfnamefont {H.~D.~J.}\ \bibnamefont {Keen}}, \bibinfo {author} {\bibfnamefont {I.}~\bibnamefont {Polyak}}, \bibinfo {author} {\bibfnamefont {I.~T.}\ \bibnamefont {Khan}}, \bibinfo {author} {\bibfnamefont {J.}~\bibnamefont {Pilipczuk}}, \bibinfo {author} {\bibfnamefont {J.~J.~M.}\ \bibnamefont {Kirsopp}}, \bibinfo {author} {\bibfnamefont {K.}~\bibnamefont {Yamamoto}}, \bibinfo
  {author} {\bibfnamefont {M.}~\bibnamefont {Tudorovskaya}}, \bibinfo {author} {\bibfnamefont {M.}~\bibnamefont {Krompiec}}, \bibinfo {author} {\bibfnamefont {M.}~\bibnamefont {Sze}}, \bibinfo {author} {\bibfnamefont {N.}~\bibnamefont {Fitzpatrick}}, \bibinfo {author} {\bibfnamefont {R.~J.}\ \bibnamefont {Anderson}},\ and\ \bibinfo {author} {\bibfnamefont {V.}~\bibnamefont {Bhasker}},\ }\href {https://www.quantinuum.com/products-solutions/inquanto} {\bibinfo {title} {{InQuanto: Quantum Computational Chemistry}}} (\bibinfo {year} {2022})\BibitemShut {NoStop}%
\bibitem [{\citenamefont {Peruzzo}\ \emph {et~al.}(2014)\citenamefont {Peruzzo}, \citenamefont {McClean}, \citenamefont {Shadbolt}, \citenamefont {Yung}, \citenamefont {Zhou}, \citenamefont {Love}, \citenamefont {Aspuru-Guzik},\ and\ \citenamefont {O’brien}}]{peruzzo2014variational}%
  \BibitemOpen
  \bibfield  {author} {\bibinfo {author} {\bibfnamefont {A.}~\bibnamefont {Peruzzo}}, \bibinfo {author} {\bibfnamefont {J.}~\bibnamefont {McClean}}, \bibinfo {author} {\bibfnamefont {P.}~\bibnamefont {Shadbolt}}, \bibinfo {author} {\bibfnamefont {M.-H.}\ \bibnamefont {Yung}}, \bibinfo {author} {\bibfnamefont {X.-Q.}\ \bibnamefont {Zhou}}, \bibinfo {author} {\bibfnamefont {P.~J.}\ \bibnamefont {Love}}, \bibinfo {author} {\bibfnamefont {A.}~\bibnamefont {Aspuru-Guzik}},\ and\ \bibinfo {author} {\bibfnamefont {J.~L.}\ \bibnamefont {O’brien}},\ }\bibfield  {title} {\bibinfo {title} {A variational eigenvalue solver on a photonic quantum processor},\ }\href@noop {} {\bibfield  {journal} {\bibinfo  {journal} {Nature communications}\ }\textbf {\bibinfo {volume} {5}},\ \bibinfo {pages} {4213} (\bibinfo {year} {2014})}\BibitemShut {NoStop}%
\bibitem [{\citenamefont {Ostaszewski}\ \emph {et~al.}(2021)\citenamefont {Ostaszewski}, \citenamefont {Grant},\ and\ \citenamefont {Benedetti}}]{Ostaszewski2021structure}%
  \BibitemOpen
  \bibfield  {author} {\bibinfo {author} {\bibfnamefont {M.}~\bibnamefont {Ostaszewski}}, \bibinfo {author} {\bibfnamefont {E.}~\bibnamefont {Grant}},\ and\ \bibinfo {author} {\bibfnamefont {M.}~\bibnamefont {Benedetti}},\ }\bibfield  {title} {\bibinfo {title} {Structure optimization for parameterized quantum circuits},\ }\href {https://doi.org/10.22331/q-2021-01-28-391} {\bibfield  {journal} {\bibinfo  {journal} {{Quantum}}\ }\textbf {\bibinfo {volume} {5}},\ \bibinfo {pages} {391} (\bibinfo {year} {2021})}\BibitemShut {NoStop}%
\bibitem [{\citenamefont {Kohda}\ \emph {et~al.}(2022)\citenamefont {Kohda}, \citenamefont {Imai}, \citenamefont {Kanno}, \citenamefont {Mitarai}, \citenamefont {Mizukami},\ and\ \citenamefont {Nakagawa}}]{PhysRevResearch.4.033173}%
  \BibitemOpen
  \bibfield  {author} {\bibinfo {author} {\bibfnamefont {M.}~\bibnamefont {Kohda}}, \bibinfo {author} {\bibfnamefont {R.}~\bibnamefont {Imai}}, \bibinfo {author} {\bibfnamefont {K.}~\bibnamefont {Kanno}}, \bibinfo {author} {\bibfnamefont {K.}~\bibnamefont {Mitarai}}, \bibinfo {author} {\bibfnamefont {W.}~\bibnamefont {Mizukami}},\ and\ \bibinfo {author} {\bibfnamefont {Y.~O.}\ \bibnamefont {Nakagawa}},\ }\bibfield  {title} {\bibinfo {title} {Quantum expectation-value estimation by computational basis sampling},\ }\href {https://doi.org/10.1103/PhysRevResearch.4.033173} {\bibfield  {journal} {\bibinfo  {journal} {Phys. Rev. Res.}\ }\textbf {\bibinfo {volume} {4}},\ \bibinfo {pages} {033173} (\bibinfo {year} {2022})}\BibitemShut {NoStop}%
\bibitem [{\citenamefont {DeCross}\ \emph {et~al.}(2023)\citenamefont {DeCross}, \citenamefont {Chertkov}, \citenamefont {Kohagen},\ and\ \citenamefont {Foss-Feig}}]{PhysRevX.13.041057}%
  \BibitemOpen
  \bibfield  {author} {\bibinfo {author} {\bibfnamefont {M.}~\bibnamefont {DeCross}}, \bibinfo {author} {\bibfnamefont {E.}~\bibnamefont {Chertkov}}, \bibinfo {author} {\bibfnamefont {M.}~\bibnamefont {Kohagen}},\ and\ \bibinfo {author} {\bibfnamefont {M.}~\bibnamefont {Foss-Feig}},\ }\bibfield  {title} {\bibinfo {title} {Qubit-reuse compilation with mid-circuit measurement and reset},\ }\href {https://doi.org/10.1103/PhysRevX.13.041057} {\bibfield  {journal} {\bibinfo  {journal} {Phys. Rev. X}\ }\textbf {\bibinfo {volume} {13}},\ \bibinfo {pages} {041057} (\bibinfo {year} {2023})}\BibitemShut {NoStop}%
\bibitem [{\citenamefont {Sivarajah}\ \emph {et~al.}(2020)\citenamefont {Sivarajah}, \citenamefont {Dilkes}, \citenamefont {Cowtan}, \citenamefont {Simmons}, \citenamefont {Edgington},\ and\ \citenamefont {Duncan}}]{Sivarajah:2020lfo}%
  \BibitemOpen
  \bibfield  {author} {\bibinfo {author} {\bibfnamefont {S.}~\bibnamefont {Sivarajah}}, \bibinfo {author} {\bibfnamefont {S.}~\bibnamefont {Dilkes}}, \bibinfo {author} {\bibfnamefont {A.}~\bibnamefont {Cowtan}}, \bibinfo {author} {\bibfnamefont {W.}~\bibnamefont {Simmons}}, \bibinfo {author} {\bibfnamefont {A.}~\bibnamefont {Edgington}},\ and\ \bibinfo {author} {\bibfnamefont {R.}~\bibnamefont {Duncan}},\ }\bibfield  {title} {\bibinfo {title} {{t|ket\ensuremath{\rangle}: a retargetable compiler for NISQ devices}},\ }\href {https://doi.org/10.1088/2058-9565/ab8e92} {\bibfield  {journal} {\bibinfo  {journal} {Quantum Sci. Technol.}\ }\textbf {\bibinfo {volume} {6}},\ \bibinfo {pages} {014003} (\bibinfo {year} {2020})},\ \Eprint {https://arxiv.org/abs/2003.10611} {arXiv:2003.10611 [quant-ph]} \BibitemShut {NoStop}%
\bibitem [{qua(2024{\natexlab{c}})}]{quantinuum_H2_1}%
  \BibitemOpen
  \href {https://www.quantinuum.com/products-solutions/system-model-h2-series} {\bibinfo {title} {Quantinuum {H2-1}}} (\bibinfo {year} {2024}{\natexlab{c}})\BibitemShut {NoStop}%
\bibitem [{\citenamefont {Crippa}\ \emph {et~al.}(2024)\citenamefont {Crippa}, \citenamefont {Romiti}, \citenamefont {Funcke}, \citenamefont {Jansen}, \citenamefont {K\"uhn}, \citenamefont {Stornati},\ and\ \citenamefont {Urbach}}]{Crippa:2024cqr}%
  \BibitemOpen
  \bibfield  {author} {\bibinfo {author} {\bibfnamefont {A.}~\bibnamefont {Crippa}}, \bibinfo {author} {\bibfnamefont {S.}~\bibnamefont {Romiti}}, \bibinfo {author} {\bibfnamefont {L.}~\bibnamefont {Funcke}}, \bibinfo {author} {\bibfnamefont {K.}~\bibnamefont {Jansen}}, \bibinfo {author} {\bibfnamefont {S.}~\bibnamefont {K\"uhn}}, \bibinfo {author} {\bibfnamefont {P.}~\bibnamefont {Stornati}},\ and\ \bibinfo {author} {\bibfnamefont {C.}~\bibnamefont {Urbach}},\ }\bibfield  {title} {\bibinfo {title} {{Towards determining the (2+1)-dimensional Quantum Electrodynamics running coupling with Monte Carlo and quantum computing methods}},\ }\href@noop {} {\  (\bibinfo {year} {2024})},\ \Eprint {https://arxiv.org/abs/2404.17545} {arXiv:2404.17545 [hep-lat]} \BibitemShut {NoStop}%
\bibitem [{\citenamefont {Farrell}\ \emph {et~al.}(2024)\citenamefont {Farrell}, \citenamefont {Illa}, \citenamefont {Ciavarella},\ and\ \citenamefont {Savage}}]{Farrell:2023fgd}%
  \BibitemOpen
  \bibfield  {author} {\bibinfo {author} {\bibfnamefont {R.~C.}\ \bibnamefont {Farrell}}, \bibinfo {author} {\bibfnamefont {M.}~\bibnamefont {Illa}}, \bibinfo {author} {\bibfnamefont {A.~N.}\ \bibnamefont {Ciavarella}},\ and\ \bibinfo {author} {\bibfnamefont {M.~J.}\ \bibnamefont {Savage}},\ }\bibfield  {title} {\bibinfo {title} {{Scalable Circuits for Preparing Ground States on Digital Quantum Computers: The Schwinger Model Vacuum on 100 Qubits}},\ }\href {https://doi.org/10.1103/PRXQuantum.5.020315} {\bibfield  {journal} {\bibinfo  {journal} {PRX Quantum}\ }\textbf {\bibinfo {volume} {5}},\ \bibinfo {pages} {020315} (\bibinfo {year} {2024})},\ \Eprint {https://arxiv.org/abs/2308.04481} {arXiv:2308.04481 [quant-ph]} \BibitemShut {NoStop}%
\bibitem [{\citenamefont {Granet}\ \emph {et~al.}(2024)\citenamefont {Granet}, \citenamefont {Ghanem},\ and\ \citenamefont {Dreyer}}]{Granet:2024kuu}%
  \BibitemOpen
  \bibfield  {author} {\bibinfo {author} {\bibfnamefont {E.}~\bibnamefont {Granet}}, \bibinfo {author} {\bibfnamefont {K.}~\bibnamefont {Ghanem}},\ and\ \bibinfo {author} {\bibfnamefont {H.}~\bibnamefont {Dreyer}},\ }\bibfield  {title} {\bibinfo {title} {{Practicality of quantum adiabatic algorithm for chemistry applications}},\ }\href@noop {} {\  (\bibinfo {year} {2024})},\ \Eprint {https://arxiv.org/abs/2407.09993} {arXiv:2407.09993 [quant-ph]} \BibitemShut {NoStop}%
\bibitem [{\citenamefont {Luescher}(1981)}]{luescher1981symmetry}%
  \BibitemOpen
  \bibfield  {author} {\bibinfo {author} {\bibfnamefont {M.}~\bibnamefont {Luescher}},\ }\bibfield  {title} {\bibinfo {title} {Symmetry-breaking aspects of the roughening transition in gauge theories},\ }\href@noop {} {\bibfield  {journal} {\bibinfo  {journal} {Nuclear Physics B}\ }\textbf {\bibinfo {volume} {180}},\ \bibinfo {pages} {317} (\bibinfo {year} {1981})}\BibitemShut {NoStop}%
\bibitem [{\citenamefont {L{\"u}scher}\ and\ \citenamefont {Weisz}(2002)}]{luscher2002quark}%
  \BibitemOpen
  \bibfield  {author} {\bibinfo {author} {\bibfnamefont {M.}~\bibnamefont {L{\"u}scher}}\ and\ \bibinfo {author} {\bibfnamefont {P.}~\bibnamefont {Weisz}},\ }\bibfield  {title} {\bibinfo {title} {Quark confinement and the bosonic string},\ }\href@noop {} {\bibfield  {journal} {\bibinfo  {journal} {Journal of High Energy Physics}\ }\textbf {\bibinfo {volume} {2002}},\ \bibinfo {pages} {049} (\bibinfo {year} {2002})}\BibitemShut {NoStop}%
\bibitem [{\citenamefont {Nogueira}\ and\ \citenamefont {Kleinert}(2005)}]{nogueira2005quantum}%
  \BibitemOpen
  \bibfield  {author} {\bibinfo {author} {\bibfnamefont {F.~S.}\ \bibnamefont {Nogueira}}\ and\ \bibinfo {author} {\bibfnamefont {H.}~\bibnamefont {Kleinert}},\ }\bibfield  {title} {\bibinfo {title} {Quantum electrodynamics in 2+ 1 dimensions, confinement, and the stability of u (1) spin liquids},\ }\href@noop {} {\bibfield  {journal} {\bibinfo  {journal} {Physical review letters}\ }\textbf {\bibinfo {volume} {95}},\ \bibinfo {pages} {176406} (\bibinfo {year} {2005})}\BibitemShut {NoStop}%
\bibitem [{\citenamefont {Chagdaa}\ \emph {et~al.}(2017)\citenamefont {Chagdaa}, \citenamefont {Galsandorj}, \citenamefont {Laermann},\ and\ \citenamefont {Purev}}]{chagdaa2017width}%
  \BibitemOpen
  \bibfield  {author} {\bibinfo {author} {\bibfnamefont {S.}~\bibnamefont {Chagdaa}}, \bibinfo {author} {\bibfnamefont {E.}~\bibnamefont {Galsandorj}}, \bibinfo {author} {\bibfnamefont {E.}~\bibnamefont {Laermann}},\ and\ \bibinfo {author} {\bibfnamefont {B.}~\bibnamefont {Purev}},\ }\bibfield  {title} {\bibinfo {title} {Width and string tension of the flux tube in su (2) lattice gauge theory at high temperature},\ }\href@noop {} {\bibfield  {journal} {\bibinfo  {journal} {Journal of Physics G: Nuclear and Particle Physics}\ }\textbf {\bibinfo {volume} {45}},\ \bibinfo {pages} {025002} (\bibinfo {year} {2017})}\BibitemShut {NoStop}%
\bibitem [{\citenamefont {Caselle}\ \emph {et~al.}(2024)\citenamefont {Caselle}, \citenamefont {Cellini},\ and\ \citenamefont {Nada}}]{caselle2024numerical}%
  \BibitemOpen
  \bibfield  {author} {\bibinfo {author} {\bibfnamefont {M.}~\bibnamefont {Caselle}}, \bibinfo {author} {\bibfnamefont {E.}~\bibnamefont {Cellini}},\ and\ \bibinfo {author} {\bibfnamefont {A.}~\bibnamefont {Nada}},\ }\bibfield  {title} {\bibinfo {title} {Numerical determination of the width and shape of the effective string using stochastic normalizing flows},\ }\href@noop {} {\bibfield  {journal} {\bibinfo  {journal} {arXiv preprint arXiv:2409.15937}\ } (\bibinfo {year} {2024})}\BibitemShut {NoStop}%
\bibitem [{\citenamefont {Peng}\ \emph {et~al.}(2024)\citenamefont {Peng}, \citenamefont {Diamantini}, \citenamefont {Funcke}, \citenamefont {Hassan}, \citenamefont {Jansen}, \citenamefont {K\"uhn}, \citenamefont {Luo},\ and\ \citenamefont {Naredi}}]{Peng:2024xbl}%
  \BibitemOpen
  \bibfield  {author} {\bibinfo {author} {\bibfnamefont {C.}~\bibnamefont {Peng}}, \bibinfo {author} {\bibfnamefont {M.~C.}\ \bibnamefont {Diamantini}}, \bibinfo {author} {\bibfnamefont {L.}~\bibnamefont {Funcke}}, \bibinfo {author} {\bibfnamefont {S.~M.~A.}\ \bibnamefont {Hassan}}, \bibinfo {author} {\bibfnamefont {K.}~\bibnamefont {Jansen}}, \bibinfo {author} {\bibfnamefont {S.}~\bibnamefont {K\"uhn}}, \bibinfo {author} {\bibfnamefont {D.}~\bibnamefont {Luo}},\ and\ \bibinfo {author} {\bibfnamefont {P.}~\bibnamefont {Naredi}},\ }\bibfield  {title} {\bibinfo {title} {{Hamiltonian Lattice Formulation of Compact Maxwell-Chern-Simons Theory}},\ }\href@noop {} {\  (\bibinfo {year} {2024})},\ \Eprint {https://arxiv.org/abs/2407.20225} {arXiv:2407.20225 [hep-th]} \BibitemShut {NoStop}%
\bibitem [{\citenamefont {Von~Neumann}(2013)}]{from2013mathematical}%
  \BibitemOpen
  \bibfield  {author} {\bibinfo {author} {\bibfnamefont {J.}~\bibnamefont {Von~Neumann}},\ }\href@noop {} {\emph {\bibinfo {title} {Mathematical foundations of quantum mechanics}}},\ Vol.~\bibinfo {volume} {38}\ (\bibinfo  {publisher} {Springer-Verlag},\ \bibinfo {year} {2013})\BibitemShut {NoStop}%
\bibitem [{\citenamefont {Zhang}\ \emph {et~al.}(2021)\citenamefont {Zhang}, \citenamefont {Kyaw}, \citenamefont {Kottmann}, \citenamefont {Degroote},\ and\ \citenamefont {Aspuru-Guzik}}]{Zhang2020}%
  \BibitemOpen
  \bibfield  {author} {\bibinfo {author} {\bibfnamefont {Z.-J.}\ \bibnamefont {Zhang}}, \bibinfo {author} {\bibfnamefont {T.~H.}\ \bibnamefont {Kyaw}}, \bibinfo {author} {\bibfnamefont {J.~S.}\ \bibnamefont {Kottmann}}, \bibinfo {author} {\bibfnamefont {M.}~\bibnamefont {Degroote}},\ and\ \bibinfo {author} {\bibfnamefont {A.}~\bibnamefont {Aspuru-Guzik}},\ }\bibfield  {title} {\bibinfo {title} {Mutual information-assisted adaptive variational quantum eigensolver},\ }\href {https://doi.org/10.1088/2058-9565/abdca4} {\bibfield  {journal} {\bibinfo  {journal} {Quantum Science and technology}\ }\textbf {\bibinfo {volume} {6}},\ \bibinfo {pages} {035001} (\bibinfo {year} {2021})}\BibitemShut {NoStop}%
\bibitem [{\citenamefont {Tkachenko}\ \emph {et~al.}(2021)\citenamefont {Tkachenko}, \citenamefont {Sud}, \citenamefont {Zhang}, \citenamefont {Tretiak}, \citenamefont {Anisimov}, \citenamefont {Arrasmith}, \citenamefont {Coles}, \citenamefont {Cincio},\ and\ \citenamefont {Dub}}]{Tkachenko2020}%
  \BibitemOpen
  \bibfield  {author} {\bibinfo {author} {\bibfnamefont {N.~V.}\ \bibnamefont {Tkachenko}}, \bibinfo {author} {\bibfnamefont {J.}~\bibnamefont {Sud}}, \bibinfo {author} {\bibfnamefont {Y.}~\bibnamefont {Zhang}}, \bibinfo {author} {\bibfnamefont {S.}~\bibnamefont {Tretiak}}, \bibinfo {author} {\bibfnamefont {P.~M.}\ \bibnamefont {Anisimov}}, \bibinfo {author} {\bibfnamefont {A.~T.}\ \bibnamefont {Arrasmith}}, \bibinfo {author} {\bibfnamefont {P.~J.}\ \bibnamefont {Coles}}, \bibinfo {author} {\bibfnamefont {L.}~\bibnamefont {Cincio}},\ and\ \bibinfo {author} {\bibfnamefont {P.~A.}\ \bibnamefont {Dub}},\ }\bibfield  {title} {\bibinfo {title} {Correlation-informed permutation of qubits for reducing ansatz depth in the variational quantum eigensolver},\ }\href {https://doi.org/10.1103/prxquantum.2.020337} {\bibfield  {journal} {\bibinfo  {journal} {PRX Quantum}\ }\textbf {\bibinfo {volume} {2}},\ \bibinfo {pages} {1} (\bibinfo {year} {2021})}\BibitemShut {NoStop}%
\bibitem [{\citenamefont {Materia}\ \emph {et~al.}(2024)\citenamefont {Materia}, \citenamefont {Ratini}, \citenamefont {Angeli},\ and\ \citenamefont {Guidoni}}]{Materia2024}%
  \BibitemOpen
  \bibfield  {author} {\bibinfo {author} {\bibfnamefont {D.}~\bibnamefont {Materia}}, \bibinfo {author} {\bibfnamefont {L.}~\bibnamefont {Ratini}}, \bibinfo {author} {\bibfnamefont {C.}~\bibnamefont {Angeli}},\ and\ \bibinfo {author} {\bibfnamefont {L.}~\bibnamefont {Guidoni}},\ }\bibfield  {title} {\bibinfo {title} {Quantum information driven ansatz (qida): Shallow-depth empirical quantum circuits from quantum chemistry},\ }\bibfield  {journal} {\bibinfo  {journal} {The Journal of Physical Chemistry A}\ }\href {https://doi.org/10.1021/acs.jpca.4c03756} {10.1021/acs.jpca.4c03756} (\bibinfo {year} {2024})\BibitemShut {NoStop}%
\bibitem [{\citenamefont {Bender}\ and\ \citenamefont {Zohar}(2020)}]{PhysRevD.102.114517}%
  \BibitemOpen
  \bibfield  {author} {\bibinfo {author} {\bibfnamefont {J.}~\bibnamefont {Bender}}\ and\ \bibinfo {author} {\bibfnamefont {E.}~\bibnamefont {Zohar}},\ }\bibfield  {title} {\bibinfo {title} {Gauge redundancy-free formulation of compact qed with dynamical matter for quantum and classical computations},\ }\href {https://doi.org/10.1103/PhysRevD.102.114517} {\bibfield  {journal} {\bibinfo  {journal} {Phys. Rev. D}\ }\textbf {\bibinfo {volume} {102}},\ \bibinfo {pages} {114517} (\bibinfo {year} {2020})}\BibitemShut {NoStop}%
\bibitem [{\citenamefont {Kaplan}\ and\ \citenamefont {Stryker}(2020)}]{PhysRevD.102.094515}%
  \BibitemOpen
  \bibfield  {author} {\bibinfo {author} {\bibfnamefont {D.~B.}\ \bibnamefont {Kaplan}}\ and\ \bibinfo {author} {\bibfnamefont {J.~R.}\ \bibnamefont {Stryker}},\ }\bibfield  {title} {\bibinfo {title} {Gauss's law, duality, and the hamiltonian formulation of u(1) lattice gauge theory},\ }\href {https://doi.org/10.1103/PhysRevD.102.094515} {\bibfield  {journal} {\bibinfo  {journal} {Phys. Rev. D}\ }\textbf {\bibinfo {volume} {102}},\ \bibinfo {pages} {094515} (\bibinfo {year} {2020})}\BibitemShut {NoStop}%
\bibitem [{\citenamefont {Clemente}\ \emph {et~al.}(2022)\citenamefont {Clemente}, \citenamefont {Crippa},\ and\ \citenamefont {Jansen}}]{PhysRevD.106.114511}%
  \BibitemOpen
  \bibfield  {author} {\bibinfo {author} {\bibfnamefont {G.}~\bibnamefont {Clemente}}, \bibinfo {author} {\bibfnamefont {A.}~\bibnamefont {Crippa}},\ and\ \bibinfo {author} {\bibfnamefont {K.}~\bibnamefont {Jansen}},\ }\bibfield  {title} {\bibinfo {title} {Strategies for the determination of the running coupling of ($2+1$)-dimensional qed with quantum computing},\ }\href {https://doi.org/10.1103/PhysRevD.106.114511} {\bibfield  {journal} {\bibinfo  {journal} {Phys. Rev. D}\ }\textbf {\bibinfo {volume} {106}},\ \bibinfo {pages} {114511} (\bibinfo {year} {2022})}\BibitemShut {NoStop}%
\end{thebibliography}%

\clearpage
\newpage

\onecolumngrid
\begin{appendix}

\section{Quantum circuit definition}
\label{app:quantumcirc}

In order to build the quantum circuit we consider an analysis of the mutual information (MI) with exact diagonalization, and with the intent of using the results in larger systems.

\begin{figure*}[htp!]
    \centering
    \includegraphics[width=1\textwidth]{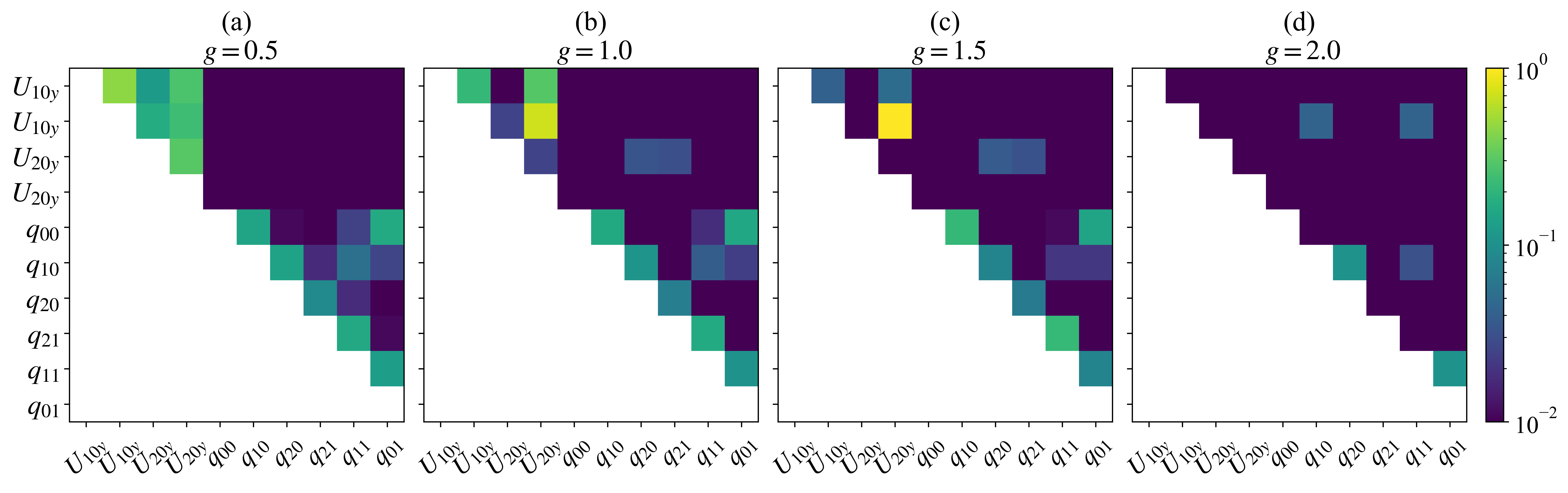}
    \caption{\textbf{Mutual information between the qubits:} The MI is computed with exact diagonalization for every pair of qubits at four values of $g=0.5,1.0,1.5,2.0$, panel $(a),(b),(c),(d)$ respectively. An explanatory example is panel $(d)$ where we have larger values ($>10^{-2}$) for qubits related to gauge field $U_{10y}$ and sites $q_{10},q_{11}$. Connection is present also between the sites $q_{01},q_{11}$ and $q_{10},q_{20}$. }
    \label{fig:mutualinfo}
\end{figure*}

\begin{figure*}[htp!]
    \centering
    \includegraphics[width=0.95\textwidth]{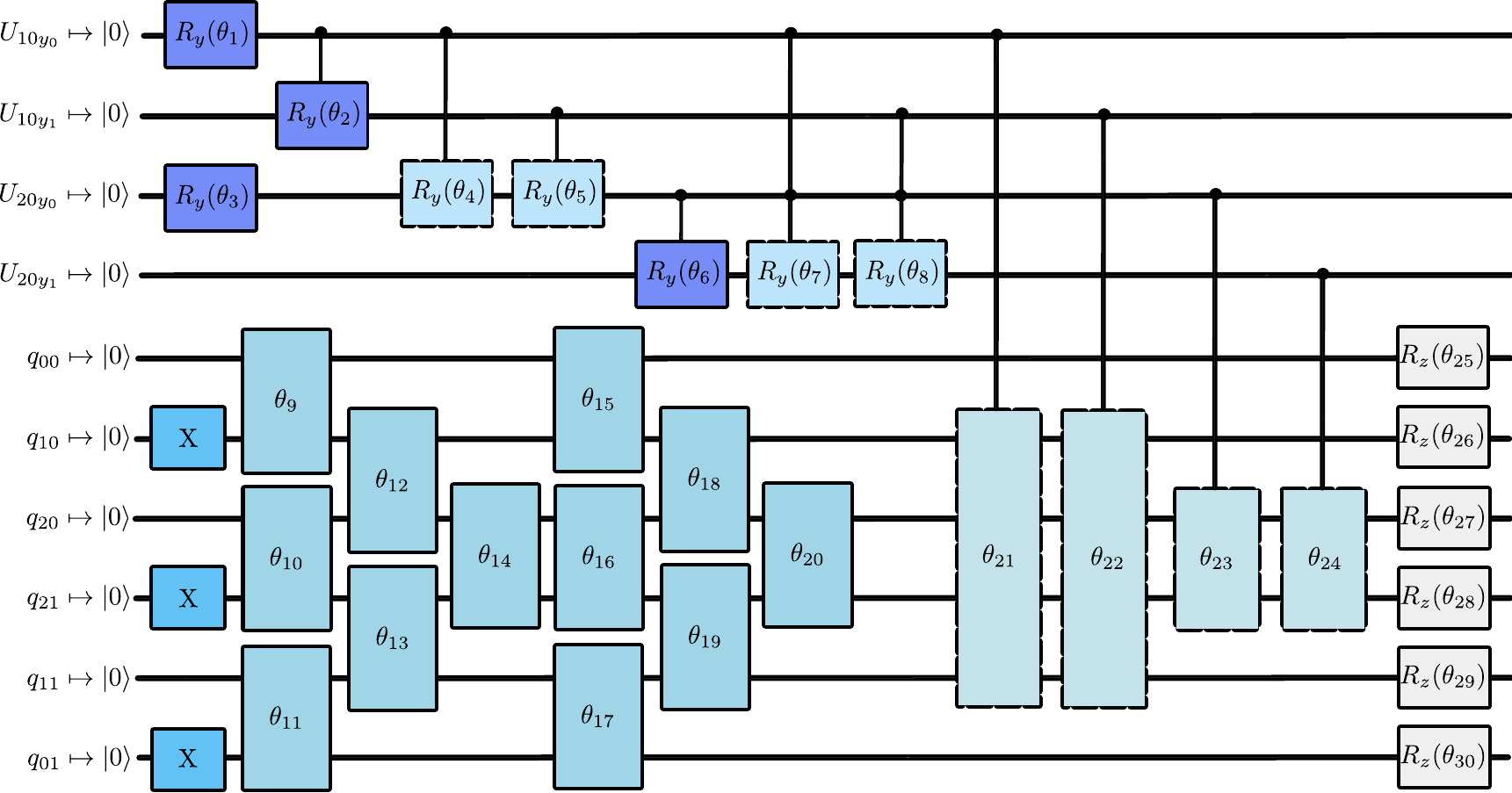}
    \caption{\textbf{Parameterized quantum circuit for $3\times 2$ lattice:} The circuit has two main parts, the upper four qubits represent the gauge fields $U_{10y},U_{20y}$ with $l=1$ and with relative entanglement structure. The lower six qubits show the fermions $q_{\vec{n}}$ ($\vec{n}=(n_x,n_y)$ site coordinates). Two layers of iSWAP gates (parameterized gates with $\theta_i$, $i\in [9,20]$ \textit{solid lines}) and the entanglement (parameterized gates with $\theta_i$, $i\in [21,24]$, \textit{dashed lines}) with the gauge fields are applied to the circuit. A final layer of rotational gates $R_z(\theta)$ correct eventual relative phases. The entangling gates are highlighted with dashed frames.}
    \label{fig:qcircuit3x2}
\end{figure*}

The mutual information quantifies the shared information between two variables, \(X\) and \(Y\). It is computed by adding the individual information content, also known as the \textit{von Neumann entropy}~\cite{from2013mathematical}, of \(X\), denoted as \(S(X)\), and \(Y\), denoted as \(S(Y)\). In this process, the joint information content of \(X\) and \(Y\), denoted as \(S(X, Y)\), is counted twice, while the unique information of each variable is counted only once. To correct for this double counting, we subtract the joint information \(S(X, Y)\). Hence, the mutual information \(I(X; Y)\) is expressed as
\begin{equation}
  I(X; Y) = S(X) + S(Y) - S(X, Y).
\end{equation}
This equation captures the amount of information shared between \(X\) and \(Y\).
By studying the mutual information, we can identify the correlated components (fermions and gauge fields) within the lattice, facilitating the design of a custom structure for the entanglement in the quantum circuit. A similar approach has been explored to study the entanglement in quantum chemistry in Refs.~\cite{Zhang2020,Tkachenko2020,Materia2024}.
Our analysis is carried out for four values of the coupling, i.e. $g=0.5,1.0,1.5,2.0$,  and the results are depicted in Fig.~\ref{fig:mutualinfo}. 
The couplings are chosen to cover the three main regions of the static potential (Coulomb, linear electric strings, string breaking) and to show the transition through these phases. 
In the first panel $(a)$ we are in the Coulomb regime and we can read the following information from the plot: the qubits that represent the gauge fields ($U_{10y}$ and $U_{20y}$) have a nonzero value of MI, thus we put entangling gates between them. The correlation decreases through panels $(b)$ and $(c)$ and becomes lower than $0.01 $ in the string breaking case, panel $(d)$.
The MI for fermionic fields shows a large correlation, particularly for the sites connected with a link, panels $(a),(b)$ and smaller values in $(c)$. 
In panel $(d)$, we have larger values ($>10^{-2}$) for qubits related to the gauge field $U_{10y}$ and the sites $q_{10}$ and $q_{11}$, and also between the sites $q_{01}\leftrightarrow q_{11}$ and $q_{10}\leftrightarrow q_{20}$. This corresponds to the coupling where two dynamical charges form onto the two fermionic sites $q_{00}$ and $q_{21}$, leaving them in a less connected position, i.e. the MI, and thus the entanglement, with respect to the rest of the lattice is small.  
The quantum circuit we consider for the analysis of a wide range of couplings is depicted in Fig.~\ref{fig:qcircuit3x2}. With this structure we can reduce the depth of the circuit while preserving its expressivity.

\section{Shot number dependence}
\label{app:noisemodel}

\begin{figure*}[htp!]
    \centering
    \includegraphics[width=0.9\linewidth]{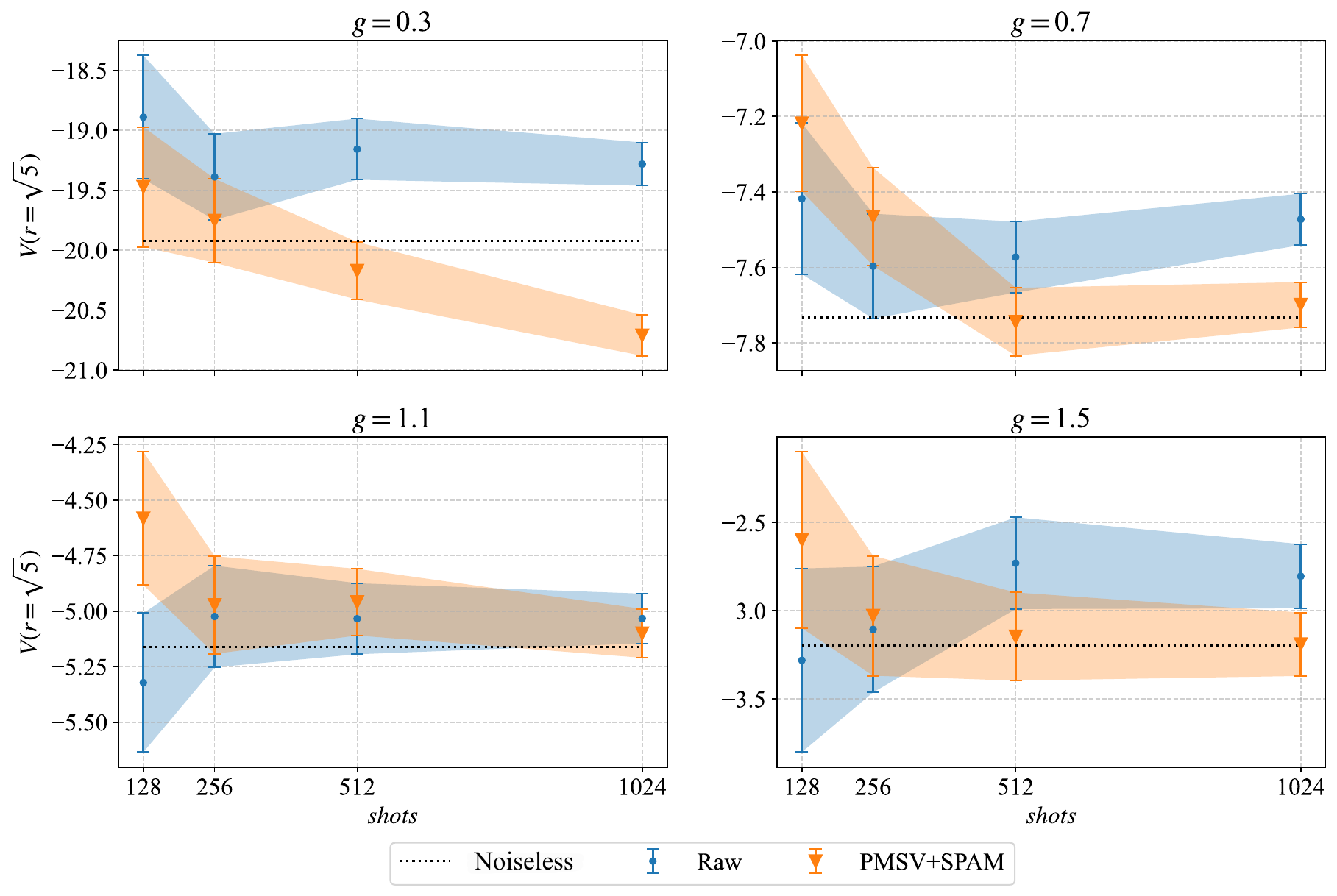}
    \caption{\textbf{Study of the number of shots:} The static potential values are computed with the emulator H1-1E and $2^i$ with $i=[7,8,9,10]$ shots. The circles represent raw data, without any error mitigation applied, while the triangles are obtained with a combination of PMSV and SPAM. The horizontal dotted lines are noiseless results.}
    \label{fig:h1e1_vqe_ed_pmsvspam}
\end{figure*}

We perform an in-depth analysis of the computation of the static potential using a selected range of shots to study the behavior of different error mitigation techniques.
For this study we use the H1-1E emulator and the PMSV and SPAM error mitigation techniques introduced in Sec.~\ref{subsec:noise_mitig}.
The results are depicted in Fig.~\ref{fig:h1e1_vqe_ed_pmsvspam} for four values of the coupling $g=0.3,0.7,1.1,1.5$. On the $x$-axis we consider the range of shots $2^i$ with $i=[7,8,9,10]$.
The dotted horizontal lines correspond to the results obtained with a state vector simulation using the optimal parameters in the VQE Ansatz. 
The \textit{circle points} are raw data, obtained without any error mitigation, and are connected by shaded bands to guide the eye.
The \textit{triangle points} are results after applying both PMSV and SPAM mitigation techniques.
Note that for each number of shots on the horizontal axis, we run a single job to compute the expectation value of the Hamiltonian, and we report the statistical shot noise as described in the main text. 
The results may deviate across multiple runs and these runs were performed with the H1-1E emulator in September 2024.

\section{Truncation and Gauss's law dependence}\label{app:truncation}

This section discusses the dependence of the results on the truncation parameter chosen. Here we consider three values of $l=1,3,7$ (cases where we need to exclude only a single unphysical state) and compute the static potential for the $3\times 2$ lattice.

\begin{figure}[htp!]
    \centering
    \includegraphics[width=0.55\textwidth]{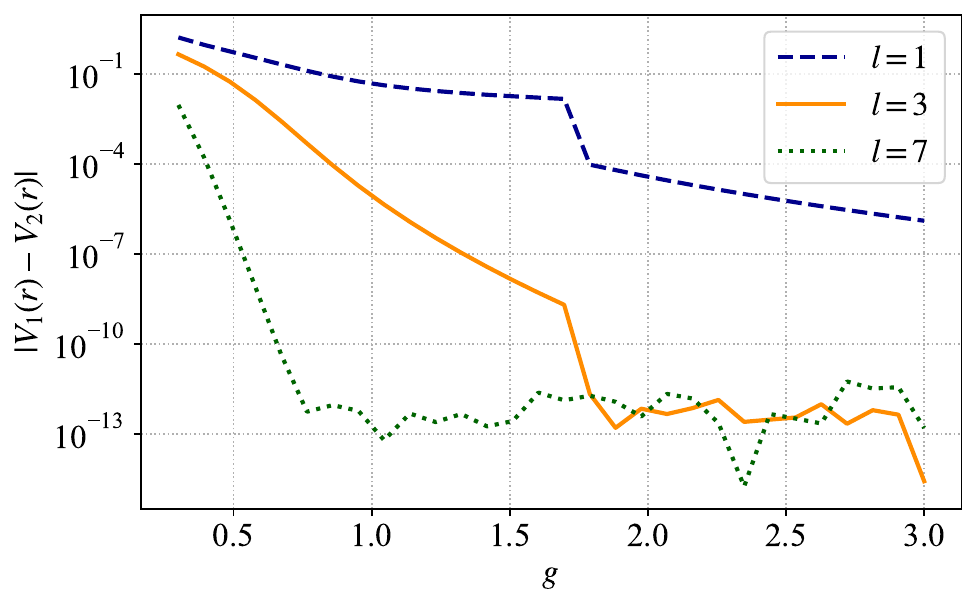}
    \caption{\textbf{Difference of static potentials:} We consider two subsets of dynamical links when Gauss's law is applied and compute $V_1(r)$ and $V_2(r)$ with ED and with the set of dynamical links $\{E_{10y},E_{20y}\}$ and $\{E_{00y},E_{20y}\}$, respectively. The difference in the results decreases when the truncation is larger.}
    \label{fig:energy_diff_dyn_links}
\end{figure}

We have seen in Sec.~\ref{sec:hamilt}, that we can reduce the resources required for the analysis when applying Gauss's law. This procedure leads to a subset of dynamical links which corresponds to two in the case of an OBC system of $3 \times 2$.
The choice of the subset is arbitrary, however, it is important to consider that it may affect the results. 
Depicted in Fig.~\ref{fig:energy_diff_dyn_links}, we show the absolute difference, as a function of the coupling, between the static potential $V_1(r)$ and $V_2(r)$ computed with ED and with the set of dynamical links $\{E_{10y},E_{20y}\}$ and $\{E_{00y},E_{20y}\}$, respectively.

The difference between the two configurations decreases when a large truncation is considered and for data at larger $g$. 
Note that, even if the numerical values of the probabilities may differ by the choice of dynamical links, the presence of electric flux tubes and the string breaking phenomenon is not affected. Thus, the physical phenomena can be studied qualitatively also at small truncations.  

\begin{figure}[htp!]
    \centering
    \includegraphics[width=0.55\columnwidth]{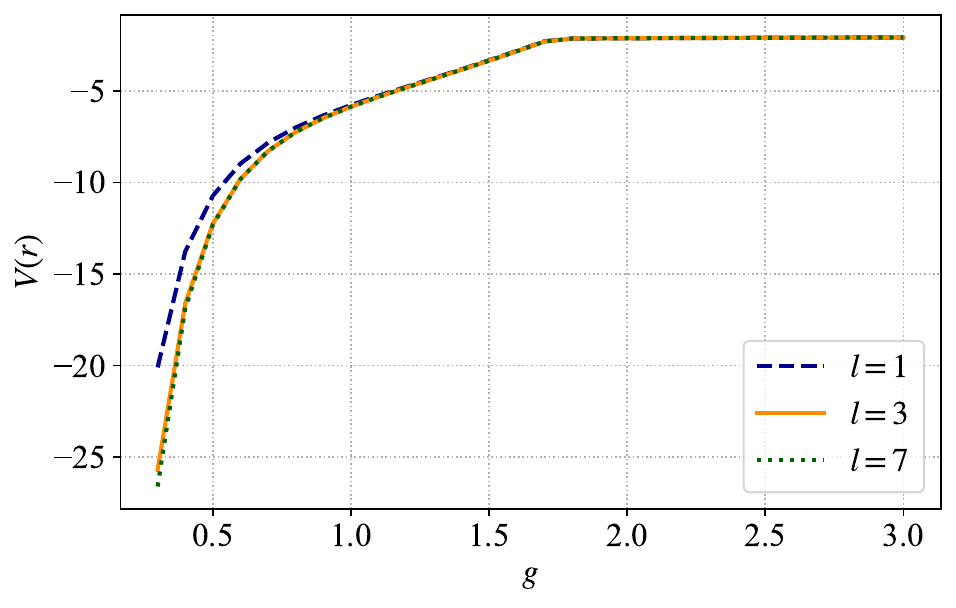}
    \caption{\textbf{Static potential with different truncation values $l$:} Exact diagonalization analysis shows that for the linear and constant part of the static potential ($g \gtrsim 1.0$), truncation $l=1$ can give accurate results. In the weak coupling regime, a higher $l$ is required to improve the accuracy. This behavior is expected, as we are working in the electric basis.}
    \label{fig:vr_truncation}
\end{figure}

As depicted in Fig.~\ref{fig:vr_truncation}, to obtain accurate results, a larger truncation is needed when we consider small values of the bare coupling $g$. At larger $g \gtrsim 1.0$, the electric flux string breaking results with $l=1$ are compatible with larger truncations.

\begin{figure}[htp!]
    \centering
    \includegraphics[width=0.3\linewidth]{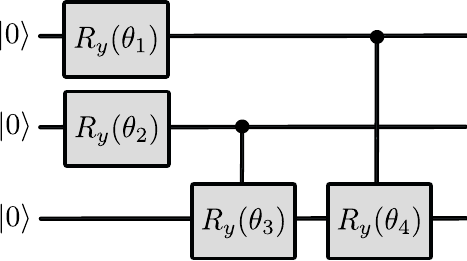}
    \caption{\textbf{Variational circuit for Gray encoding, truncation $l=3$:} $\ket{010}$ represents the \textit{vacuum state} and the state $\ket{100}$ excluded.}
    \label{fig:grayl3}
\end{figure}

\begin{figure}[htp!]
    \centering
    \includegraphics[width=0.6\columnwidth]{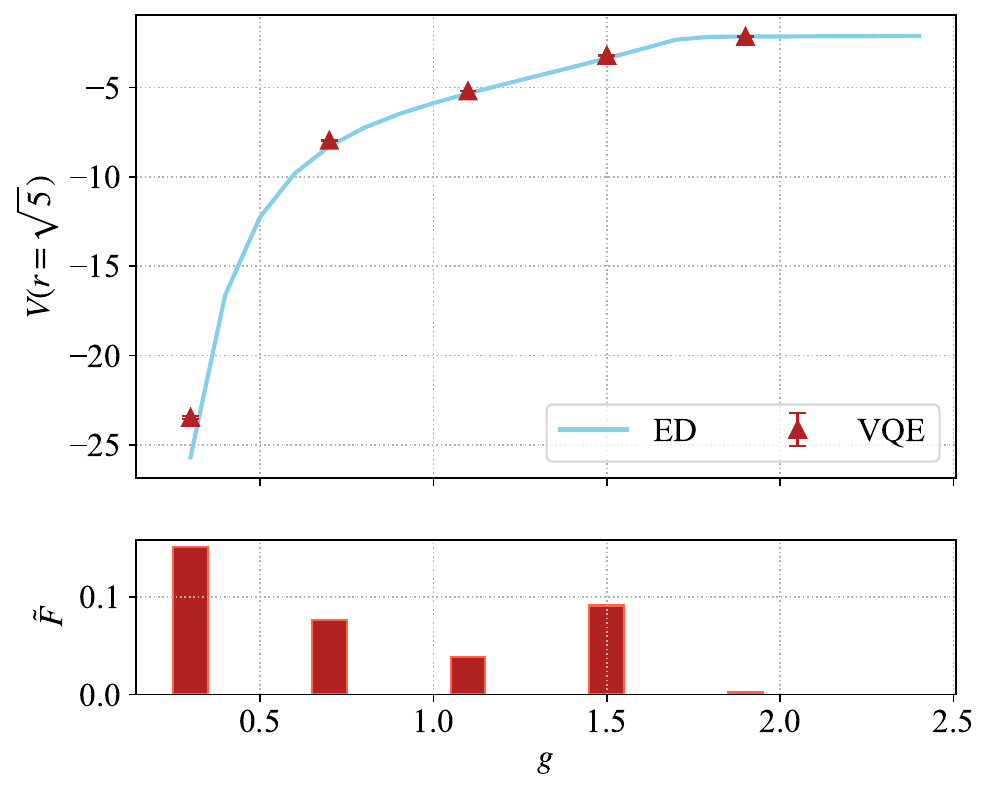}
    \caption{\textbf{Variational quantum results $3\times 2$ system, $l=3$:}(\textit{top panel}) Static potential at different coupling $g$ with ED (\textit{solid line}) and quantum variational results (\textit{triangles}), performed with NFT optimizer and $10^4$ shots. (\textit{bottom panel}) Infidelity ($1-$fidelity) between variational quantum data and ED. The error bars are smaller than the markers.}
    \label{fig:vr_l3}
\end{figure}

We also apply the VQE for a larger truncation, $l=3$. The resources required are described in the main text in Table~\ref{tab:res3x2obc}, where we can see that we need 3 qubits for each gauge field. We utilize the parameterized circuit of Fig.~\ref{fig:grayl3} to exclude the unphysical states for each gauge field with this truncation. In particular, we exclude the unphysical state $\ket{100}$.
The \textit{upper panel} in Fig.~\ref{fig:vr_l3} illustrates the comparison between the variational quantum results and exact diagonalization, while the \textit{lower panel} shows the infidelity, Eq.~\eqref{eq:infidelity}. 
The variational quantum results can reproduce the general trend of the static potential curve. However, at small $g$, they do not align accurately with the ED value, with a fidelity of $\sim 85 \%$. Since in this region the ground state superposition is large, we expect to need more entanglement between the qubits.
A possible method to reduce the state superposition can be the application of a magnetic basis formulation, see e.g. Ref.~\cite{Haase2021resourceefficient,PhysRevD.102.114517,PhysRevD.102.094515,PhysRevD.106.114511}.

\section{Variational quantum circuit for $4\times 3$ lattice}\label{app:4x3circuit}
In this section we report the explicit structure of the quantum circuit used to compute the quantum variational results for the $4\times 3$ system.

\begin{figure}[H]
    \centering
    \includegraphics[width=0.85\linewidth]{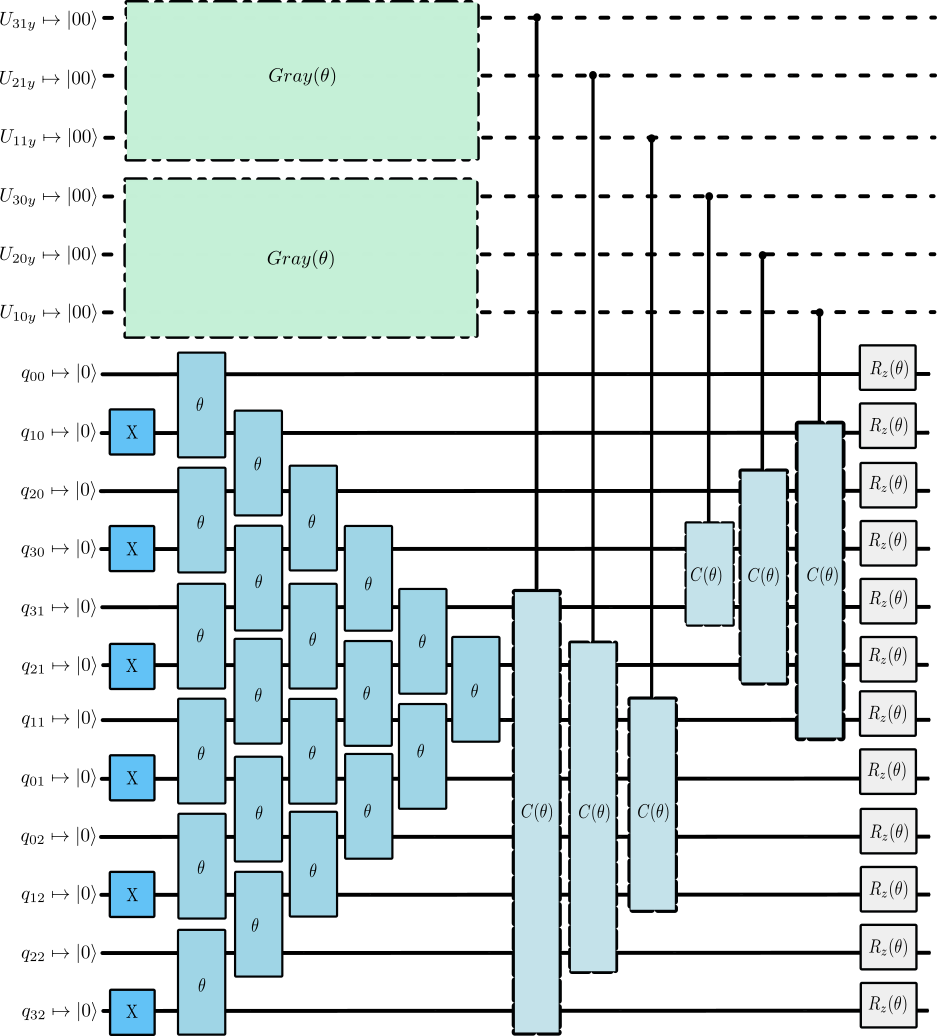}
    \caption[Parameterized quantum circuit for $4\times 3$ lattice]{\textbf{Parameterized quantum circuit for $4\times 3$ lattice:} The circuit has two main parts, the upper 12 qubits represent the six gauge fields $U_{31y},U_{21y},U_{11y},U_{30y},U_{20y},U_{10y}$ (each \textit{dashed horizontal line} is equal to two qubits) with relative entanglement structure $Gray(\theta)$, i.e. extended version of the structure applied to the $3\times 2$ lattice, Fig.~\ref{fig:qcircuit3x2}, up to 6 qubits. The lower 12 qubits show the fermions $q_{\vec{n}}$ ($\vec{n}=(n_x,n_y)$ site coordinates), starting in the vacuum configuration ($X$ gates). A cone shape layer of iSWAP gates (parameterized $\theta$ gates with \textit{solid lines}) and the entanglement (parameterized control gates $C(\theta)$ with \textit{dashed lines}) with the gauge fields are applied to the circuit. A final layer of rotational gates $R_z(\theta)$ correct eventual relative phases. Here, for simplicity, we have not numbered the $\theta$ parameters. }
    \label{fig:qcircuit4x3}
\end{figure}

The quantum circuit, depicted in Fig.~\ref{fig:qcircuit4x3}, is built with a similar structure of the smaller lattice $3\times 2$ (in Appendix~\ref{app:quantumcirc}). Specifically, we use the knowledge from the mutual information (MI) of smaller systems and we apply entangling gates accordingly.
We have tested that, for smaller systems, the higher values of MI were between the links belonging to the same plaquettes. Thus, to reduce the number of CNOT gates, we entangle the links $\{U_{11y},U_{21y},U_{31y}\}$ and $\{U_{10y},U_{20y},U_{30y}\}$, within the Gray encoding structure ($Gray(\theta)$ in the figure).

For the fermionic part, we start with the vacuum state, i.e. we apply $X$ \textit{gates} to have $\ket{1}$ on the odd fermionic sites ($\ket{0}$ on even sites). Then we apply the $i$SWAP structure in Fig.~\ref{fig:qcircuit4x3} (\textit{gates with $\theta$}). A set of C$i$SWAP gates ($C(\theta)$) is considered to entangle the links with the two sites at their edges: $U_{31y}\leftrightarrow \{ q_{31},q_{32}\}$, $U_{21y}\leftrightarrow \{ q_{21},q_{22}\}$, $U_{11y}\leftrightarrow \{ q_{11},q_{12}\}$, $U_{30y}\leftrightarrow \{ q_{30},q_{31}\}$, $U_{20y}\leftrightarrow \{ q_{20},q_{21}\}$ and $U_{10y}\leftrightarrow \{ q_{10},q_{11}\}$.
Lastly, a final layer of $R_z(\theta)$ is applied to the fermionic qubits to correct eventual relative phases.

\end{appendix}

\clearpage
\newpage
\widetext

\end{document}